\documentclass[aps,a4paper,twocolumn,11pt]{quantumarticle}
\pdfoutput=1
\usepackage[utf8]{inputenc}
\usepackage[english]{babel}
\usepackage[T1]{fontenc}
\usepackage{amsmath,amsfonts,amsbsy,amssymb,graphicx,enumerate,latexsym,bm,ulem,multirow,listings,here,qcircuit} 
\usepackage{hyperref}
\usepackage{tikz}
\usepackage{lipsum}
\usepackage{color}

\begin{document}                 
\title{Quantum-Error-Mitigation Circuit Groups for Noisy Quantum Metrology }   

\author{Yusuke~Hama}  
\affiliation{Quemix Inc., 2-11-2 Nihombashi, Chuo-ku, Tokyo 103-0027, Japan}  
\author{Hirofumi~Nishi}  
\affiliation{
Laboratory for Materials and Structures, Institute of Innovative Research, Tokyo Institute of Technology, Yokohama 226-8503, Japan}
\affiliation{Quemix Inc., 2-11-2 Nihombashi, Chuo-ku, Tokyo 103-0027, Japan} 


\begin{abstract}
Quantum technologies work by utilizing properties inherent in quantum systems such as quantum coherence and quantum entanglement and are expected to be superior to classical counterparts 
for solving certain problems in science and engineering. The quantum technologies are, however,  fragile against an interaction with an environment (decoherence) and in order to utilize them with high accuracy
we need to develop error mitigation techniques which reduce decoherence effects. In this work, we analyze quantum error mitigation (QEM) protocol for quantum metrology in the presence of quantum noise. 
We demonstrate the effectiveness of our QEM protocol by analyzing three types of quantum Fisher information (QFI), ideal (error-free) QFI, noisy (erroneous) QFI, and quantum-error-mitigated QFI, and show both analytically and numerically that the scaling behaviors of quantum-error-mitigated QFI with respect to the number of probes become restored to the those exhibited in the ideal quantum metrology.    
Our QEM protocol is constructed by an ensemble of quantum circuits, namely QEM circuit groups, 
and has advantages such that it can be applied to noisy quantum metrology for any type of initial state as well as any type of the probe-system Hamiltonian, and it can be physically implemented in any type of quantum device. Furthermore, the quantum-error-mitigated QFI become approximately equal to the ideal QFI for almost any values of physical quantities to be sensed. Our protocol enables us to use quantum entanglement as a resource to perform high-sensitive quantum metrology even under the influence of quantum noise. 
\end{abstract}

\maketitle

\section{Introduction}\label{Intro}   
Over the last few decades, our technologies for engineering and controlling  quantum systems including solid-state systems (superconducting circuits, nitrogen-vacancy (NV) centers in diamonds) 
as well as atomic-molecular and optical systems (trapped ions, cold atoms) have advanced rapidly \cite{linke2017experimental,SCQRPP2017,SCQNISQ20191,SCQARCMP2020,ZhugroupSQC2020,TsaigroupSCCQC2021,trappedionNISQ2019,rondin2014magnetometry}. 
Meanwhile, many types of quantum information processing protocols and technologies such as quantum computing \cite{QCFeynman,QCDeutsch,QCLloyd,
DiVincenzoQC,QCQINandC,QSRMP2014,linke2017experimental,SCQRPP2017,acin2018quantum,SCQNISQ20191,SCQARCMP2020,ZhugroupSQC2020,TsaigroupSCCQC2021,trappedionNISQ2019,QCchemistryRMP2020}, 
quantum communication and network \cite{acin2018quantum,gisin2007quantum,chen2021review,wei2022towards}, and quantum metrology (sensing) \cite{acin2018quantum,braunstein1994statistical,braunstein1996generalized,leibfried2004toward,giovannetti2004quantum,giovannetti2006quantum,giovannetti2011advances,ma2011quantum,toth2014quantum,degen2017quantum,pezze2018quantum,braun2018quantum,liu2019quantum,meyer2021fisher,danilin2021quantum},
have been proposed and intensively investigated. We are now in the era where we can demonstrate such quantum technologies using many kinds of physical systems. The essential feature of the quantum technologies is that they work by properties inherent in quantum systems such as quantum coherence and quantum entanglement, 
 and for solving certain problems in science and engineering they are expected to outperform the classical counterparts.   
 For instance, quantum computers can conduct prime factorization and database retrieval more efficiently than classical computers do \cite{QCQINandC}. 
 In quantum metrology  quantum entangled states like  GHZ states are used to create  sensitivities of parameters to be estimated (phase, magnetic field, etc.)   
 which are higher than those in the classical regime: the scaling behavior with respect to the number of probes is called the Heisenberg-limit scaling \cite{leibfried2004toward,giovannetti2004quantum,giovannetti2006quantum,giovannetti2011advances,ma2011quantum,toth2014quantum,degen2017quantum,pezze2018quantum,braun2018quantum,liu2019quantum,meyer2021fisher}. 
 
Although the quantum technologies exhibit such advantages, they are weak against quantum noise effects (decoherence) \cite{EkertgroupQCdissipation,resch2021benchmarking}. 
To overcome this difficulty,  in quantum computing many types of quantum error correcting (QEC) codes have been explored \cite{SCQRPP2017,SCQARCMP2020,ZhugroupSQC2020,TsaigroupSCCQC2021,ShorPRAQEC1995,QCQINandC,NemotogroupQEC,lidar2013quantum,QECRoffe}. 
Near-term quantum computers, however, have still being built as intermediate-scale devices yet and are fragile against quantum noise so-called noisy intermediate-scale quantum (NISQ) devices \cite{SCQARCMP2020,trappedionNISQ2019,PreskillNISQ2018}.
The QEC codes are not  harnessed in these machines and recently another type of technique for reducing quantum noise effects called quantum error mitigation (QEM) has been studied intensively \cite{QCchemistryRMP2020,hybridQCalgorithmJPSJ2021,EMPRL2017,EMNature2019,EMPRX2017,EMPRX2018,EMarxiv2018,PhysRevA.98.062339,song2019quantum,zhang2020error,mcardle2019error,jattana2020general,xiong2020sampling,zlokapa2020deep,EMPRA2021,CandSQEMPRAp2021,OttenGrayQEM1,OttenGrayQEM2,QSEQEM, CliffordQEM,LearningBasedQEM,VirtualDistillationQEM,koczor2021exponential,PRXQuantum.2.010316,piveteau2021error,lostaglio2021error,suzuki2022quantum,piveteau2022quasiprobability, pascuzzi2022computationally, takagi2021optimal, larose2022mitiq, koczor2021dominant,hama2022quantum,cai2022quantum}.
As similar to quantum computing, the investigation of quantum metrology in the presence of quantum noise is an important issue \cite{toth2014quantum,pezze2018quantum,degen2017quantum,huelga1997improvement,escher2011general,matsuzaki2011magnetic,demkowicz2012elusive,chin2012quantum,chaves2013noisy,kolodynski2013efficient,demkowicz2014using,alipour2014quantum,alipour2014quantum,ozaydin2014phase,jeske2014quantum,macieszczak2015zeno,brask2015improved,smirne2016ultimate,sekatski2017quantum,demkowicz2017adaptive,hou2017quantum,matsuzaki2018quantum,koczor2020variational,he2021quantum,long2022entanglement}.
For instance, when the GHZ state is subject to phase damping (dephasing) 
the sensitivity gets worse such that 
it does not show the Heisenberg-limit scaling but instead  the standard-quantum-limit (shot-noise-limit) scaling  \cite{pezze2018quantum,degen2017quantum,huelga1997improvement}.   
Another important theme is the reduction of quantum noise effects via QEC codes  \cite{degen2017quantum,dur2014improved,kessler2014quantum,arrad2014increasing,unden2016quantum,zhou2018achieving,shettell2021practical}.  
As both the QEC codes and the QEM have been utilized to reduce the noise effects in quantum computing, 
it is a natural attempt to apply QEM methods to improve sensitivities of noisy quantum metrology \cite{zhao2021error,yamamoto2022error}. 
The investigation of QEM protocols for noisy quantum metrology is important as follows.   
In quantum computing, to implement QEC codes we need large-scale quantum devices with having small error rates of gates below threshold so-called fault-tolerant quantum computers \cite{SCQRPP2017,SCQARCMP2020,ZhugroupSQC2020,TsaigroupSCCQC2021,ShorPRAQEC1995,QCQINandC,NemotogroupQEC,lidar2013quantum,QECRoffe}.
Currently, it is high challenging to build such devices and so does the engineering of large-scale quantum metrological devices in which QEC are feasible.    
On the other hand, as demonstrated in quantum computing QEM can be performed with near-term quantum (NISQ) devices \cite{song2019quantum,zhang2020error}, and similarly QEM protocol are expected to become implemented in quantum metrological devices and its validity can be tested experimentally.      
\begin{figure*}[!t] 
\centering
\includegraphics[width=0.9\textwidth]{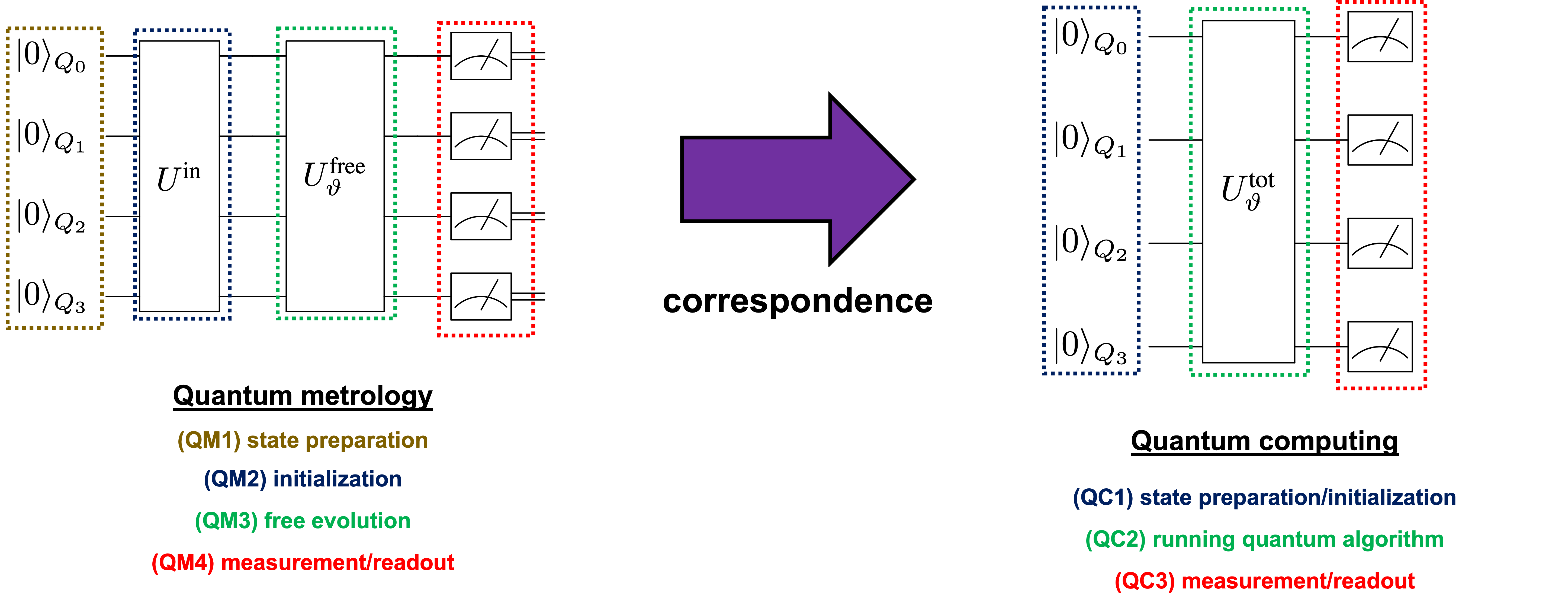}
\caption{Left schematic represents the whole procedure of quantum metrology which consists of four parts, (QM1) state preparation (preparing all-$|0\rangle$ states), (QM2)
initialization described by the unitary operation $U^\text{in}$, (QM3) free evolution given by $U^\text{free}_\vartheta,$
and (QM4) measurement/readout. The parameter of physical quantity to be estimated, $\vartheta,$ is encoded in the unitary operation $U^\text{free}_\vartheta.$        
The sensitivity of $\vartheta$ is described as its variance which is given by a quantum mechanical expectation value of a certain physical observable taken with respect to the quantum state generated by the total unitary operation $U^\text{tot}_\vartheta=U^\text{free}_\vartheta\cdot U^\text{in}$. Mathematically, the task of quantum metrology comprised of these four elementary procedures is equivalent to that of quantum computing given by the unitary operation $U^\text{tot}_\vartheta$ (quantum algorithm to be ran), the right figure.  }   
\label{Qmetrologyscheme} 
\end{figure*} 
In this work, we investigate our QEM protocol for noisy quantum metrology.
We discuss both analytically and numerically the validity of our QEM protocol by analyzing three types of quantum Fisher information (QFI),
ideal (noiseless) QFI,  noisy (erroneous) QFI, and  quantum-error-mitigated QFI. 
We do this for three types of initial states, coherent spin states (CSS), the GHZ states, and symmetric Dicke states (SDS), and examine various kinds of quantum noise channels which influence single-qubit states. Here we choose Markovian and non-Markovian phase damping (we abbreviate them as MPD and NMPD, respectively) and Markovian amplitude damping (MAD).
We show that the quantum-error-mitigated QFI approximately exhibit the same scaling behaviors with the ideal QFI with respect to the number of probes.
Namely, owing to our QEM protocol the sensitivities or the scaling behaviors of QFI become restored to those for the ideal quantum metrology cases.
 The characteristic of our QEM protocol is that it is composed of an ensemble of quantum circuits which we call QEM circuit groups.
 Since it is composed of quantum gates and quantum measurements, we can apply our QEM protocol to noisy quantum metrology for any type of initial probe state as well as any type of probe-system Hamiltonian and reduce any type of quantum noise, and it can be exploited by any type of quantum device. Furthermore, in contrast to the noisy QFI the quantum-error-mitigated QFI is approximately equal to the ideal QFI for almost any value of the physical quantity to be sensed. 
 The big advantage of our protocol is that quantum entanglement can be used as a resource for conducting high-sensitivity quantum metrology even under the influence of quantum noise. 

The structure of this paper is given as follows. 
It begins by Sec. \ref{basic} with presenting the basic theory of quantum metrology.
 In Sec. \ref{NQMQEM} we explain the formulation of noisy quantum metrology with presenting the details of quantum noise channels, the MPD, the MAD, and the NMPD, and give the explanation of our QEM protocol.    
In Sec. \ref{nss},  we show the numerical simulations of our QEM protocol and discuss in detail the scaling behavior of QFI with respect to the number of probes and the validity of our protocol.
In Sec. \ref{discussions}, we discuss characteristics of our QEM protocol. First, we discuss the three QFI dependencies on the parameter to be estimated and describe the advantages of our protocol.
Second, we make comparisons between our protocol and other QEM methods \cite{QCchemistryRMP2020,hybridQCalgorithmJPSJ2021,EMPRL2017,EMPRX2018,song2019quantum,zhang2020error,xiong2020sampling,CandSQEMPRAp2021,LearningBasedQEM,piveteau2021error,piveteau2022quasiprobability,takagi2021optimal,larose2022mitiq,cai2022quantum,song2019quantum,zhang2020error}.  
Third, we examine the efficacy of our protocol under NISQ-device parameters such as decay rates and gate times. 
Sec. \ref{conclusions} is devoted to the conclusion of this paper. 

\section{Basic Theory}\label{basic}  
\subsection{Modeling}\label{model}  
We introduce $N_q$ identical probe quantum systems for quantum metrology and assume all of them to be qubits. In the following, we just call the probes as qubits. 
Let us denote the $j$-th qubit by $Q_j$ ($j=0,1,\ldots,N_q-1$) and we describe its quantum state by the computational basis $\{  |0\rangle_j,  |1\rangle_j \}$,
where $|0\rangle = (1,0)^\text{T}, |1\rangle = (0,1)^\text{T}$ and the superscript ``T" denotes the transpose.  
The $|0\rangle$ state is taken to be the ground state. 
The Hamiltonian describing the $N_q$ two-level system is given by
$H_0 = -\frac{\hbar \omega_{01}}{2} \sum_{j=0}^{N_q-1} Z_j,$ where $\omega_{01}$ is the frequency describing the energy-level splitting between the $|0\rangle$ and $|1\rangle$ states and  $Z_j$ is the $Z$ gate acting on the $j$-th qubit.
In this work, we formulate our theory in the interaction picture with respect to the Hamiltonian $H_0$ and $H_0$ does not explicitly appear in the formulation of  time evolution of the system. 
Next, we describe $d$ real parameters to be estimated by a vector $\vec{\vartheta} = (\vartheta_1, \vartheta_2, \ldots, \vartheta_d)$. They could be, for instance, magnetic field, frequency, or phase. 
For simplicity, in this work we consider the single parameter estimation ($d=1$) and hereinafter we just write the parameter to be estimated by $\vartheta.$
Let us explain the procedures of (ideal) quantum metrology with using the schematic of them presented in Fig. \ref{Qmetrologyscheme}. 
First, we prepare all-$|0\rangle$ states (procedure (QM1)). Second, we prepare initial quantum states of the qubits to be desired, e.g., coherent spin states (CSS) or quantum entangled states such as GHZ states or symmetric Dicke states (SDS) (procedure (QM2)), and let us denote the unitary operation for generating the initial state by $U^{ \text{in}}$ and write the time for completing the initialization by $T^\text{in}$.  The initial state is described by the state vector $| \psi^\text{in} \rangle = U^{ \text{in}} |0\rangle ^{\otimes N_q}  $ or by the density matrix $\rho^{\text{in}} =| \psi^\text{in} \rangle \langle \psi^\text{in}|$.  The unitary operation $U^\text{in} $ can be composed by quantum gates and assume that all gate operation times are equivalent and write it by $\Delta t$.  Namely, the unitary transformation $U^\text{in} $ is composed of $d^\text{in}$ quantum gates such that $T^\text{in} = d^\text{in} \Delta t.$
When the initial state $\rho^\text{in} $ is created, third, we freely time evolute the qubits for a time $T^\text{free}$ by a Hermitian operator (Hamiltonian) $H^\text{free}$ which includes the parameter $\vartheta$ (procedure (QM3)). This free time evolution is represented by the unitary operation $U^\text{free}_\vartheta = \exp\left( -i\vartheta H^\text{free}\right)$.
Like $U^ \text{in}$, we can express the free-evolution operator $U^\text{free}_\vartheta$ by quantum gates and let us denote the number of the gate operations for doing this by $d^\text{free}$  and hence $T^\text{free} = d^\text{free} \Delta t.$ 
For instance, when the free Hamiltonian is given by the the Zeeman-interaction Hamiltonian, $H^\text{free}_\vartheta = -\frac{\hbar \gamma_g B}{2} \sum_{j=0}^{N_q-1} Z_j$, where $\gamma_g$ and $B$ are the gyromagnetic ratio of the qubit systems and the magnitude of the magnetic field to be sensed, respectively. 
In this case, $H^\text{free} = \sum_{j=0}^{N_q-1} \frac{Z_j}{2}$ and $\vartheta= \gamma_g B T^\text{free},$ and essentially the estimation of the parameter $\vartheta$ is equivalent to that of the magnetic-field strength $B$, i.e., magnetic-field quantum sensing.   
The unitary operation $U^{ \text{free}}_{\vartheta}$ is expressed as the tensor product of rotation gates as   $U^\text{free}_{\vartheta} = \bigotimes_{j=0}^{N_q-1} R^z_j(-\vartheta)$, where $R^z_j(-\vartheta)$ describes the rotation about $z$ axis (quantization axis of the qubit) with the rotation angle $-\vartheta.$
As a result, we obtain the output state $\rho^\text{out}_\vartheta =  |\psi^\text{out}_\vartheta\rangle \langle \psi^\text{out}_\vartheta|=U^\text{free}_\vartheta \rho^\text{in} (U^\text{free}_\vartheta)^\dagger$.
In other words, the quantum metrological task is performed by the total unitary operation $U^\text{tot}_\vartheta = U^\text{free}_\vartheta \cdot U^\text{in}$ which is constructed by the $  d^\text{tot} = d^\text{free} + d^\text{in}$ gate operations and the total time of the quantum metrological task is $T^\text{tot} = T^\text{free} + T^\text{in} = d^\text{tot}  \Delta t $. 
The parameter $\vartheta$ is encoded in the output state $\rho^\text{out}_\vartheta $.
Finally, we perform a quantum measurement (readout) represented by positive operator-valued measurement (POVM) $E_k$, where $k=1,2,\ldots, N_E$ with $N_E$ denoting the number of different POVM  (procedure QM4), and repeat these four procedures. Consequently, we obtain the probability distribution (eigenspectrum) associated with the POVM $E_k$.
As shown in the right-hand side of Fig. \ref{Qmetrologyscheme} such a way of the formulation of quantum metrology 
is equivalent to the mathematical treatment of gate-based quantum computing. Namely, the quantum metrological task given by the unitary operations  $U^\text{in}$ and $U^\text{free}_\vartheta$ is equivalent to the quantum computing with the quantum algorithm described by the unitary operation $U^\text{tot}_\vartheta$. As we demonstrate in Secs. \ref{NQMQEM} and \ref{nss} such a mathematical description enables us to treat both analytically and numerically noisy quantum metrology as well as quantum error mitigation by mathematical methods used in noisy quantum computing \cite{QCchemistryRMP2020,hybridQCalgorithmJPSJ2021,EMPRL2017,EMNature2019,EMPRX2017,EMPRX2018,EMarxiv2018,PhysRevA.98.062339,song2019quantum,zhang2020error,mcardle2019error,jattana2020general,xiong2020sampling,zlokapa2020deep,EMPRA2021,CandSQEMPRAp2021,OttenGrayQEM1,OttenGrayQEM2,QSEQEM, CliffordQEM,LearningBasedQEM,VirtualDistillationQEM,koczor2021exponential,PRXQuantum.2.010316,piveteau2021error,lostaglio2021error,suzuki2022quantum,piveteau2022quasiprobability, pascuzzi2022computationally, takagi2021optimal, larose2022mitiq, koczor2021dominant,hama2022quantum,cai2022quantum}. 

\subsection{Parameter Estimation and QFI}
Let us now discuss the way to estimate the parameter $\vartheta$.
 We denote POVM acting on the $j$-th probe and the associated measurement outcome by  ${E_{k_j}}$ ($k_j =1,\ldots, N_{\text{E}})$  and $x_{k_j}$, respectively,
 and write the sequence of the measurement outcome by the vector $\boldsymbol{x}_{\boldsymbol{k}} = (x_{k_0},\ldots, x_{k_{N_q-1}})^{\text{T}}$.
 Here we have introduced the notations $\boldsymbol{k}=(k_0,\ldots,k_{N_q-1})$ and $\boldsymbol{E}_{\boldsymbol{k}}=( E_{k_0},\ldots, E_{k_{N_q-1}})$.
 By taking into account of the probabilistic characteristic of the quantum theory, we regard the measurement outcome of the POVM ${E_{\boldsymbol{k}}}$  as the sequence of random variables
 $\boldsymbol{X}^{ (N_q) }_{\boldsymbol{E}_{\boldsymbol{k}}} = \left(X^{E_{k_0}}_{0}, \ldots, X^{E_{k_{N_q-1}}}_{{N_q}-1}\right)$, where $X^{E_{k_j}}_j=0,1$ with $j=0,1,\ldots,N_q-1.$
 Here the measurement outcome describes the observed quantum state of the $j$th qubit  which is either $|0\rangle$ or $|1\rangle$.
 In other words, the observed quantum state is represented as the quantum state labeled by the $N_q$-bit strings as $\Big{|} X^{E_{k_0}}_{0}, \ldots, X^{E_{k_{N_q-1}}}_{{N_q}-1}      \Big{\rangle}$ and consider them as the sequence of random variables. 
 We write the probability such that the outcome of the $j$-th probe is $x_{k_j}$ by 
 $p_{\vartheta}\left[ X^{E_{k_j}}_j=x_{k_j}\right] $, which is given by $p_{\vartheta}\left[ X^{E_{k_j}}_j=x_{k_j}\right]  = \text{Tr} \left[  \rho^{\text{out}} E_{k_j}  \right]$
 with $\rho^{\text{out}} $ denoting an output state.
 Note that we write the random variables with capital letters while we write the outcome (output data) with lowercase letters. 
  The probability of getting the outcome $\boldsymbol{x}_k$ is given by 
   \begin{align}
P_{\vartheta}\left[\boldsymbol{X}^{(N_q)}_{\boldsymbol{E}_{\boldsymbol{k}}} = \boldsymbol{x}_k\right] = \text{Tr}\left[  \rho^{\text{out}}_{\vartheta}   \bigotimes_{j=0}^{ N_q-1 }    E_{k_j}     \right].
 \label{jointprob1}
\end{align}
When the output state  $\rho^\text{out}_\vartheta $  is  a separable state we can write it as $\rho^{\text{out}}_{ \vartheta } = \bigotimes_{j=0}^{N_q-1} \rho^{\text{out}}_{\vartheta, j}$
 and the probability $ P_{\vartheta}\left[\boldsymbol{X}^{(N_q)}_{\boldsymbol{E}_{\boldsymbol{k}}} = \boldsymbol{x}_k\right]  $ can be expressed as
  \begin{align}
P_{\vartheta}\left[\boldsymbol{X}^{(N_q)}_{\boldsymbol{E}_{\boldsymbol{k}}}  = \boldsymbol{x}_k\right]  &=  \text{Tr}\left[  \bigotimes_{j=0}^{N_q-1 }  \rho^{\text{out}}_{ \vartheta,j}   E_{k_j}     \right] \notag\\
&= \prod_{j=0}^{N_q-1 }  p_{\vartheta}\left[ X^{E_{k_j}}_j =x_{k_j}\right].
 \label{jointprob2} 
\end{align}\normalsize
The above equation is going to be used when we analyze the quantum metrology for the initial state taken to be CSS. 

 Next, we introduce an estimator which is a functional of the output state $\rho^{\text{out}}_{ \vartheta } $ and assumed to be an unbiased estimator denoted by
 $\Theta^{\text{est}} \left[ \boldsymbol{X}^{ (N_q) }_{\boldsymbol{E}_{\boldsymbol{k}}}; \rho^{\text{out}}_{\vartheta} \right]$. By definition, we have 
 \begin{widetext}
  \begin{align}
 &  \text{Exp}\left[ \Theta^{\text{est}} \left[ \boldsymbol{X}^{ (N_q) }_{\boldsymbol{E}_{\boldsymbol{k}}}; \rho^{\text{out}}_{\vartheta} \right ] \right]  =
  \sum_{ \boldsymbol{X}^{ (N_q) }   }  P_{\vartheta} \left[ \boldsymbol{X}^{ (N_q) }_{\boldsymbol{E}_{\boldsymbol{k}}} \right]  \cdot 
 \Theta^{\text{est}} \left[ \boldsymbol{X}^{ (N_q) }_{\boldsymbol{E}_{\boldsymbol{k}}}; \rho^{\text{out}}_{\vartheta} \right ]  
 = \vartheta,  \label{unbiasedaverage} \\
 & \text{Var} [\vartheta] =   \sum_{ \boldsymbol{X}^{ (N_q) }_{\boldsymbol{E}_{\boldsymbol{k}}}   }  P_{\vartheta} \left[\boldsymbol{X}^{ (N_q) }_{\boldsymbol{E}_{\boldsymbol{k}}} \right]  
  \cdot  \left( 
   \Theta^{\text{est}} \left[ \boldsymbol{X}^{ (N_q) }_{\boldsymbol{E}_{\boldsymbol{k}}}; \rho^{\text{out}}_{\vartheta} \right ]    - \vartheta  \right)^2,
  \label{variancedef1}
   \end{align} \end{widetext}
where ``Exp" and ``Var" are the notations for denoting an expectation value and a variance, respectively.  
 We describe that the sensitivity (accuracy) of estimating the parameter $\vartheta$ is high when the variance $ \text{Var} [\vartheta]$ in Eq. \eqref{variancedef1} is small and is represented by a quantum mechanical expectation value 
 via the probability $P_{\vartheta} \left[ \boldsymbol{X}^{ (N_q) }_{\boldsymbol{E}_{\boldsymbol{k}}} \right] $.  It is lower bounded by the quantum Cramer-Rao inequality \cite{braunstein1994statistical,braunstein1996generalized,leibfried2004toward,giovannetti2004quantum,giovannetti2006quantum,
giovannetti2011advances,ma2011quantum,toth2014quantum,degen2017quantum,pezze2018quantum,braun2018quantum,liu2019quantum,meyer2021fisher},
\begin{align}
 \text{Var} [\vartheta] \geq  \frac{1}{ \nu I_{\text{CF}} \left[ \boldsymbol{E}_{\boldsymbol{k}}; \rho^{\text{out}}_{\vartheta}\right] }   
\geq   \frac{1}{\nu I_{\text{QF}} \left[ \rho^{\text{out}}_{ \vartheta}  \right]}, \label{QCRineq}
\end{align} 
 where $ I_\text{CF} \left[\boldsymbol{E}_{\boldsymbol{k}}; \rho^\text{out}_\vartheta \right]$ and $I_\text{QF} \left[ \rho^\text{out}  \right]$ are
 the classical Fisher information (CFI) and the quantum Fisher information (QFI) \cite{braunstein1994statistical,braunstein1996generalized,leibfried2004toward,giovannetti2004quantum,giovannetti2006quantum, 
giovannetti2011advances,ma2011quantum,toth2014quantum,degen2017quantum,pezze2018quantum,braun2018quantum,liu2019quantum,meyer2021fisher,rath2021quantum,yu2022quantum}, respectively.   
  We have presented both $\boldsymbol{E}_{\boldsymbol{k}}$ and  $\rho^\text{out}_\vartheta $ for writing the argument of $ I_\text{CF} \left[\boldsymbol{E}_{\boldsymbol{k}}; \rho^\text{out}_\vartheta \right]$ 
since it depends on both of these quantities while for that of $I_\text{QF} \left[ \rho^\text{out}_\vartheta  \right]$ we have just written the output state  $ \rho^\text{out}_ \vartheta $ since it does not depend on  $\boldsymbol{E}_{\boldsymbol{k}}$.
Note that when the output state $\rho^\text{out}_\vartheta $  is  a separable state the CFI of the $j$-th qubit $I_\text{CF} \left[E_{k_{j}};  \rho^\text{out}_{ \vartheta,j}  \right]$  as well as the QFI of the $j$-th qubit
 $I_\text{QF} \left[ \rho^\text{out}_{ \vartheta,j }  \right]$  are both equivalent for any $j$. Hence, we obtain
 $ I_\text{CF} \left[\boldsymbol{E}_{ \boldsymbol{k} }; \rho^\text{out}_{\vartheta}\right] = N_q  I_\text{CF} \left[E_{k_j};  \rho^\text{out}_{ \vartheta,j}  \right]$ 
 for $E_{k_0} = E_{k_1} = \cdots = E_{k_{N_q-1}} $
 and $I_\text{QF} \left[ \rho^\text{out}_\vartheta  \right] = N_q  I_{\text{QF}} \left[  \rho^\text{out}_{ \vartheta,j}  \right]$ .
 The mathematical representations of $ I_\text{CF} \left[E_{\boldsymbol{k}};  \rho^\text{out} \right]$ and $I_\text{QF} \left[  \rho^\text{out}_\vartheta  \right]$ for such a case are given by \cite{braunstein1994statistical,braunstein1996generalized,
giovannetti2011advances,ma2011quantum,toth2014quantum,degen2017quantum,pezze2018quantum,braun2018quantum,liu2019quantum,meyer2021fisher}
   \footnotesize  \begin{align}
 & I_\text{CF} \left[\boldsymbol{E}_{\boldsymbol{k}};  \rho^\text{out}_\vartheta  \right] =\sum_{j=0}^{ N_q-1  } \sum_{k_j}
  \frac{1}{  p_{\vartheta}\left[ X^{E_{k_j}}_j = x_{k_j}\right] } 
  \notag\\ & \times \left(  \frac{d   }{d\vartheta}   p_{\vartheta}\left[ X^{E_{k_j}}_j = x_{k_j}\right]  \right)^2, \label{defCFI} \\
 &  I_{\text{QF}} \left[ \rho^\text{out}_ \vartheta  \right] = \text{Tr} \left[  \rho^\text{out}_\vartheta 
   L^2\left[  \rho^\text{out}_\vartheta \right]     \right],  \label{defQFI} 
\end{align}\normalsize
 where $L\left[ \rho^\text{out}_\vartheta  \right]$ is the symmetric logarithmic derivative.
It is Hermitian and satisfies 
 \begin{align}
  \frac{\partial \rho^\text{out}_\vartheta} {\partial \vartheta}  = \frac{1}{2}\left[
 \rho^\text{out}_\vartheta    L \left[  \rho^\text{out}_\vartheta \right]  +
 \rho^\text{out}_\vartheta    L \left[  \rho^\text{out}_\vartheta  \right] \right].  \label{SLDprop} 
\end{align}
\begin{figure*}
\centering
\includegraphics[width=0.70\textwidth]{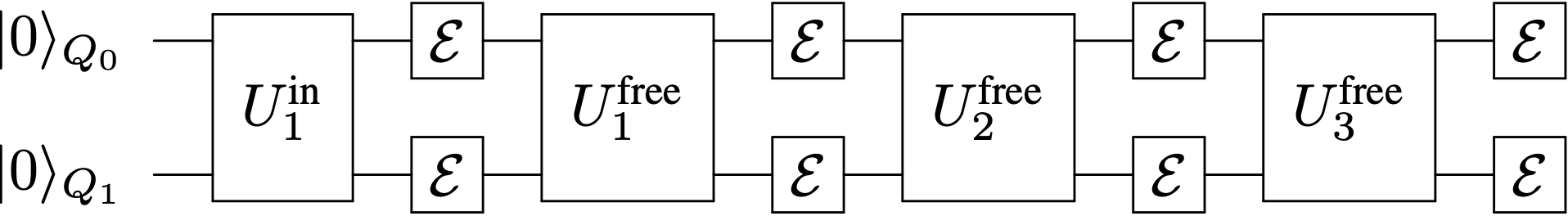}
\caption{Illustration of noisy quantum metrology represented as a quantum circuit. Here we show it for $d^\text{in}=1,d^\text{free}=3$ and $N_q =2.$ 
The symbol $\mathcal{E}$ denotes the action of quantum noise.   }  
\label{noisyQmetrologycircuit} 
\end{figure*} 
 Let us represent  $ \rho^\text{out}_\vartheta $ in a spectral decomposition form,  
 $ \rho^\text{out}_\vartheta  =  \sum_{j_\vartheta} \lambda _{j_\vartheta} | \lambda _{j_\vartheta} \rangle \langle \lambda _{j_\vartheta} |$,
 where  $\lambda _{j_\vartheta}$ and $ | \lambda _{j_\vartheta} \rangle$ denote the eigenvalues and the associated eigenvectors, respectively.
The  QFI $ I_\text{QF} \left[  \rho^\text{out} _\vartheta \right]$ can be expressed in terms of these quantities as \cite{braunstein1994statistical,ma2011quantum,toth2014quantum,degen2017quantum,pezze2018quantum,braun2018quantum,liu2019quantum,meyer2021fisher}, \begin{align}
I_\text{QF}\left[  \rho^\text{out}_\vartheta   \right]& =\sum_{\substack{ (j_\vartheta, k_\vartheta)\\ \lambda _{j_\vartheta}+\lambda _{k_\vartheta}>0}} 
\frac{2}{\lambda _{j_\vartheta}+\lambda _{j_\vartheta}}  \big{|}  \langle  \lambda _{k_\vartheta} |  \partial_\vartheta  \rho^\text{out}_\vartheta  |
  \lambda _{j_\vartheta}   \rangle \big{|}   ^2 . \label{QFIexpression1} 
\end{align}
For a pure state $ | \psi_\vartheta \rangle$, we have $ I_\text{QF} \left[  \rho^\text{out}_\vartheta  \right]  =  4 \left(
 \big{|} \partial_\vartheta \psi_\vartheta \rangle \big{|}^2-  \big{|} \langle  \psi_\vartheta | \partial_\vartheta  \psi_\vartheta \rangle \big{|}^2
  \right), $  where $ \big{|} \partial_\vartheta \psi_\vartheta \rangle =  \partial_\vartheta \big{|} \psi_\vartheta \rangle.$
  Furthermore, it is represented by the Hermitian operator $H$ (recall that $U^\text{free}_\vartheta = \exp\left( -i\vartheta H\right)$)  as  
  $ I_\text{QF} \left[  \rho^\text{out}_\vartheta   \right]  =  4\big{(} \langle H^2 \rangle_{\rho^\text{in}} - \langle H \rangle_{\rho^\text{in}}^2 \big{)}, $
  where $\langle A \rangle_{\rho^\text{in}} = \text{Tr}(\rho^\text{in} \cdot A)$ with $A=H,H^2.$
  In addition to Eq. \eqref{QFIexpression1}, the QFI $I_\text{QF} \left[  \rho^\text{out}_\vartheta   \right]$ is described in an alternative way as \cite{pezze2018quantum,braun2018quantum,koczor2020variational}
 \begin{align}
I_\text{QF} \left[  \rho^\text{out}_\vartheta  \right] = \lim_{\vartheta \to 0} \frac{8\left(1- \mathcal{F}\left[  \rho^\text{out}_{ \vartheta=0} , \rho^\text{out}_ \vartheta  \right]\right)}{\vartheta^2} , \label{QFIexpression2} 
\end{align}
where 
 \begin{align}
\mathcal{F}\left[  \rho^\text{out}_{ \vartheta=0} , \rho^\text{out}_\vartheta  \right]= \text{Tr} \left[ \left( (\rho^\text{out}_{ \vartheta=0} )^{\frac{1}{2}}
    \rho^\text{out}_\vartheta      (\rho^\text{out}_{ \vartheta=0} )^{\frac{1}{2}} \right)^{\frac{1}{2}}\right],   \label{defFidelity} 
\end{align}
     is the 
  fidelity of  the quantum states $ \rho^\text{out}_{ \vartheta=0}  $ and $ \rho^\text{out}_\vartheta.  $
  
  Before ending this subsection, we note that the quantum metrology for estimating a parameter $\omega = \frac{\vartheta}{T^\text{free}}$ can also be done, and its sensitivity
  Var$[\omega]$ is related to Var$[\vartheta]$ as  Var$[\vartheta]= (T^\text{free})^2$Var$[\omega]$ and correspondingly we have 
  $I_\text{QF} \left[  \rho^\text{out}_\omega  \right] =  (T^\text{free})^2I_\text{QF} \left[  \rho^\text{out}_\vartheta  \right],$
  where both the QFI $I_\text{QF} \left[  \rho^\text{out}_\omega  \right]$ (or the CFI $I_\text{CF} \left[\boldsymbol{E}_{\boldsymbol{k}};  \rho^\text{out}_\omega  \right] $) and $I_\text{QF} \left[  \rho^\text{out}_\vartheta  \right]$ (or the CFI $I_\text{CF} \left[\boldsymbol{E}_{\boldsymbol{k}};  \rho^\text{out}_\vartheta  \right] $)  is calculated with the same quantum state, i.e., 
  $\rho^\text{out}_\omega \equiv \rho^\text{out}_\vartheta.$ 
  As we discuss in  Sec. \ref{nss} and  Sec. \ref{thetadependence}, 
 the QFI (CFI) of $\vartheta$ is described as the scaling in $N_q$ while the scaling of the QFI (or CFI) of $\omega$ is expressed by both $N_q$ and $T^\text{free}$. 
 In this work we focus on the quantum metrology of $\vartheta$ (phase estimation) and explore the quantum-error mitigation protocol which enables to restore the ideal scaling of QFI in $N_q$.  

\section{Noisy Quantum Metrology and QEM Protocol}\label{NQMQEM}   
\subsection{Noisy Quantum Metrology}\label{noisyQM} 
Since we have given above the modeling of the ideal quantum metrology, let us now discuss a modeling of noisy quantum metrology.  
To do this, we introduce a notation for the elementary gate operations which constitute the unitary operation $U^\text{tot}_{\vartheta} $  and write them by $U^\text{tot}_{\vartheta,l} $ with $l=1,\ldots,N^\text{tot}:$
$U^\text{tot}_{\vartheta}=  U^\text{tot}_{\vartheta,d^\text{tot}} \cdots  U^\text{tot}_{\vartheta,1}=    \prod_{l=1}^{d^\text{tot}}   U^\text{tot}_{\vartheta,l} $.
 For simplicity, we consider that the qubits are subject to a single kind of quantum noise which acts on them  independently and homogeneously and call it single-qubit-state quantum noise. 
The examples include dephasing (PD), amplitude damping (AD), and depolarizing channel. 
Let us mathematically express the noisy quantum dynamics (time evolution under quantum noise) of the qubits such that for every time step $\Delta t$ 
they are subject to the single-qubit-state quantum noise under consideration. By introducing the notation $ \mathcal{T}[\rho, U]= U\rho U^\dagger$ and denoting the superoperator describing the action of the quantum noise during a time interval $[t, \tilde{t}]$ by $\mathcal{E}_{\tilde{t}, t}$, the noisy output state of the qubits is represented by the density matrix \cite{hybridQCalgorithmJPSJ2021,CandSQEMPRAp2021,OttenGrayQEM1, koczor2021dominant,Qiskit,Ais2014QuantumIA}
\newpage
\begin{widetext}
\begin{align}
\rho^\text{out}_\vartheta  =            \mathcal{E}_{t_{d_\text{tot}},t_{d_\text{tot}-1}} \Big{[}  \mathcal{T}\big{[}  \mathcal{E}_{ t_{d_\text{tot}-1},t_{d_\text{tot}-2} } \Big{[} \mathcal{T} \big{[} \cdots  \mathcal{E}_{t_2,t_1} \Big{[} \mathcal{T}\big{[}\mathcal{E}_{t_1,t_0} \Big{[}   \mathcal{T}\big{[} \rho(t_0), U^\text{tot}_{\vartheta,1}\big{]} \Big{]}, U^\text{tot}_{\vartheta,2} \big{]} \Big{]}\cdots, U^\text{tot}_{\vartheta, d_\text{tot}-1 }\big{]} \Big{]}, U^{ \text{free}}_{\vartheta,N_{\text{free}} } \big{]} \Big{]}  ,
\label{noisyoutput}
\end{align}\end{widetext}
where $t_\alpha = \alpha\Delta t$ ($\alpha=0,\ldots, d_\text{tot}$) and $\rho(t_0)= (|0\rangle \langle 0|)^{\otimes N_q} $.
In Fig \ref{noisyQmetrologycircuit},  we show the schematic of noisy quantum metrology described as the quantum circuit based on Eq. \eqref{noisyoutput}. In the following we discuss in detail  three types of quantum noise channels which we choose for our noisy quantum noisy simulations, Markovian phase damping (MPD), Markovian amplitude damping (MAD), and non-Markovian phase damping (NMPD): for  examples of non-Markovian AD see for instance \cite{ban2005decoherence}.  
Note that all the quantum master equations given below are described in the interaction picture.
\subsubsection{Markov Phase Damping}\label{MPD} 
The MPD process is described by the quantum master equation 
 \begin{widetext}\begin{align}
 \frac { \partial \rho_\text{MPD} (t)}{\partial t }  =  \frac{\gamma_\text{PD}}{2} \sum_{j=0}^{N_q-1} \mathcal{L} _{\text{PD},j} [\rho_\text{MPD} (t)] =  \frac{\gamma_\text{PD}}{2} \sum_{j=0}^{N_q-1}  \left[ Z_j    \rho_\text{MPD}(t)   Z_j   -  \rho_\text{MPD} (t)   \big{\}} \right],
 \label{MPDQME1}
\end{align}  \end{widetext}
where $\gamma_\text{PD}$ denotes the decay rate and $ \mathcal{L} _{\text{PD},j}$ is the Liouvillian operator of the MPD acting on the $j$th qubit. 
The density matrix $\rho_\text{MPD} (t)$ is the solution of the above equation and represents the noisy quantum state induced by the MPD  at time $t$. 
The quantum master equation  \eqref{MPDQME1} can be formally solved as
\begin{footnotesize}
\begin{widetext}  \begin{align}
 \rho_\text{MPD} (t)  & = \mathcal{E}^\text{MPD}_{t,t_0} [\rho (t_0)] =  \exp\left[ \frac{\gamma_\text{PD}}{2}t \sum_{j=0}^{N_q-1} \mathcal{L} _{\text{PD},j} \right] [\rho (t_0)] \notag\\
 &=  \sum_{\alpha_0=0}^1    \cdots \sum_{\alpha_{N_q-1}=0}^1  
  \left( \mathcal{M}^\text{MPD}_{\alpha_0}(t,t_0) \otimes \cdots \otimes \mathcal{M}^\text{MPD}_{\alpha_{N_q}-1} (t,t_0) \right)
  \rho(t_0)  \left( \mathcal{M}^\text{MPD}_{\alpha_0}(t,t_0) \otimes \cdots \otimes \mathcal{M}^\text{MPD}_{\alpha_{N_q-1}} (t,t_0) \right)^\dagger, 
 \label{MPDQME2}
\end{align} \end{widetext}\end{footnotesize}
where $\mathcal{M}^\text{MPD}_{\alpha_j}(t,t_0) $ ($j=0,\ldots,N_q-1 $) is the Kraus operator associated with the MPD acting on the $j$-th qubit:
$\mathcal{M}^\text{MPD}_0(t,t_0)  = \sqrt{\frac{1+e^{-\gamma_\text{PD}t}}{2}} \boldsymbol{1}_{2\times2}, \mathcal{M}^\text{MPD}_1(t,t_0)  = \sqrt{\frac{1-e^{-\gamma_\text{PD}t}}{2}} Z. $
The Kraus operators $\mathcal{M}^\text{MPD}_{\alpha_j}(t) $ ($j=0,\ldots, N_q-1$)  satisfy
 $\sum_{\alpha_j=0} ^1    \big{(}\mathcal{M}^\text{MPD}_{\alpha_j}\big{)}^\dagger(t)  \mathcal{M}_{\alpha_j} (t) =\boldsymbol{1}_{2\times2, j}$ with $\boldsymbol{1}_{2\times2, j}$ denoting the identity operator acting on the $j$-th qubit.
 We denote a density matrix of the $j$th qubit describing a quantum state under the influence of the MPD by $ \rho_{\text{MPD},j} (t)$ and the  matrix representation is given by
\begin{widetext} \begin{align}
 \rho_{\text{MPD},j} (t) = \sum_{\alpha_j=0}^1     \left( \mathcal{M}^\text{MPD}_{\alpha_j}(t,t_0) \right)  \rho_j(t_0) \left( \mathcal{M}^\text{MPD}_{\alpha_j} (t,t_0) \right)^\dagger
  =  \left (
		\begin{array}{cc} 
		 \left[ \rho_{j}(t_0)  \right]_{00}  &  \left[ \rho_{j}(t_0)  \right]_{01}  e^{-\gamma_\text{PD}t }  \\
		 \left[ \rho_{j}(t_0)  \right]_{10}  e^{-\gamma_\text{PD}t } & \left[ \rho_{j}(t_0)  \right]_{11}
		\end{array}
	\right ),
\label{MPDQS1}
\end{align} \end{widetext}
where $\rho_{j}(t_0) = | 0\rangle_{j}\langle0 |$ and $ \left[ \rho_{j}(t_0)  \right]_{n_jn^\prime_{j}}$ ($n_j,n^\prime_{j}=0,1$)
is the $n_{j},n^\prime_{j}$-th element of the density matrix $\rho_{j}(t_0) $.  
The MPD (or more broadly Markovian quantum noises) satisfies the divisibility $ \mathcal{E}^\text{MPD}_{t_1+t_2,t_0} = \mathcal{E}^\text{MPD}_{t_1+t_2,t_1}  \cdot \mathcal{E}^\text{MPD}_{t_1,t_0} $.
Owing to this property, the relation between $\rho_\text{MPD}\big{(} t_{l+1} \big{)}$ and $\rho_\text{MPD}(t_l)$ ($l=0,\ldots,d_\text{tot}-1 $) is given by  \begin{footnotesize}
\begin{widetext} \begin{align}
\rho_\text{MPD}\big{(} t_{l+1}  \big{)} = \sum_{\alpha_0=0}^1  \cdots \sum_{\alpha_{N_q-1}=0}^1  
  \left( \mathcal{M}^\text{MPD}_{\alpha_0}(\Delta t) \otimes \cdots \otimes \mathcal{M}^\text{MPD}_{\alpha_{N_q}-1} (\Delta t) \right)
  \rho_\text{MPD}(t_l)  \left( \mathcal{M}^\text{MPD}_{\alpha_0}(\Delta t) \otimes \cdots \otimes \mathcal{M}^\text{MPD}_{\alpha_{N_q-1}} (\Delta t) \right)^\dagger.
\label{MPDQS2}
\end{align} \end{widetext}\end{footnotesize}
From Eqs. \eqref{noisyoutput} and \eqref{MPDQS2} we obtain the output quantum state influenced by the MPD.
\subsubsection{Markov Amplitude Damping}\label{MAD} 
The quantum master equation describing the MAD process is
\begin{widetext}\begin{align}
\frac { \partial \rho_\text{MAD} (t)}{\partial t }  =  \gamma_\text{AD} \sum_{j=0}^{N_q-1} \mathcal{L} _{\text{AD},j} [\rho_\text{MAD} (t)] 
 =  \gamma_\text{AD}  \sum_{j=0}^{N_q-1} \Big{[}  \tilde{\sigma}^-_j    \rho_\text{MAD} (t)   \tilde{\sigma}^+_j  
 - \frac{1}{2} \big{\{}   \tilde{\sigma}^+_j  \tilde{\sigma}^-_j,  \rho_\text{MAD} (t)   \big{\}} \Big{]} ,
 \label{MADQME1}
\end{align}  \end{widetext}
where the second term of the right-hand side of Eq. \eqref{MADQME1} is the aniti-commutator between $\tilde{\sigma}^+_j\tilde{\sigma}^-_j $ and $  \rho_\text{MAD} (t) $
and it is defined by $\big{\{} A, B \big{\}} = AB+BA$ with $A$ and $B$ are operators.
$ \tilde{\sigma}^\mp_j = \frac{X_j \pm iY_j}{2}$ are the raising and lowering operators with $X_j$ and $Y_j$ denoting the $X$- and $Y$-gate operations acting on $Q_j$, respectively.
Note that $ \tilde{\sigma}^-\tilde{\sigma}^+ = P^0, \tilde{\sigma}^+\tilde{\sigma}^- =P^1 $, where $P^0$ and $P^1$ are the projection operators of $|0\rangle$ and $|1\rangle$, respectively.
As similar to Eq. \eqref{MPDQME2}, the solution of the quantum master equation \eqref{MADQME1} is given by
\begin{footnotesize}
\begin{widetext}  \begin{align}
 \rho_\text{MAD} (t)  & = \mathcal{E}^\text{MAD}_{t,t_0} [\rho (t_0)] =  \exp\left[ \gamma_\text{AD}t  \sum_{j=0}^{N_q-1} \mathcal{L} _{\text{AD},j} \right] [\rho (t_0)] \notag\\
 &=  \sum_{\alpha_0=0}^1    \cdots \sum_{\alpha_{N_q-1}=0}^1  
  \left( \mathcal{M}^\text{MAD}_{\alpha_0}(t,t_0) \otimes \cdots \otimes \mathcal{M}^\text{MAD}_{\alpha_{N_q}-1} (t,t_0) \right)
  \rho(t_0)  \left( \mathcal{M}^\text{MAD}_{\alpha_0}(t,t_0) \otimes \cdots \otimes \mathcal{M}^\text{MAD}_{\alpha_{N_q-1}} (t,t_0) \right)^\dagger, 
 \label{MADQME2}
\end{align} \end{widetext}\end{footnotesize}
where $\mathcal{M}^\text{MAD}_{\alpha_j}(t,t_0) $ ($j=0,\ldots,N_q-1 $) is the Kraus operator describing the MAD acting on $Q_j$:
$\mathcal{M}^\text{MAD}_0(t,t_0)  = P_0 +  \sqrt{e^{-\gamma_\text{AD}t}}P^1, \mathcal{M}^\text{MAD}_1(t,t_0)  = \sqrt{1-e^{-\gamma_\text{AD}t}}\tilde{\sigma}^-$.
These Kraus operators satisfy  $\sum_{\alpha_j=0} ^1   \big{(}\mathcal{M}^\text{MAD}_{\alpha_j}\big{)}^\dagger(t) \mathcal{M}^\text{MAD}_{\alpha_j} (t) =\boldsymbol{1}_{2\times2, j}.$ 
The matrix representation of a density matrix $ \rho_{\text{MAD},j} (t) $ which describes a quantum state of $Q_j$ under the affect of the MAD at time $t$ is
 \begin{widetext}  \begin{align}
 \rho_{\text{MAD},j} (t)   = \sum_{\alpha_j=0}^1     \left( \mathcal{M}^\text{MAD}_{\alpha_j}(t,t_0) \right)  \rho(t_0) \left( \mathcal{M}^\text{MAD}_{\alpha_j} (t,t_0) \right)^\dagger =
 \left (
		\begin{array}{cc} 
		 1- \left[ \rho_{j} (t_0)    \right]_{11}  e^{-\gamma_\text{MAD} t}&  \left[ \rho_{j}  (t_0)   \right]_{01} e^{- \frac{\gamma_\text{MAD} t}{2}}  \\
		 \left[ \rho_{j}  (t_0)   \right]_{10} \cdot    e^{- \frac{\gamma_\text{MAD} t}{2}} &  \left[ \rho_{j} (t_0)    \right]_{11}  e^{-\gamma_\text{MAD} t}
		\end{array}
	\right ). 
\label{MADQS1}  
\end{align}\end{widetext}
Furthermore, as similar to Eq. \eqref{MPDQS2} due to the divisibility of the MAD we have
\begin{footnotesize} \begin{widetext} \begin{align}
\rho_\text{MAD}\big{(} t_{l+1} \big{)} = \sum_{\alpha_0=0}^1  \cdots \sum_{\alpha_{N_q-1}=0}^1  
  \left( \mathcal{M}^\text{MAD}_{\alpha_0}(\Delta t) \otimes \cdots \otimes \mathcal{M}^\text{MAD}_{\alpha_{N_q}-1} (\Delta t) \right)
  \rho_\text{MAD}(t_l)  \left( \mathcal{M}^\text{MAD}_{\alpha_0}(\Delta t) \otimes \cdots \otimes \mathcal{M}^\text{MAD}_{\alpha_{N_q-1}} (\Delta t) \right)^\dagger.
\label{MADQS2}
\end{align} \end{widetext} \end{footnotesize}
The output state under the influence of the MAD is given by Eqs. \eqref{noisyoutput} and \eqref{MADQS2}.
\subsubsection{Non-Markov Phase Damping}\label{NMPD} 
 In this work, we focus on NMPD represented by the quantum master equation \cite{koczor2020variational,yu2010entanglement,kumar2018non,utagi2020temporal}
\begin{widetext}  \begin{align}
 \frac { \partial \rho_\text{NMPD} (t)}{\partial t }  =   \frac{\gamma_\text{PD} }{2} f_\text{NMPD}(t) \sum_{j=0}^{N_q-1} \mathcal{L} _{\text{PD},j} [\rho (t)]  
 =  \frac{\gamma_\text{PD} }{2} f_\text{NMPD}(t) \sum_{j=0}^{N_q-1}  \Big{[} Z_j    \rho_\text{NMPD}(t)   Z_j  -  \rho_\text{NMPD} (t)   \big{\}} \Big{]},
 \label{NMPDQME1}
\end{align} \end{widetext}
where  $f_\text{NMPD}(t) = \left( 1 - e^{-\gamma_\text{C}t}\right)$ is the function which characterizes the non-Markovian behavior in terms of the two quantities, the decay rate of the qubit $\gamma_\text{PD}$ (see also Eq.  \eqref{MPDQME1}) and 
the environment decay rate $\gamma_\text{C} = \tau^{-1}_\text{C}$ with $\tau_\text{C}$ denoting correlation time of the environment. 
When $\gamma_\text{C} \gg \gamma_\text{PD}$ we call that PD is Markovian whereas when $\gamma_\text{C} \ll \gamma_\text{PD}$ we call non-Markovian.
The quantum master equation \eqref{NMPDQME1} is formally solved as
\begin{footnotesize} \begin{widetext}  \begin{align}
 \rho_\text{NMPD} (t)  =\exp\left( \frac{\gamma_\text{PD} }{2} \int_{t_0}^t d\tilde{t}  \left(f_\text{NMPD}(\tilde{t}) \sum_{j=0}^{N_q-1} \mathcal{L} _{\text{PD},j}  \right)   \right)  \rho_\text{NMPD} (t_0)
 = \exp\left(\frac{\gamma_\text{PD} }{2} F_\text{NMPD}(t) \sum_{j=0}^{N_q-1} \mathcal{L} _{\text{PD},j}   \right)  \rho_\text{NMPD} (t_0), 
 \label{NMPDQME2}
\end{align} \end{widetext} \end{footnotesize}
where $F_\text{NMPD}(t) = \left( t+\frac{e^{-\gamma_\text{C}t}-1}{\gamma_\text{C}} \right)  $. The function $F_\text{NMPD}(t) $ satisfies
$F^\prime_\text{NMPD}(t)=\frac{dF_\text{NMPD}(t) }{dt}= f_\text{NMPD}(t)$.
Note that in the Markovian limit we obtain $f_\text{NMPD}(t) \to 1.$  
The denstiy matrix $\rho_\text{NMPD} (t)$ in Eq.  \eqref{NMPDQME2} can be expressed in terms of the initial state $\rho (t_0)$ and  Kraus operators $\mathcal{M}^\text{NMPD}_{0} (t,t_0)= \sqrt{\frac{1+e^{-\gamma_\text{PD}F_\text{NMPD}(t)}}{2}} \boldsymbol{1}_{2\times2} $ and $\mathcal{M}^\text{NMPD}_{1} (t,t_0)=\sqrt{\frac{1-e^{-\gamma_\text{PD}F_\text{NMPD}(t)}}{2}} Z $ as
\begin{footnotesize} \begin{widetext}  \begin{align}
 \rho_\text{NMPD} (t)  =  \sum_{\alpha_0=0}^1    \cdots \sum_{\alpha_{N_q-1}=0}^1  
  \left( \mathcal{M}^\text{NMPD}_{\alpha_0}(t,t_0) \otimes \cdots \otimes \mathcal{M}^\text{NMPD}_{\alpha_{N_q}-1} (t,t_0) \right)
  \rho(t_0)  \left( \mathcal{M}^\text{NMPD}_{\alpha_0}(t,t_0) \otimes \cdots \otimes \mathcal{M}^\text{NMPD}_{\alpha_{N_q-1}} (t,t_0) \right)^\dagger.
 \label{NMPDQME3}
\end{align} \end{widetext} \end{footnotesize}
Next, let us analyze the map describing the time evolution of quantum state which occur during a time interval $[t_a,t_b]$ with $t_0<t_a<t_b$ and denote it by $\mathcal{E}^\text{NMPD}_{t_b,t_a}$, i.e., 
$ \rho_\text{NMPD} (t_b)=\mathcal{E}^\text{NMPD}_{t_b,t_a}[ \rho_\text{NMPD}(t_a)]$.
By using the commutativity,  $\left[ f_\text{NMPD}(t_b) \sum_{j_1=0}^{N_q-1} \mathcal{L} _{\text{PD},j_1} , f_\text{NMPD}(t_a) \sum_{j_2=0}^{N_q-1} \mathcal{L} _{\text{PD},j_2}       \right][\rho]=0$,
from the quantum master equation  \eqref{NMPDQME1} we obtain
  \footnotesize \begin{widetext} 
 \begin{align}
\rho_\text{NMPD}\big{(} t_{b} \big{)} &= \mathcal{E}^\text{NMPD}_{t_b,t_a}[ \rho_\text{NMPD}(t_a)]= \exp\left(\frac{\gamma_\text{PD} }{2} \left(F_\text{NMPD}(t_b)-F_\text{NMPD}(t_a)\right) \sum_{j=0}^{N_q-1} \mathcal{L} _{\text{PD},j}   \right)  \rho_\text{NMPD} (t_a) \notag\\
&=\sum_{\alpha_0=0}^1    \cdots \sum_{\alpha_{N_q-1}=0}^1  
  \left( \mathcal{M}^\text{NMPD}_{\alpha_0}(t_{b},t_{a}) \otimes \cdots \otimes \mathcal{M}^\text{NMPD}_{\alpha_{N_q}-1} (t_{b},t_{a}) \right)
  \rho(t_a)  \left( \mathcal{M}^\text{NMPD}_{\alpha_0}(t_{b},t_{a}) \otimes \cdots \otimes \mathcal{M}^\text{NMPD}_{\alpha_{N_q}-1} (t_{b},t_{a})  \right)^\dagger, 
\label{NMPDQS1}
\end{align} 
  \end{widetext}\normalsize
  where 
 \begin{footnotesize} \begin{widetext} 
 \begin{align}
\mathcal{M}^\text{NMPD}_{0}(t_{b},t_{a})    =    \sqrt{\frac{1+e^{-\gamma_\text{PD}[F_\text{NMPD}(t_{b})-F_\text{NMPD}(t_{a})]}}{2}} \mathbf{1}_{2 \times 2}, \quad
   \mathcal{M}^\text{NMPD}_{1}(t_{b},t_{a})    =    \sqrt{\frac{1-e^{-\gamma_\text{PD}[F_\text{NMPD}(t_{b})-F_\text{NMPD}(t_{a})]}}{2}} Z.
\label{NMPDQS2}
\end{align} \end{widetext} \end{footnotesize}
 Like Eqs. \eqref{MPDQS2} and \eqref{MADQS2}, by using  Eqs. \eqref{NMPDQS1} and \eqref{NMPDQS2} with setting $t_a=t_l, t_b=t_{l+1}$ ($l=0,\ldots,N^\text{tot}-1$), we become able to simulate the noisy time evolution which occur during the time interval 
 $[t_l, t_{l+1}],$ i.e., the noisy time evolution of quantum states represented by Eq. \eqref{noisyoutput} for the NMPD. 
 At the end, we note that other types of NMPD and noisy time evolution driven by them can be represented by changing the functional form of $F_\text{NMPD}(t)$, and furthermore QEM of them can be established.
\subsection{QEM Protocol}\label{QEM} 
Let explain our QEM protocol. Here we discuss the case of the quantum metrology under the MAD. We treat the MAD since it affects both diagonal and off-diagonal components of density matrices, which means that the AD effect influences every part of the density matrices. 
Once a QEM protocol for the MAD effect is established then it is straightforward to establish QEM protocols for other quantum noise effects such as the MPD and the NMPD effects (they only affect off-diagonal components of density matrices) and generalized amplitude damping effect (AD effect at finite temperature): we discuss them in Appendix \ref{appendix1}. 
In the following, we use "ideal" as  the subscript to describe density matrices describing ideal (noise-free) output states like $ \rho^\text{out}_{\text{ideal},\vartheta}$ whereas for noisy output states we write $ \rho^\text{out}_{\text{noisy},\vartheta}$. 

\begin{figure*}[!t] 
\centering
\includegraphics[width=0.8 \textwidth]{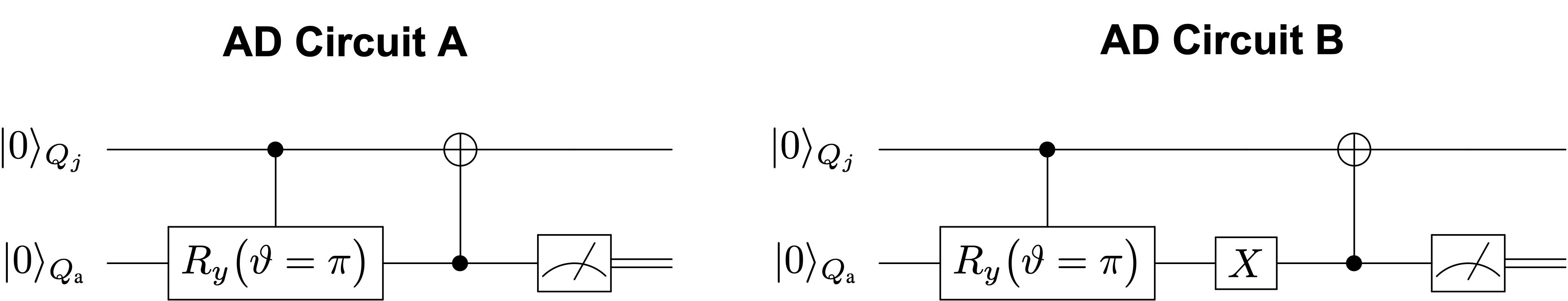}
\caption{AD-effect circuits A (a) and  B (b).   }  
\label{ADcircuits} 
\end{figure*}

Our QEM protocol is established as follows. First, we rewrite the AD Liouvillian operatior $ \mathcal{L} _{ \text{AD}} [\rho]$ (see Eq. \eqref{MADQME1}) as 
  \begin{align}
&  \mathcal{L} _{ \text{AD},j} [\rho(t)]  =   \frac{- \rho (t)  +\rho_{Z_j} (t)  }{4}  +  \rho_{\tilde{\sigma}^-_j} (t) -  \rho_{P^1_j} (t), \notag\\
 & \rho_{Z_j} (t) = Z_j   \rho (t)   Z_j, \quad  \rho_{\tilde{\sigma}^-_j} (t) = \tilde{\sigma}^-_j \rho (t)   \tilde{\sigma}^+_j, \notag\\
 &   \rho_{P^1_j} (t) = P^1_j   \rho (t)    P^1_j.
\label{LindbladMADforqcirc}
\end{align} 
 As described in the above equation, all four terms,  $\rho (t), \rho_{Z_j} (t),  \rho_{\tilde{\sigma}^-_j} (t),$ and  $\rho_{P^1_j} (t)$,
are represented in the form $\mathcal{A} \rho(t)\mathcal{A}^\dagger$, where $\mathcal{A}^\dagger$ stands for any gate operations or quantum measurements:
here $\mathcal{A} = \boldsymbol{1}, Z_j, \tilde{\sigma}^-_j ,  P^1_j  .$
 Second, we introduce the two quantum circuits shown in Fig. \ref{ADcircuits} and we call them AD-effect circuits A (Fig. \ref{ADcircuits} (a)) and B (Fig. \ref{ADcircuits} (b)) \cite{hama2022quantum}.
The AD-effect circuit A is composed of the controlled-rotational gate $\text{C}R_y(\vartheta)[Q_j;Q_\text{a}]$,  
(the qubit $Q_j$ is the control bit while an ancilla bit $Q_\text{a}$ is the target bit), 
which is the controlled-rotational operation around the $y$ axis with the rotational angle $\vartheta$, and the CNOT (C$X$) gate $\text{C}X[Q_\text{a};Q_{_j}]$ ($Q_\text{a}$ is the control bit while $Q_j$ is the target bit). 
On the other hand, the AD-effect circuit B is composed of the controlled-rotational gate $\text{C}R_y(\vartheta)[Q_j;Q_\text{a}] $, the single-qubit operation $\boldsymbol{1}_j \otimes X_\text{a}$, 
and the CNOT operation $\text{C}X[Q_\text{a};Q_j]$.
Note that for the AD circuit A by writing $U_{\text{ADA},j}(\vartheta)=\text{C}X[Q_\text{a};Q_j]\cdot\text{C}R_y(\vartheta)[Q_j;Q_\text{a}] $ and setting $\vartheta = \vartheta^{\text{AD}}_t$ such that $\cos^2 \left( \frac{\vartheta^{\text{AD}}_t}{2} \right) = e^{-\gamma_\text{AD} t},$ 
we obtain  $ \rho_{\text{ADA},j}(\vartheta^{\text{AD}}_t) =
\text{Tr}_{Q_\text{a}}\left[  U_{\text{ADA},j}(\vartheta)\cdot  |00\rangle_{j\text{a}}\langle00|   \cdot U_{\text{ADA},j}^\dagger(\vartheta)  \right] = 
\sum_{\alpha=0}^1  \mathcal{M}^\text{MAD}_\alpha(t) \cdot  |0\rangle_j\langle0| \cdot  \big{(}  \mathcal{M}^\text{MAD}_\alpha(t)\big{)}^\dagger$, which is equivalent to  the single-qubit quantum state $ \rho_{\text{MAD},j} (t) $ in Eq. \eqref{MADQS1}.
By using Eq. \eqref{LindbladMADforqcirc} and the AD circuits A and B,  our QEM protocol is established and is represented by the quantum gates and the quantum measurements.
In addition, we introduce two types of quantities, a dimensionless time $\tau_\text{AD}$ which is defined by $\tau_\text{AD}= \gamma_\text{AD} \Delta t$ and density matrices
$\rho_{k\cdots1,\vartheta} =  \left(\prod_{l=1}^{k } U^\text{tot}_{\vartheta,l} \right) \rho(0) \left(\prod_{l=1}^{k } U^\text{tot}_{\vartheta,l} \right)^\dagger,$ where $k=1,\ldots,d_\text{tot}$.
We reformulate the noisy quantum state $ \rho^\text{out}_{\text{AD},\vartheta} $ in Eq. \eqref{noisyoutput} with a perturbation theory with respect to $\tau_\text{AD}$ as 
$ \rho^\text{out}_{\text{AD},\vartheta}  =  \rho^\text{out}_{\text{ideal},\vartheta} + \sum_{p=1}^\infty \frac{\tau_\text{AD}^p}{p!} \Delta^\text{AD}_p  \big{[} \rho^\text{out}_{\text{ideal},\vartheta} \big{]},  $
where $\rho^\text{out}_{\text{ideal},\vartheta} =  \left(\prod_{l=1}^{d_\text{tot} } U^\text{tot}_{\vartheta,l}\right) \rho(0) \left(\prod_{l=1}^{d_\text{tot} } U^\text{tot}_{\vartheta,l} \right)^\dagger$.
The quantity $ \Delta^\text{AD}_p  \big{[} \rho^\text{out}_{\text{ideal},\vartheta} \big{]}  $ describes the $p$th order theoretically-estimated AD effect on the ideal quantum state $\rho^\text{out}_{\text{ideal},\vartheta}.$
In the following, we demonstrate the derivation of our QEM protocol in the first-order regime in $\tau_\text{AD}$. 
We present the details of the second-order QEM protocol in Appendix \ref{appendix1}.
By using Eq. \eqref{LindbladMADforqcirc}, the first-order AD effect $  \Delta^\text{AD}_1  \big{[} \rho^\text{out}_{\vartheta,\text{ideal}}  \big{]} $ is evaluated as
 \footnotesize \begin{align}
& \Delta^{\text{AD}}_1 \big{[} \rho^\text{out}_{\vartheta,\text{ideal}} \big{]}  =
 \sum_{k=1}^{d_\text{tot} } \Delta^{\text{AD}}_{1,k}  \big{[}  \rho^\text{out}_{\vartheta,\text{ideal}}  \big{]}, \notag\\
& \Delta^{\text{AD}}_{1,k}  \big{[}  \rho^\text{out}_{\vartheta,\text{ideal}}  \big{]} =
 \left(\prod_{l=k+1}^{N_{\text{tot}} } U^\text{tot}_{\vartheta,l} \right)  \mathcal{L}^{\text{AD}}_1\big{[} \rho_{k\cdots1} \big{]} 
 \left(\prod_{l=k+1}^{N_{\text{tot}} } U^\text{tot}_{\vartheta,l}\right)^\dagger. \label{evalMADeffect}
\end{align}\normalsize
To compute $ \Delta^{\text{AD}}_1 \big{[} \rho^\text{out}_{\vartheta,\text{ideal}} \big{]}$ in Eq. \eqref{evalMADeffect}, 
we need $3N_qd_\text{tot}+1$ quantum circuits and they consist of four types of quantum circuits,
 the quantum circuit which is solely composed by the unitary operations $U^\text{tot}_{\vartheta,l} $ (the original quantum circuit for the quantum metrology under consideration), 
the quantum circuits consist of $U^\text{tot}_{\vartheta,l}$ and additional $Z$-gate operations and the number of them is $N_qd_\text{tot}$, and the AD-effect circuits A and B in Fig. \ref{ADcircuits}.
The number of the elementary circuits comprising the AD-effect circuit A is $N_qd_\text{tot}$ and similarly for the AD circuit B. 
The non-unitary operations $\{\tilde{\sigma}^\pm, P^0,P^1\}$ can be created by the AD circuits A and B with setting $\vartheta^\text{AD}_t  = \pi$.
By writing $U_\text{ADB}(\vartheta^\text{AD}_t )=\text{C}X[Q_\text{a};Q_j]\cdot   X_\text{a}   \cdot\text{C}R_y(\vartheta^\text{AD}_t )[Q_j;Q_\text{a}] $
and  $ \rho^{\text{ADA(B)}} _{k\cdots1,\vartheta,j} = U_\text{ADA(B)}(\vartheta^\text{AD}_t )\cdot \left(\prod_{l=1}^{k } U^\text{tot}_{\vartheta,l} \right) \rho(0) \left(\prod_{l=1}^{k } U^\text{tot}_{\vartheta,l}  \right)^\dagger\cdot U^\dagger_\text{ADA(B)}(\vartheta^\text{AD}_t )$,
the non-unitary operations $\{\tilde{\sigma}^\pm, P^0,P^1\}$ are obtained as
\begin{align} 
  & \rho^{\text{ADA}} _{k\cdots1,\vartheta,j}   \quad  \underset{ \substack{|0 \rangle_\text{a}, \\ \vartheta^\text{AD}_t \to \pi} }{\longrightarrow} \quad   P^0_j \rho_{k\cdots1,\vartheta} P^0_j, \notag\\
   &  \rho^{\text{ADA}} _{k\cdots1,\vartheta,j}  \quad  \underset{ \substack{|1 \rangle _\text{a}, \\ \vartheta^\text{AD}_t \to \pi} }{\longrightarrow} \quad   \tilde{\sigma}^-_j \rho_{k\cdots1,\vartheta} \tilde{\sigma}^+_j, \notag\\
    &  \rho^{\text{ADB}} _{k\cdots1,\vartheta,j}    \quad  \underset{ \substack{|0 \rangle _\text{a}, \\ \vartheta^\text{AD}_t \to \pi} }{\longrightarrow} \quad  P^1_j \rho_{k\cdots1,\vartheta} P^1_j , \notag\\
    &  \rho^{\text{ADB}} _{k\cdots1,\vartheta,j}   \quad  \underset{ \substack{|1 \rangle _\text{a}, \\ \vartheta^\text{AD}_t \to \pi} }{\longrightarrow} \quad  \tilde{\sigma}^+_j \rho_{k\cdots1,\vartheta} \tilde{\sigma}^-_j,    \label{ancillapostselection}
  \end{align} 
  where $|0 \rangle_\text{a}$ ($|1 \rangle_\text{a}$) is the $|0 \rangle$ ($|1 \rangle$) state of $Q_\text{a}$.
To construct the QEM protocol for the AD effect we use two types of operations, the AD-effect quantum circuits A with the post selection of $|1 \rangle _\text{a}$ which generate the additional $ \tilde{\sigma}^-_j$ operations, 
and the AD-effect quantum circuits B with the post selection of $|0 \rangle _\text{a}$ which create the additional $P^1_j$ operations.
As a result, we obtain the quantum circuits for $\Delta^{\text{AD}}_1 \big{[} \rho^\text{out}_{\vartheta,\text{ideal}} \big{]} $ in Eq. \eqref{evalMADeffect},  namely, quantum-error-mitigation circuit group,
 and we are now ready to derive the basic formula of our QEM protocol. Let us write the output quantum state generated in a real experiment of quantum metrology by 
 $ \rho^\text{out,real}_{\text{AD},\vartheta} $ (more generally we write $ \rho^\text{out,real}_{\text{noisy},\vartheta} $ to express the noisy quantum state generated in a real device) 
 so as to distinguish from the theoretically-evaluated noisy output quantum state $\rho^\text{out}_{\text{AD},\vartheta}$ ($\rho^\text{out}_{\text{noisy},\vartheta}$).   
 As similar to $\rho^\text{out}_{\text{AD},\vartheta}$, we describe  $ \rho^\text{out,real}_{\text{AD},\vartheta}  $ in the perturbation series in $\tau_{\text{AD}}$ as 
$ \rho^\text{out,real}_{\text{AD},\vartheta} = \rho^\text{out}_{\text{ideal},\vartheta} + \sum_{p=1}^\infty  \frac{\tau^p_{\text{AD}}}{p!} \cdot \delta^{\text{AD}}_{p} \big{[}   \rho^\text{out}_{\text{ideal},\vartheta}   \big{]}   $. 
Here we have used the symbol $\delta^{\text{AD}}_{p}$ (lowercase delta) to describe the $p$th-order AD effect in a real system in contrast to the theoretically-evaluated one $\Delta^{\text{AD}}_{p}$ (capital delta). 
By using Eqs. \eqref{LindbladMADforqcirc}-\eqref{ancillapostselection}, the basic formula of our first-order QEM protocol is derived and is given by
\begin{footnotesize}
 \begin{align}
 \rho^\text{out}_{\text{QEM}_{\text{1st}},\vartheta=0}  &=\rho^\text{out,real}_{\text{AD},\vartheta=0} 
  - \tau_\text{AD}\cdot \Delta^{\text{AD}}_1 \big{[} \rho^\text{out}_{\text{ideal},\vartheta=0} \big{]} \notag\\
 &=   \rho^\text{out}_{\text{ideal},\vartheta=0}
\notag\\&  + \tau_\text{AD} \left( \delta^{\text{AD}}_{1} \big{[}   \rho^\text{out}_{\text{ideal},\vartheta=0}   \big{]} 
 -\Delta^{\text{AD}}_{1} \big{[}   \rho^\text{out}_{\text{ideal},\vartheta=0}   \big{]}   \right) \notag\\& + \mathcal{O}(\tau_\text{AD}^2), \notag\\
  \rho^\text{out}_{\text{QEM}_{\text{1st}},\vartheta}  &=\rho^\text{out,real}_{\text{AD},\vartheta} 
  - \tau_\text{AD}\cdot \Delta^{\text{AD}}_1 \big{[} \rho^\text{out}_{\text{ideal},\vartheta} \big{]} \notag\\
 &=   \rho^\text{out}_{\text{ideal},\vartheta}
  + \tau_\text{AD} \left( \delta^{\text{AD}}_{1} \big{[}   \rho^\text{out}_{\text{ideal},\vartheta}   \big{]} 
 -\Delta^{\text{AD}}_{1} \big{[}   \rho^\text{out}_{\text{ideal},\vartheta}   \big{]}   \right)
 \notag\\ &+ \mathcal{O}(\tau_\text{AD}^2).
\label{QEMDM}
\end{align}\end{footnotesize}
Let us call $ \rho^\text{out}_{\text{QEM}_{\text{1st}},\vartheta=0}, \rho^\text{out}_{\text{QEM}_{\text{1st}},\vartheta} $ in the above equation as quantum-error-mitigated states.
By regarding $\delta^{\text{AD}}_{p} \equiv \Delta^{\text{AD}}_{p}$, the lowest-order terms in $\tau_\text{AD}$ contained in $ \rho^\text{out}_{\text{QEM}_{\text{1st}},\vartheta=0}, \rho^\text{out}_{\text{QEM}_{\text{1st}},\vartheta} $ are $\mathcal{O}(\tau_\text{AD}^2).$ Consequently, in the first-order perturbation theory by adapting our QEM protocol the QFI becomes approximately restored to that of the ideal quantum states: $ I_{\text{QF}} \left[  \rho^\text{out}_{\text{QEM}_{\text{1st}},\vartheta=0}   \right]  \simeq  I_{\text{QF}} \left[ \rho^\text{out}_{\text{ideal},\vartheta=0}  \right]$, and
 $ I_{\text{QF}} \left[  \rho^\text{out}_{\text{QEM}_{\text{1st}},\vartheta}   \right]  \simeq  I_{\text{QF}} \left[ \rho^\text{out}_{\text{ideal},\vartheta}  \right]$, 
 where  $ I_{\text{QF}} \left[  \rho^\text{out}_{\text{QEM}_{\text{1st}},\vartheta=0}   \right] $ and
 $ I_{\text{QF}} \left[  \rho^\text{out}_{\text{QEM}_{\text{1st}},\vartheta}   \right] $ are the quantum-error-mitigated QFI. 
In other words, we have improved the sensitivity of the estimation of $\vartheta$ such that its quality has become restored to that of the ideal one via our QEM protocol.
 As we have described in the end of Sec. \ref{model}, the key ingredient of our QEM protocol is that we describe the quantum metrological procedures, the composition of the initialization $U^\text{in}$ and the free evolution $U^\text{free}$ as the quantum circuit. 
 Once this is done, our QEM protocol is established which is constructed by the quantum-error-mitigation circuit group and enables us to compute
  $\Delta^{\text{AD}}_{1} \big{[}   \rho^\text{out}_{\vartheta,\text{ideal}}   \big{]}.  $  Several comments are in order.  
Firstly, when we compute the eigenvalues of quantum-error-mitigated states they include negative values in general. This means that we cannot use both  Eqs. \eqref{QFIexpression1} and \eqref{QFIexpression2}. 
  In order to solve this problem, instead of using these two equations we use the error-propagation formula \cite{ma2011quantum,
toth2014quantum}
 \begin{align}
   \Delta^2 _O(\vartheta) &= \frac{\langle  O^2   \rangle_{\rho_\vartheta}-\langle  O   \rangle_{\rho_\vartheta}^2}{(\partial_\vartheta \langle  O   \rangle_{\rho_\vartheta})^2} ,  \label{errorpropagation}
\end{align}
  where $\langle  A   \rangle_{\rho_\vartheta} = \text{Tr}(A\rho_\vartheta)$ with $A=O,O^2$ and $O$ denoting a physical operator.
  By choosing properly the operator $O$ according to the chosen initial state,
  the variance $\Delta^2 _O(\vartheta)$ becomes the inverse of CFI,
  $I_\text{CF} \left[O;  \rho_\vartheta  \right]$.
  In the case of noisy quantum metrology, the CFI $I_\text{CF} \left[O;  \rho_\vartheta  \right]$ becomes equivalent to the QFI, $I_\text{QF}, \left[\rho_\vartheta  \right]$, by doing the optimization (maximization) with respect to $\vartheta$.
  In this work, such an optimization is done by fixing $T^\text{free}$.     
  When we use the error-propagation formula \eqref{errorpropagation} numerically, the derivative $\partial_\vartheta$  is numerically performed by the formula \cite{burden2015numerical}
  \begin{footnotesize} \begin{align}
  \frac{\partial f }{\partial \vartheta } \simeq \frac{f(\vartheta-2h)-8f(\vartheta-h)+8f(\vartheta+h)-f(\vartheta+2h)                      }
  {12h}, \label{numericalderivative}
  \end{align} \end{footnotesize}
  where $f$ is a continuous and differentiable function and $h \ll1$.
  For computing quantum-error-mitigated QFI, we use the  error-propagation formula \eqref{errorpropagation} by first introducing normalized quantum-error-mitigated states 
  $ \tilde{\rho}^\text{out}_{\text{QEM}_{\text{1st}},\vartheta(\vartheta=0)}= 
  \frac{\rho^\text{out}_{\text{QEM}_{\text{1st}},\vartheta(\vartheta=0)}}{\text{Tr}\left( \rho^\text{out}_{\text{QEM}_{\text{1st}},\vartheta(\vartheta=0)} \right)}$ and then use
  Eq. \eqref{errorpropagation} with taking $\rho_\vartheta =   \tilde{\rho}^\text{out}_{\text{QEM}_{\text{1st}},\vartheta}$ 
  and $O$ to be the physical operator such that $I_\text{CF} \left[O;  \rho^\text{out}_{\text{ideal},\vartheta}  \right]=I_\text{QF} \left[\rho^\text{out}_{\text{ideal},\vartheta}  \right]$.
   Such an approach is also going to be taken for the second-order QEM and denote the second-order normalized quantum-error-mitigated states by $ \tilde{\rho}^\text{out}_{\text{QEM}_{\text{2nd}},\vartheta(\vartheta=0)}, \tilde{\rho}^\text{out}_{\text{QEM}_{\text{2nd}},\vartheta(\vartheta)}$. The specific choices of $O$ is discussed in the next section.
  Secondly, to describe the closeness of quantum states we need measures besides fidelity which can be used not only for ideal and noisy states but also for quantum-error-mitigated states which exhibit negative eigenvalues in general.
  Here we choose the trace distance $D\left(\rho_1,\rho_2 \right)=\frac{1}{2}\text{Tr}|| \rho_1-\rho_2 ||_1$,
  where $|| \hat{A} ||_1$ is the trace norm of an operator $\hat{A}$ defined by $|| \hat{A} ||_1 = \sqrt{\hat{A}^\dagger \hat{A}}$ \cite{QCQINandC,wilde2013quantum}. 
  We choose  $\rho^\text{out}_{\text{ideal},\vartheta} $ for  $\rho_1$ while we choose $\rho^\text{out}_{\text{noisy},\vartheta}$ or $\tilde{\rho}^\text{out}_{\text{QEM}_{\text{1st/2nd}},\vartheta}$
  for $\rho_2.$ In the following, we numerically examine whether  $D\left(\rho^\text{out}_{\text{ideal},\vartheta},\tilde{\rho}^\text{out}_{\text{QEM}_{\text{1st/2nd}},\vartheta} \right)$
  is smaller than $D\left(\rho^\text{out}_{\text{ideal},\vartheta},\rho^\text{out}_{\text{noisy},\vartheta} \right)$ or not:
 If $D\left(\rho^\text{out}_{\text{ideal},\vartheta},\tilde{\rho}^\text{out}_{\text{QEM}_{\text{1st/2nd}},\vartheta} \right) < D\left(\rho^\text{out}_{\text{ideal},\vartheta},\rho^\text{out}_{\text{noisy},\vartheta} \right)$, then the quantum-error-mitigated states are closer to the ideal state $\tilde{\rho}^\text{out}_{\text{QEM}_{\text{1st/2nd}},\vartheta}$
 compared to the noisy state $\rho^\text{out}_{\text{noisy},\vartheta} $ and is the  indication such that our QEM protocol is valid.   
\section{Numerical Simulations}\label{nss}   
In this section, we show the noisy quantum simulation results of our QEM protocol.
We compute three types of normalized QFI (QFI per qubit), the ideal QFI  $ \frac{I_{\text{QF}} \left[ \rho^\text{out}_{\text{ideal},\vartheta}  \right]}{N_q}$, 
the noisy (erroneous)  QFI $ \frac{I_{\text{QF}} \left[ \rho^\text{out}_{\text{noisy},\vartheta,}  \right]}{N_q}$, and the quantum-error-mitigated QFI $ \frac{I_{\text{QF}} \left[  \rho^\text{out}_{\text{QEM}_{\text{1st/2nd}},\vartheta}   \right]}{N_q}$ 
with performing both the first- and the second-order QEM. We write the normalized QFI by a symbol $\mathcal{I}_\text{QF}$ and discuss the numerical behaviors  with respect to the number $N_q$ and the noise strength. 
Furthermore, we compute three types of trace distance, $D\left(\rho^\text{out}_{\text{ideal},\vartheta},\rho^\text{out}_{\text{noisy},\vartheta} \right),D\left(\rho^\text{out}_{\text{ideal},\vartheta},\tilde{\rho}^\text{out}_{\text{QEM}_{\text{1st}},\vartheta} \right),$ and $D\left(\rho^\text{out}_{\text{ideal},\vartheta},\tilde{\rho}^\text{out}_{\text{QEM}_{\text{2nd}},\vartheta} \right)$.
For the initial states we choose the coherent spin states (CSS), the GHZ states, and  the symmetric Dicke states (SDS) for even $N_q$, while for the quantum noise channels we choose the MPD, the MAD, and the NMPD. 
The noisy evolution is performed by using Eq. \eqref{noisyoutput} and Eqs. \eqref{MPDQS2}, \eqref{MADQS2}, and \eqref{NMPDQS1} with Eq. \eqref{NMPDQS2} for the MPD, the MAD, and the NMPD, respectively.
 For the CSS and the GHZ states the free Hamiltonian is chosen to be the Zeeman interaction  $H^\text{free}_\vartheta = -\frac{\hbar\omega}{2} \sum_{j=0}^{N_q-1} Z_j$ with $\omega= \gamma_g B$ and the associated
free unitary evolution is described as $U^\text{free}_{\vartheta} = \prod_{l=1}^{N^\text{free}} \bigotimes_{j=0}^{N_q-1} R^z_j(-\omega\Delta t)$ whereas for the SDS we take $H^\text{free}_\vartheta = -\frac{\hbar\omega}{2} \sum_{j=0}^{N_q-1} Y_j$ and the associated unitary operation is $\prod_{l=1}^{N^\text{free}} \bigotimes_{j=0}^{N_q-1} R^y_j(-\omega\Delta t)$.  Every time we apply  $R^z_j(-\omega\Delta t)$ or $R^{y}_j(-\omega\Delta t)$ the qubits are subject to the quantum noise under consideration as described by Eq. \eqref{noisyoutput}.  Note that for simulating the quantum metrology under the AD effect
the ancilla bits are treated exactly in the same way as the $N_q$ qubits do such that the ancilla bits are subject to the AD effect which the $N_q$ qubits experience: the AD effect on the ancilla bits are represented by the Kraus operators in Eq.  \eqref{MADQS2}.
Moreover, both the AD effects on the ancilla bits and  the $N_q$ qubits are mitigated via Eq. \eqref{QEMDM}. 
In other words, the simulations of the quantum metrology under AD effect is performed as the simulation of noisy quantum metrology for $N_q+1$ (first-order QEM) or $N_q+2$ (second-order QEM) qubit systems.
Namely, we perform the simulations of noisy quantum magnetic field sensing and our QEM protocol. 
To describe the efficacy of our QEM protocol, we introduce a measure defined by  
 \begin{align}
\text{RT}_\text{QEM,QFI}=
\frac{\Big{|} \mathcal{I}_{\text{QF}} \left[ \rho^\text{out}_{\text{ideal},\vartheta} \right]-\mathcal{I}_{\text{QF}} \left[ \rho^\text{out}_{\text{noisy},\vartheta} \right]\Big{|} }
{\Big{|} \mathcal{I}_{\text{QF}} \left[ \rho^\text{out}_{\text{ideal},\vartheta} \right]-\mathcal{I}_{\text{QF}} \left[  \rho^\text{out}_{\text{QEM}_{\text{1st/2nd}},\vartheta}   \right]\Big{|} } .
\label{QEMmeasureQFI}
\end{align}
By definition, $\text{RT}_\text{QEM,QFI}$ exceeding one implies that $ \mathcal{I}_{\text{QF}} \left[  \rho^\text{out}_{\text{QEM}_{\text{1st/2nd}},\vartheta}   \right]$ is numerically closer to  $\mathcal{I}_{\text{QF}} \left[ \rho^\text{out}_{\text{ideal},\vartheta} \right]$
than $\mathcal{I}_{\text{QF}} \left[ \rho^\text{out}_{\text{noisy},\vartheta} \right]$ and can be consider as the indication of the effectiveness of our QEM protocol.   
\begin{figure}[!b] 
\centering
\includegraphics[width=0.3\textwidth]{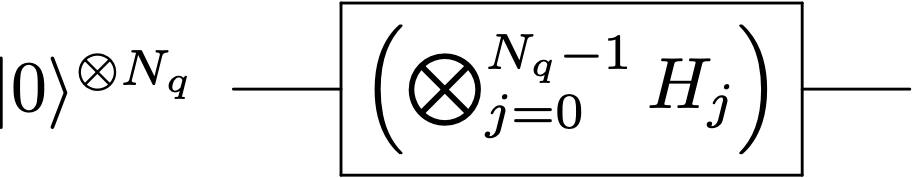}
\caption{Quantum circuit for the generation of $| \text{CSS}_{\frac{\pi}{2},0} \rangle $.   }  
\label{CSScircuit} 
\end{figure} 
\subsection{CSS}\label{CSS}
\begin{figure*}[!t] 
\centering
\includegraphics[width=0.85\textwidth]{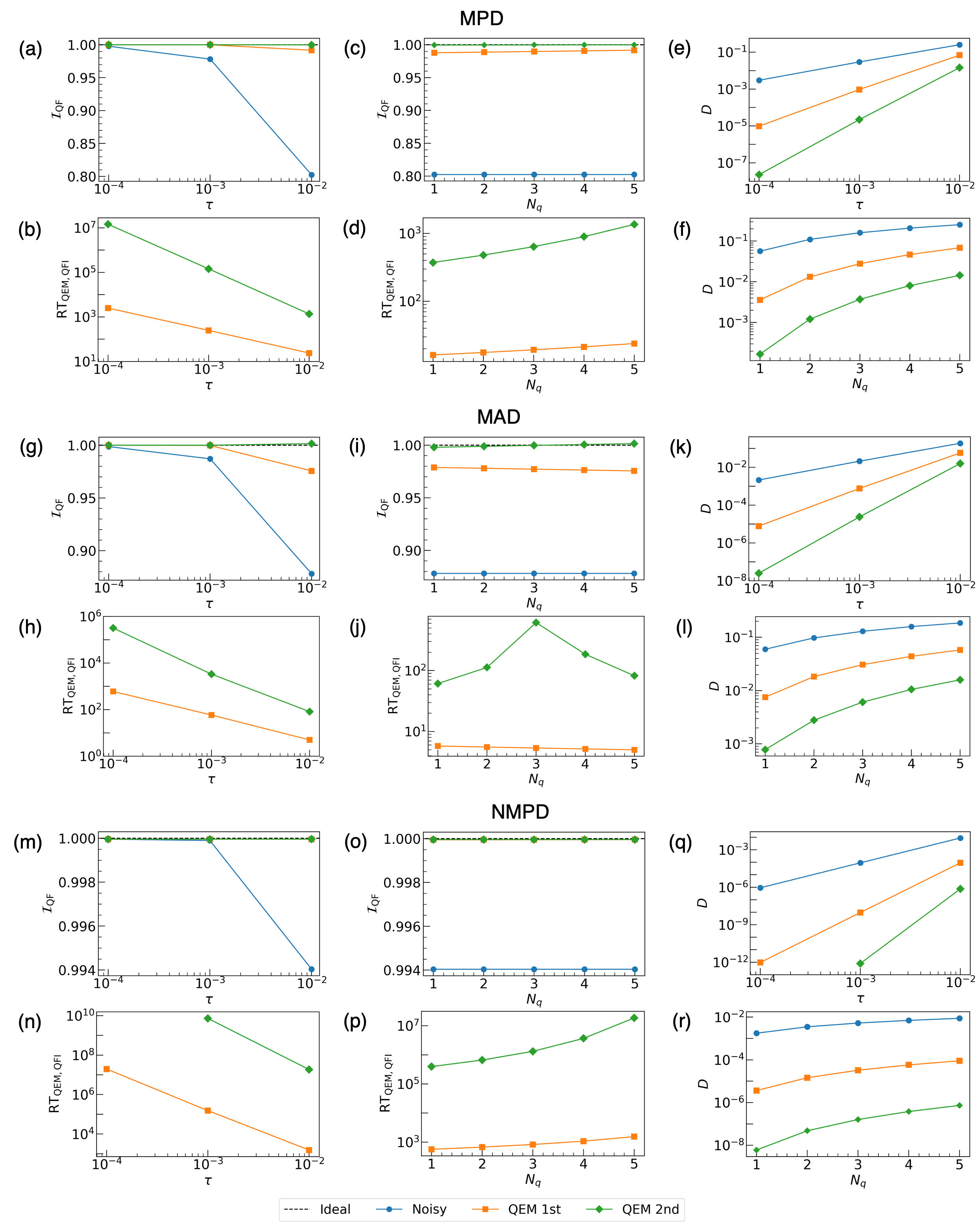}
\caption{Simulation results of  magnetic field quantum sensing for the CSS.   
In (a)-(f), (g)-(l), and (m)-(r) we present the results for the cases of the MPD, the MAD, and the NMPD, respectively.
In (a) and (m) we plot the CFI  by taking the horizontal axes to be $\tau^\text{PD}$ 
while in (g) we take the horizontal axis to be $\tau^\text{AD}$ and set $N_q=5$.
In (c), (i), and (o) we take $N_q$ for the horizontal axes with $\tau^\text{PD}=\tau^\text{AD}=10^{-2}.$
In (b)  and (n) we plot $\text{RT}_\text{QEM,QFI}$  by taking the horizontal axes to be $\tau^\text{PD}$ whereas in (h) we take the horizontal axis to be $\tau^\text{AD}$ 
with $N_q=5$. In (d), (j), and (p) we take $N_q$ for the horizontal axes with $\tau^\text{PD}=\tau^\text{AD}=10^{-2}.$ 
In (e), (k), and (q) we plot the trace distance. The horizontal axes 
 in (e) and (q) represent $\tau^\text{PD}$ while the horizontal axis for (k) describes $\tau^\text{AD}$ and $N_q=5$. In  (f), (l), and (r) we take $N_q$ for the horizontal axes with $\tau^\text{PD}=\tau^\text{AD}=10^{-2}.$ The black dashed lines, the blue, orange, and green curves represent the ideal CFI (QFI), the noisy CFI (QFI), the first-order quantum-error-mitigated CFI (QFI),
and the second-order quantum-error-mitigated CFI (QFI), respectively.}  
\label{resultsCSS} 
\end{figure*} 
First, let us discuss from the results of our simulations for the CSS. It is mathematically represented as \cite{ma2011quantum,Agarwalltxb}
 \begin{align}
| \text{CSS}_{\theta,\phi} \rangle  = 
\bigotimes_{j=0}^{N_q-1} \cos \left( \frac{\theta}{2}  \right) | 0 \rangle_{j} +  \sin \left( \frac{\theta}{2}  \right) e^{-i\phi} | 1 \rangle_{j},
\label{cssstatedef}
\end{align}
where $\theta \in [0,\pi]$ and  $\phi \in [0,2\pi]$.
Here we take $\theta=\frac{\pi}{2},\phi=0: | \text{CSS}_{\frac{\pi}{2},0} \rangle  = | \psi^\text{in} \rangle =
 (H | 0 \rangle)^{\otimes N_q} = \left( \frac{| 0 \rangle + | 1 \rangle}{\sqrt{2}}      \right)^{\otimes N_q}$, with $H$ denoting the Hadamard gate.  
After the free evolution for the time interval $T^\text{free}$, we obtain the ideal output state given by
 \begin{align}
\rho^{\text{out,CSS}}_{\text{ideal},\vartheta} &=  \left[
 \left( \frac{|0\rangle + e^{-i \vartheta} |1\rangle }{\sqrt{2}}\right) \left( \frac{\langle 0| + e^{i \vartheta} \langle1|}{\sqrt{2}}\right) \right]^{\otimes N_q }.   \label{sepoutput}
 \end{align}
 We show the quantum circuit which generates $\rho^{\text{out,CSS}}_{\vartheta_a,\text{ideal}}$ in Fig. \ref{CSScircuit}. 
 From Eqs. \eqref{QFIexpression1}, \eqref{QFIexpression2}, or the error-propagation formula  \eqref{errorpropagation} with setting $O=J^x=\sum_{j=0}^{N_q-1}X_j$
  we obtain the ideal normalized QFI
\begin{align}
 \mathcal{I}_{\text{QF}} \left[ \rho^{\text{out,CSS}}_{\text{ideal},\vartheta} \right] = 1,      \label{CSSQFI1}
 \end{align}
which is  the scaling behavior in the standard quantum limit.  Note that we compute the expectation value of $J^x$ with respect to $\rho$ as
$\text{Tr}\left( \rho J^x \right) = \text{Tr}\left( \rho_H J^z \right) $, where $ \rho_H=H\rho H$ and $J^z=\sum_{j=0}^{N_q-1}Z_j$: changing the measurement basis from the computational basis to 
$\big{\{}|+\rangle=H|0\rangle, |-\rangle=H|1\rangle\big{\}} $. Thus, we perform $H^{\otimes N_q}$ after the operation of $U^\text{free}_\vartheta$ and the quantum noise effect associated with  this final Hadamard operation is taken into account and mitigated by Eq. \eqref{QEMDM}. When the qubits are subject to the MPD or the MAD, the normalized QFI becomes \cite{liu2019quantum}
\begin{align}
\mathcal{ I}_{\text{QF}} \left[\rho^{\text{out,CSS}}_{\text{MPD},\vartheta} \right] & = e^{-2\gamma_\text{PD}T^\text{tot}},   \label{CSSQFIMPD} \\
 \mathcal{I}_{\text{QF}} \left[ \rho^{\text{out,CSS}}_{\text{MAD},\vartheta} \right] & = e^{-\gamma_\text{AD} T^\text{tot}}, \label{CSSQFIMAD}
 \end{align}
where  $\mathcal{ I}_{\text{QF}} \left[\rho^{\text{out,CSS}}_{\text{MPD},\vartheta} \right] $ and $\mathcal{ I}_{\text{QF}} \left[\rho^{\text{out,CSS}}_{\text{MAD},\vartheta} \right] $ are the normalized QFI under the MPD and the MAD, respectively. 
 We show the derivation of $\rho^{\text{out,CSS}}_{\text{MPD},\vartheta}$ in Eq. \eqref{CSSQFIMPD} and $\rho^{\text{out,CSS}}_{\text{MAD},\vartheta} $ in Eq. \eqref{CSSQFIMAD} in Appendices \ref{appendix2} and \ref{appendix3}, respectively:
The formulas of  $\rho^{\text{out,CSS}}_{\text{MPD},\vartheta}$ and $\rho^{\text{out,CSS}}_{\text{MAD},\vartheta} $ are given by Eqs.  \eqref{outputGHZMPD1} and \eqref{outputCSSAD}, respectively, and by using Eqs. \eqref{QFIexpression1} or \eqref{QFIexpression2} we obtain Eqs.  \eqref{CSSQFIMPD} and \eqref{CSSQFIMAD}. Besides using Eqs. \eqref{QFIexpression1} or \eqref{QFIexpression2}, 
Eqs.  \eqref{CSSQFIMPD} and \eqref{CSSQFIMAD} are obtained by the error-propagation formula  \eqref{errorpropagation} as follows. By choosing $O=X$, we obtain 
\begin{footnotesize} \begin{align}
I_\text{CF} \left[ \rho^{\text{out,CSS}}_{\text{MPD},\vartheta}; P_{|x\rangle, |x\rangle \in \{ |+\rangle, |-\rangle  \}} \right] = \frac{e^{-2\tau_\text{PD}N^\text{tot}}\sin^2\vartheta}{1-e^{-2\tau_\text{PD}N^\text{tot}}\cos^2\vartheta},\label{CFIMPDCSS}\\
I_\text{CF} \left[\rho^{\text{out,CSS}}_{\text{MAD},\vartheta}; P_{|x\rangle, |x\rangle \in \{ |+\rangle, |-\rangle  \}}\right] = \frac{e^{-\tau_\text{AD}N^\text{tot}}\sin^2\vartheta}{1-e^{-\tau_\text{AD} N^\text{tot}}\cos^2\vartheta}. \label{CFIMADCSS}
\end{align} \end{footnotesize}
In the quantum metrology for the CSS the measurement is done in the basis $ \{ |+\rangle, |-\rangle  \}$ and correspondingly the POVM are the projection operators 
 $P_{|\pm\rangle}$ as described in Eqs. \eqref{CFIMPDCSS} and \eqref{CFIMADCSS}. 
The difference between the ideal case and the noisy case is that the ideal CFI does not depend on $\vartheta$ whereas the noisy CFI does, which was also investigated in \cite{pezze2018quantum,degen2017quantum,huelga1997improvement}.
The noisy CFI in Eqs. \eqref{CFIMPDCSS} and  \eqref{CFIMADCSS} are optimized when $\vartheta=\frac{\pi}{2}$ and in this case they match with the QFI in Eqs. \eqref{CSSQFIMPD} and \eqref{CSSQFIMAD}.
\begin{figure}[!t] 
\centering
\includegraphics[width=0.3\textwidth]{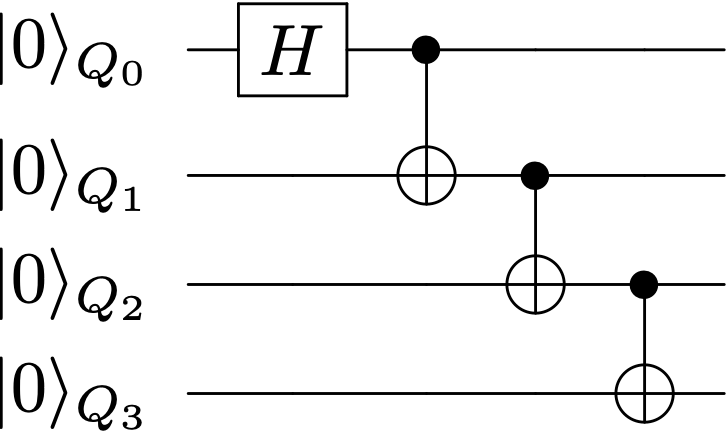}
\caption{Quantum circuit for the generation of $| \text{GHZ}_{N}\rangle $ state for $N=4.$ 
}  
\label{GHZcircuit} 
\end{figure} 
Let us now discuss our numerical results of the magnetic field quantum sensing plotted in in Fig. \ref{resultsCSS}.
In this simulation we set $\vartheta=\frac{\pi}{2}$, $\Delta t=0.1$, and $\gamma_C = \frac{\gamma_{\mathrm{PD}}}{2}$ and the CFI becomes equivalent to the QFI for all the cases, the ideal, the noisy, and the quantum-error-mitigated cases.
The plots in Figs. \ref{resultsCSS} (a)-(f), (g)-(l), and (m)-(r) are the results for the cases of the MPD, the MAD, and the NMPD, respectively.
The black dashed lines, the blue curves, the orange curves, and the green curves represent the ideal CFI (QFI), the noisy CFI (QFI), and the quantum-error-mitigated CFI (QFI). 
We plot in Figs. \ref{resultsCSS}(a), (g), and (m) the results of the CFI (QFI) for the MPD, the MAD, and the NMPD, respectively. 
We take the horizontal axes in (a), (g), and (m) to be $\tau_\text{PD}, \tau_\text{AD},$ and $\tau_\text{PD},$ respectively, such that 
$\tau_\text{PD},\tau_\text{AD}=10^{-n}$ with $n=2,3,4$  and $N_q=5$. Correspondingly, in (b), (h), and (n) we plot the results of the ratios $\text{RT}_\text{QEM,QFI}$ 
for the MPD, the MAD, and the NMPD, respectively. Namely, these results are the simulation results for verifying the efficacy of our protocol under the variation of $\tau_\text{PD}, \tau_\text{AD}.$
We see that for every noise channel  the quantum-error-mitigated QFI 
$\tilde{\rho}^\text{out}_{\text{QEM}_{\text{1st/2nd}},\vartheta}$
are closer to the ideal QFI $\mathcal{ I}_{\text{QF}} \left[\rho^{\text{out,CSS}}_{\text{ideal},\vartheta} \right]$ compared to the noisy QFI $\mathcal{ I}_{\text{QF}} \left[\rho^{\text{out,CSS}}_{\text{noisy},\vartheta} \right]$, and correspondingly the ratios $\text{RT}_\text{QEM,QFI}$ exceeds one such that its minimum value is about 5.0.
Next we discuss the results shown in Figs. (c), (d), (i), (j), (o), and (p). 
These figures show the simulation results under the variation of the number of qubits $N_q$ which is taken to be $1\sim5$
while we fix $\tau_\text{PD}$ and  $\tau_\text{AD}$  to be $10^{-2}$. As similar to the previous cases, the quantum-error-mitigated QFI are close to the ideal ones compared to the noisy QFI, and correspondingly $\text{RT}_\text{QEM,QFI}$ are greater than one which are at least about 5.0. Therefore, our protocol also works under these settings.
Finally, let us discuss the results of trace distance shown in Figs. (e), (f), (k), (l), (q), and (r).
In Figs. (e), (k), and (q) we plot the simulation results of trace distance under the variation of $\tau_\text{PD}, \tau_\text{AD}$ while $N_q$ is fixed to be five and 
(f), (l), and (r) are the results under the variation of $N_q$ while $\tau_\text{PD},\tau_\text{AD}=10^{-2}$.
Both $D\left(\rho^\text{out}_{\text{ideal},\vartheta},\tilde{\rho}^\text{out}_{\text{QEM}_{\text{1st}},\vartheta} \right)$ and $D\left(\rho^\text{out}_{\text{ideal},\vartheta},\tilde{\rho}^\text{out}_{\text{QEM}_{\text{2nd}},\vartheta} \right)$ are smaller than $D\left(\rho^\text{out}_{\text{ideal},\vartheta},\rho^\text{out}_{\text{noisy},\vartheta} \right)$ for every $N_q$ and $\tau_\text{PD}, \tau_\text{AD}$. These results imply that the quantum-error-mitigated states are closer to the ideal CSS compared to the noisy states. 
Note that the ratio $\text{RT}_\text{QEM,QFI}$ and the trace distance $D\left(\rho^\text{out}_{\text{ideal},\vartheta},\tilde{\rho}^\text{out}_{\text{QEM}_{\text{2nd}},\vartheta} \right)$ of the NMPD for $\tau_{\mathrm{PD}} = 10^{-4}$ have  not been plotted in Figs. \ref{resultsCSS} (n) and (q), respectively. The reason is that the quantum -error-mitigated density matrix $\tilde{\rho}^\text{out}_{\text{QEM}_{\text{2nd}},\vartheta}$ of the NMPD for $\tau^{\mathrm{PD}}=10^{-4}$ and the ideal density matrix $\rho^\text{out}_{\text{ideal},\vartheta}$ are too close that the loss of significant digits occur in the double-precision floating-point arithmetic.
Consequently, our QEM protocol works for the noisy magnetic field quantum sensing for the initial states taken to be the CSS.
\subsection{GHZ State}\label{GHZstate}
\begin{figure*}[!t] 
\centering
\includegraphics[width=0.85\textwidth]{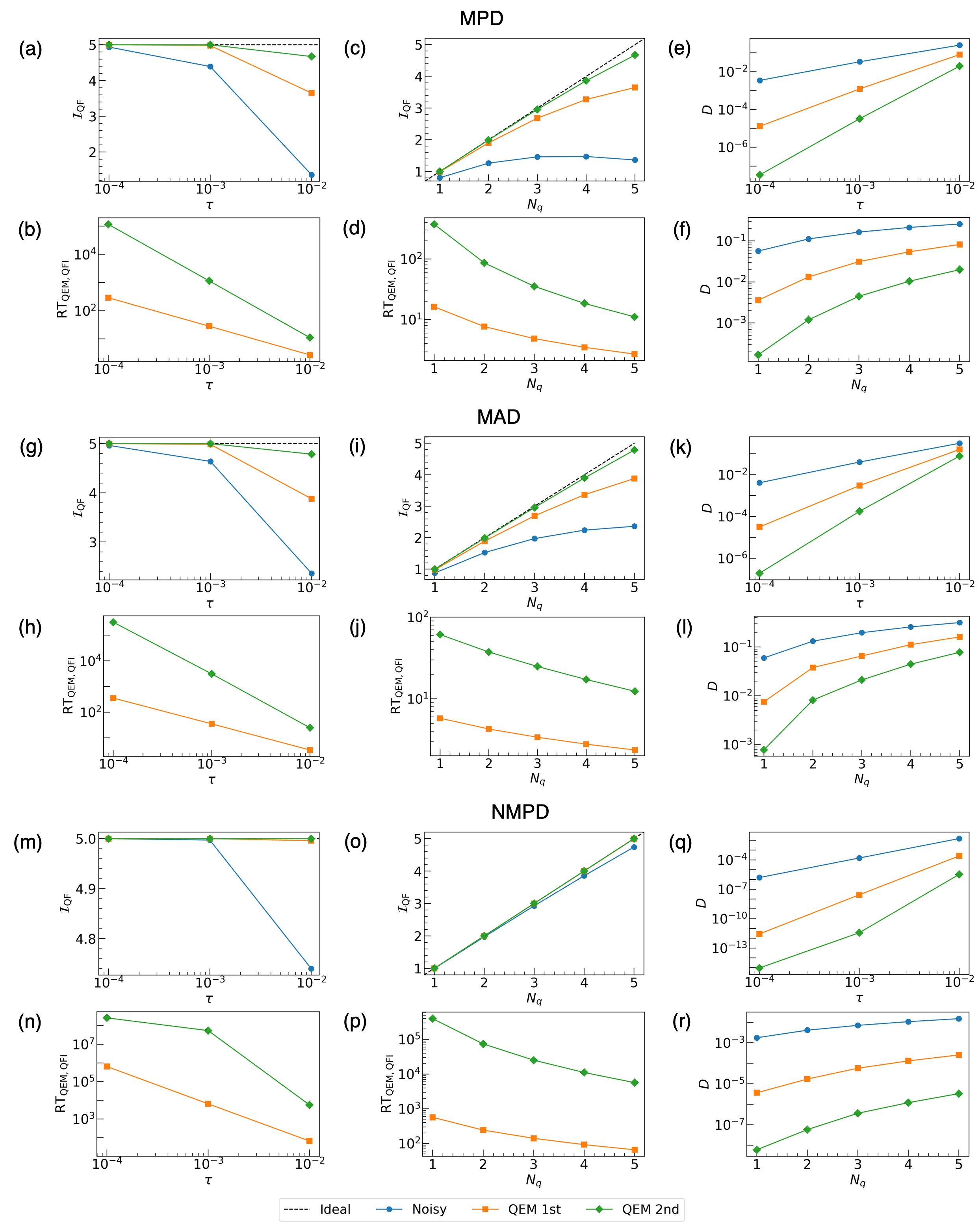}
\caption{Simulation results of  magnetic field quantum sensing for the GHZ states.   
In (a)-(f), (g)-(l), and (m)-(r) we present the results for the cases of the MPD, the MAD, and the NMPD, respectively.
In (a) and (m) we plot the CFI  by taking the horizontal axes to be $\tau^\text{PD}$ 
while in (g) we take the horizontal axis to be $\tau^\text{AD}$ and set $N_q=5$.
In (c), (i), and (o) we take $N_q$ for the horizontal axes with $\tau^\text{PD}=\tau^\text{AD}=10^{-2}.$
In (b)  and (n) we plot $\text{RT}_\text{QEM,QFI}$  by taking the horizontal axes to be $\tau^\text{PD}$ whereas in (h) we take the horizontal axis to be $\tau^\text{AD}$ 
with $N_q=5$. In (d), (j), and (p) we take $N_q$ for the horizontal axes with $\tau^\text{PD}=\tau^\text{AD}=10^{-2}.$ 
In (e), (k), and (q) we plot the trace distance. The horizontal axes 
represent $\tau^\text{PD}$ for (e) and (q) while the horizontal axis for (k) denotes
$\tau^\text{AD}$ and $N_q=5$. In  (f), (l), and (r) we take $N_q$ for the horizontal axes with $\tau^\text{PD}=\tau^\text{AD}=10^{-2}.$ The black dashed lines, the blue, orange, and green curves represent the ideal CFI (QFI), the noisy CFI (QFI), the first-order quantum-error-mitigated CFI (QFI),
and the second-order quantum-error-mitigated CFI (QFI), respectively.}  
\label{resultsGHZ} 
\end{figure*} 
Let us discuss the simulation results of the quantum metrology for the initial states taken to be the GHZ states \cite{leibfried2004toward,ma2011quantum,toth2014quantum,degen2017quantum,pezze2018quantum,braun2018quantum,meyer2021fisher,bouwmeester1999observation,monz201114,song2019generation,omran2019generation,bradley2019ten,wei2020verifying}.
To do this, we partition the $N_q$ qubits into $\mu$ groups as $N_q  = \mu N$ ($N\geq2$).
Namely, there are  $\mu$  systems of $N$ qubits.
The GHZ state of the $N$ qubits is 
\begin{align}
| \text{GHZ}_{N}\rangle = \frac{|0\rangle^{\otimes N} + |1\rangle^{\otimes N} }{\sqrt{2}},\label{GHZdef}
 \end{align}
and the overall quantum state is $| \text{GHZ}_{N}\rangle_\text{tot} = | \text{GHZ}_{N}\rangle^{\otimes \mu}$.
The gate-operation representation of $U^{ \text{in}}$ for this case is $U^\text{in,GHZ}=  
 \bigotimes_{l=0}^{N-2 } \text{C}X[Q_l;Q_{l+1}]
 \cdot  H_ {Q_0} $, where $ \text{C}X[Q_l;Q_{l+1}]$ is the CNOT gate comprised of the control bit $Q_l$ and the target bit $Q_{l+1}.$
 In Fig. \ref{GHZcircuit} we present the quantum circuit for  $U^\text{in,GHZ}$ for $N=4$.
The ideal output state is given by
\footnotesize \begin{align}
\rho^{\text{out,GHZ}_{N}}_{\text{ideal},\vartheta}  &= \bigotimes_{\alpha=1}^{\mu }  \left( \frac{ |0\rangle_\alpha^{\otimes N } + e^{-iN\vartheta} |1\rangle_\alpha^{\otimes N }}   {\sqrt{2}} \right) \notag\\
&\times  \left( \frac{ |0\rangle_\alpha^{\otimes N}  +  e^{iN\vartheta} |1\rangle_\alpha^{\otimes N} } {\sqrt{2}} \right)^\dagger,
   \label{GHZstates}
 \end{align} \normalsize
where we have used the subscript $\alpha$ for the quantum states such as $ |0\rangle_\alpha^{\otimes N }$ and $ |1\rangle_\alpha^{\otimes N }$ 
to describe that they are the quantum states of the $\alpha$-th group ($\alpha=1,\ldots,\mu$) of the $N$-qubit system.
By using Eqs. \eqref{QFIexpression1}, \eqref{QFIexpression2}, or the error-propagation formula  \eqref{errorpropagation} with setting $O=X^{\otimes N}$
we obtain the normalized QFI
 \begin{align}
 \mathcal{I}_{\text{QF}} \left[  \rho^{\text{out,GHZ}_{N}}_{\text{ideal},\vartheta}  \right] &= \mu N,  \label{GHZQFI1}
\end{align}
which is so-called the Heisenberg limit scaling. Like the case of the CSS, we compute the expectation value of $X^{\otimes N}$ such that
$\text{Tr}\left( \rho X^{\otimes N}\right) = \text{Tr}\left( \rho_H Z^{\otimes N} \right) $, and thus we perform $H^{\otimes N}$ after operating $U^\text{free}_\vartheta$, and furthermore
we mitigate the quantum noise effect associated with the final Hadamard operation $H^{\otimes N}$. When the $N$ qubits are subject to the MPD, the QFI becomes
 \begin{align}
 \mathcal{I}_{\text{QF}} \left[  \rho^{\text{out,GHZ}_{N}}_{\vartheta,\text{PD}}  \right] = \mu N e^{-\big{(} N^\text{GHZ}_\text{PD}(N) +N_\text{free}N \big{)}\tau^\text{PD}}.    \label{GHZQFI2}
\end{align}
where $N^\text{GHZ}_\text{PD}(N) = \frac{N(N+1)}{2}, $ which is the number of frequencies that qubits are affected by the MPD effect during the generation of the GHZ state.  
We show the derivation of Eq.  \eqref{GHZQFI2} in Appendix \ref{appendix2}. When we use the error-propagation formula  \eqref{errorpropagation}  by choosing $O=X^{\otimes N}$, we obtain \cite{pezze2018quantum,degen2017quantum,huelga1997improvement}.
\begin{widetext}\begin{align}
I_\text{CF} \left[ \rho^{\text{out,GHZ}_N}_{\text{MPD},\vartheta}; P_{|x\rangle, |x\rangle \in \{ |+\rangle, |-\rangle  \}^{\otimes N}} \right] = \frac{ N e^{-2\big{(} N^\text{GHZ}_\text{PD}(N) +N_\text{free}N \big{)}\tau^\text{PD}} \sin^2 N\vartheta}
{1-e^{-2\big{(} N^\text{GHZ}_\text{PD}(N) +N_\text{free}N \big{)}\tau^\text{PD}} \cos^2 N\vartheta}.\label{CFIMPDGHZ}
\end{align}\end{widetext}
In this case, the measurement is performed in the tensor-product states of $| \pm \rangle $ and the POVM are the associated projection operators as described in Eq. \eqref{CFIMPDGHZ},
$P_{|x\rangle, |x\rangle \in \{ |+\rangle, |-\rangle  \}^{\otimes N}} $.
The noisy CFI $I_\text{CF} \left[ \rho^{\text{out,GHZ}_N}_{\text{MPD},\vartheta}; P_{|x\rangle, |x\rangle \in \{ |+\rangle, |-\rangle  \}^{\otimes N}} \right]$ in Eq. \eqref{CFIMPDGHZ} is optimized when $\vartheta=\frac{\pi}{2N}$ and  becomes equivalent to the noisy QFI in Eq. \eqref{GHZQFI2}.
\begin{figure*}[!t] 
\centering
\includegraphics[width=0.9\textwidth]{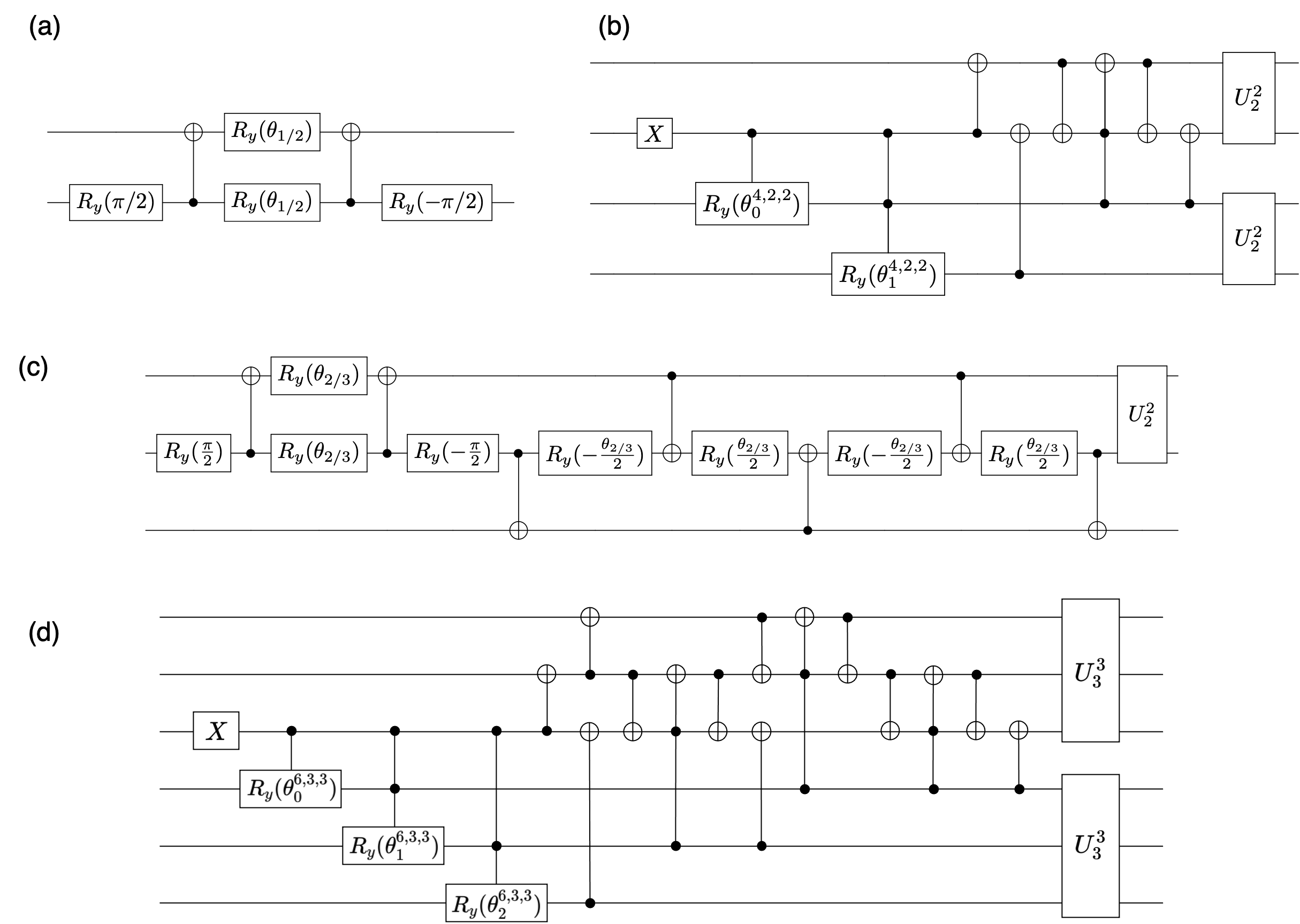}
\caption{Quantum circuits for the SDS. (a) and (b) are the quantum circuits for $N=2$ and 4, respectively.
We write the unitary operation of the quantum circuit in (a) by  $U^{2}_{2}.$ 
(c) presents the quantum circuit for the unitary operation $U^{3}_{3}$, and (d) is the one for the SDS for $N=6.$ 
}  
\label{SDScircuit} 
\end{figure*} 
Let us discuss our simulation results presented in Fig. \ref{resultsGHZ}.
Here we set $\vartheta=\frac{\pi}{2N}$ with $\mu=1$ ($N=N_q$)
and all the CFI computed in this simulation become equivalent to the QFI.
We take the same numerical settings in terms of $\tau_\text{PD},\tau_\text{AD},$ and $N$ used in the simulation for the CSS,
and we plot the results by using the same labels from (a) to (r) with those used for plotting the simulation results for the CSS which are obtained under the same numerical settings. For instance,
the numerical setting used for computing the result in Fig. \ref{resultsGHZ}(c) is equivalent to that for the result in  Fig. \ref{resultsCSS}(c).
As a result, our protocol works for any noise channel as well as for any values of $\tau_\text{PD},\tau_\text{AD},$ and $N$, and therefore our protocol is effective for the noisy magnetic field quantum sensing for the initial states taken to be the GHZ states. Note that the minimum value of the ratio $\text{RT}_\text{QEM,QFI}$ in Fig. \ref{resultsGHZ} is about 2.4. 
\begin{figure*}[!t] 
\centering
\includegraphics[width=0.85\textwidth]{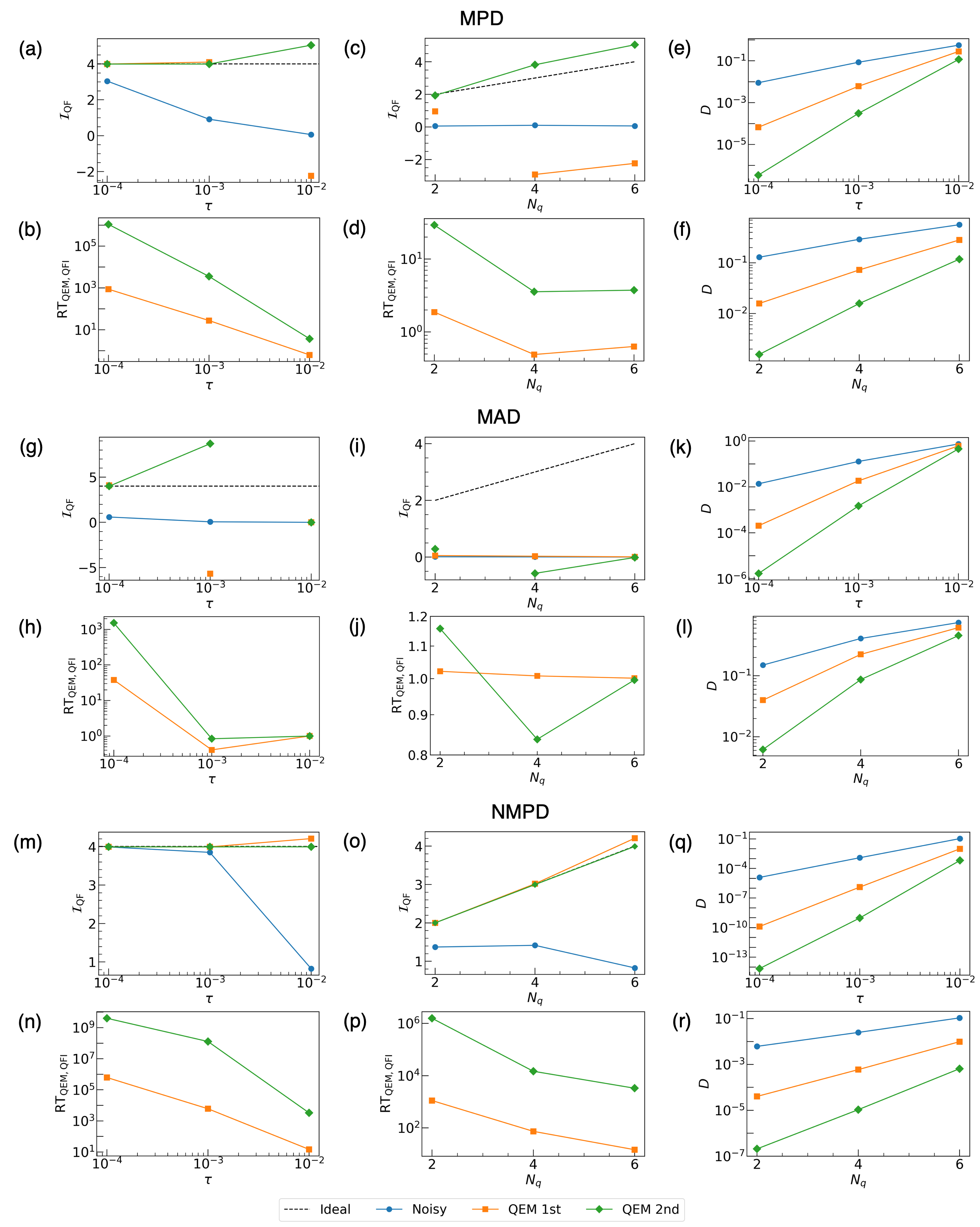}
\caption{Simulation results of  magnetic field quantum sensing for the SDS.   
In (a)-(f), (g)-(l), and (m)-(r) we present the results for the cases of the MPD, the MAD, and the NMPD, respectively.
In (a) and (m) we plot the CFI  by taking the horizontal axes to be $\tau^\text{PD}$ 
while in (g) we take the horizontal axis to be $\tau^\text{AD}$ and set $N_q=6$.
In (c), (i), and (o) we take $N_q$ for the horizontal axes with $\tau^\text{PD}=\tau^\text{AD}=10^{-2}.$
In (b)  and (n) we plot $\text{RT}_\text{QEM,QFI}$  by taking the horizontal axes to be $\tau^\text{PD}$ whereas in (h) we take the horizontal axis to be $\tau^\text{AD}$ 
with $N_q=6$. In (d), (j), and (p) we take $N_q$ for the horizontal axes with $\tau^\text{PD}=\tau^\text{AD}=10^{-2}.$ 
In (e), (k), and (q) we plot the trace distance. The horizontal axes 
 in (e) and (q) are taken to be $\tau^\text{PD}$ while the horizontal axis for (k) denotes
$\tau^\text{AD}$ and $N_q=6$. In  (f), (l), and (r) we take $N_q$ for the horizontal axes with $\tau^\text{PD}=\tau^\text{AD}=10^{-2}.$ The black dashed lines, the blue, orange, and green curves represent the ideal CFI, the noisy CFI, the first-order quantum-error-mitigated CFI,
and the second-order quantum-error-mitigated CFI, respectively.}  
\label{resultsSDS} 
\end{figure*} 
\subsection{SDS}\label{SDS}
As our final simulation, we show the results for the  symmetric Dicke states (SDS) \cite{toth2014quantum,pezze2018quantum,Agarwalltxb}.
As similar to the case of the GHZ states, we partition the $N_q$ qubits into the $\mu$ groups of the $N$-qubit systems and the mathematical representation of the SDS is given by
\begin{align}
 |  \text{SDS}_{N} \rangle = \left[ \left (
		\begin{array}{c} 
		 N  \\
		\frac{N}{2}
		\end{array}
	\right )^{-\frac{1}{2}}  \sum_{k}	\mathcal{P}_k  \left( |0 \rangle^{ \otimes \frac{N}{2}} |1 \rangle^{ \otimes \frac{N}{2}} \right)\right]^{\otimes \mu}, 
\label{SDSdef}
 \end{align}
 where the index $k$ describes all possible permutations and  there are $\left (
		\begin{array}{c} 
		 N  \\
		\frac{N}{2}
		\end{array}
	\right )$ different ways, and $\mathcal{P}_k $ is the permutation operation based on the permutation $k$.
 We note again that we take $N_q$ as well as $N$ to be  even numbers. 
 The Dicke state $|  \text{SDS}_{N} \rangle$ in Eq. \eqref{SDSdef} is equivalent to the collective spin state $|j,m \rangle$ with $j=\frac{N}{2},m=0$. The collective spin state
 $|j,m \rangle$ satisfies $\boldsymbol{J}^2|j,m \rangle = j(j+1)|j,m \rangle, J^z|j,m \rangle = m|j,m \rangle$, where  $m=-j,-j+1,\ldots,j-1,j$ \cite{Agarwalltxb}.
 The operator $\boldsymbol{J}^2$ is defined by $\boldsymbol{J}^2=\sum_{\alpha=x,y,z}(J^\alpha)^2$ with $J^\alpha=\sum_{j=0}^{N_q-1}\frac{\sigma^\alpha_j}{2}$.
 
 Let us explain the quantum circuits which generate the SDS for $N=2,4,6$
 and we present them in Fig. \ref{SDScircuit}: for quantum circuits generating  $|\text{SDS}_{N} \rangle$ for $N\geq 6$, see for instance \cite{aktar2022divide,bartschi2022short}.
   $ |  \text{SDS}_2 \rangle $  is equivalent to the Bell state $ | \Phi^+\rangle = \frac{|00\rangle+|11\rangle}{\sqrt{2}}$, 
 and it can be created by the unitary operation $U^{2}_{2}\cdot X_{Q_0}$ and its quantum-circuit expression is given in Fig. \ref{SDScircuit} (a); another way to generate $  | \Phi^+\rangle$  is the unitary operation $U^\text{SDS}_2 = \text{C}X[Q_0;Q_1]\cdot H_{0}.$  
The quantum circuit for generating the SDS for $N=4$ and $N=6$ are shown in Figs. \ref{SDScircuit}(b) and (d), respectively,
and for the creation of $ |  \text{SDS}_6 \rangle $ we use a unitary operation $U^3_3$ whose quantum circuit is depicted in
Fig. \ref{SDScircuit}(c). The rotation angles appearing in these circuits are defined by \cite{aktar2022divide,bartschi2022short} 
\begin{widetext}
\footnotesize
\begin{align}
\theta_{2 / 3}  =\arccos \sqrt{\frac{2}{3}}  \quad \theta_{1 / 2}  =\arccos \sqrt{\frac{1}{2}}  \quad
 \theta_i^{n, m, \ell}  =2 \arccos \sqrt{\frac{x_i^{n, m, \ell}}{s_i^{n, m, \ell}}}, \quad s_i^{n, m, \ell} = \sum_{j=i}^{\ell} x_j^{n, m, \ell} \quad
x_i^{n, m, \ell} 
= \left(\begin{array}{c}
    m \\
    i
\end{array}\right)\left(\begin{array}{c}
    n-m \\
    \ell-i
\end{array}\right).
\end{align} \normalsize\end{widetext}
The quantum circuit in Fig. \ref{SDScircuit} (b) is composed by the rotation angle $\theta_i^{n, m, \ell} $ for $(i,n,m,l)=(0,4,2,1),(0,4,2,2),(1,4,2,2)$ whereas
that in Fig. \ref{SDScircuit} (d) by $\theta_i^{n, m, \ell} $ for $(i,n,m,l)=(0,6,3,1),(0,6,3,2),(0,6,3,3),(1,6,3,2),(2,6,3,3).$  By using the error-propagation formula \eqref{errorpropagation} with setting $O=(J^z)^2$ the ideal value of CFI is calculated as \cite{pezze2018quantum,lucke2011twin}
\begin{widetext} \begin{align}
 \mathcal{I}_{\text{CF}} \left[  \rho^{\text{out},\text{SDS}_{N}}_{\text{ideal},\vartheta}; P_{|x\rangle, |x\rangle \in \{ |0\rangle, |1\rangle  \}^{\otimes N}} \right] =  \mu\left[ \frac{2}{N(N+2)} \left( \left(
        \frac{N(N+2)}{16} - \frac{1}{2} \right)  \tan^{2} \vartheta +1\right) \right]^{-1},
 \label{SDSQFI1}
\end{align}\end{widetext}
where $\rho^{\text{out},\text{SDS}_{N}}_{\text{ideal},\vartheta}=U^\text{free}_\vartheta|\text{SDS}_{N} \rangle(U^\text{free}_\vartheta|\text{SDS}_{N} \rangle)^\dagger$ with $H^\text{free}=J^x$ or $J^y$.
The POVM which yields the CFI in Eq. \eqref{SDSQFI1} is the projection operators of the computational basis states of the $N$-qubit system, $P_{|x\rangle, |x\rangle \in \{ |0\rangle, |1\rangle  \}^{\otimes N}}$.
Namely, the magnetic of the $x$- or $y$-component is inferred in the magnetic field quantum sensing with the usage of the SDS. 
  $\mathcal{I}_{\text{CF}} \left[  \rho^{\text{out},\text{SDS}_{N}}_{\text{ideal},\vartheta}; P_{|x\rangle, |x\rangle \in \{ |0\rangle, |1\rangle  \}^{\otimes N}} \right]$ takes maximum at $\vartheta=n\pi$ with $n$ denoting an integer and we obtain $ \mathcal{I}_{\text{QF}} \left[  \rho^{\text{out},\text{SDS}_{N}}_{\text{ideal},\vartheta}\right] = \mu\frac{N(N+2)}{2}.$
  Note that by choosing the initial state to be $|j,m\rangle$ and
 using the formula $ I_\text{QF} \left[  \rho^\text{out}_\vartheta   \right]  =  4\big{(} \langle (H^\text{free})^2 \rangle_{\rho^\text{in}} - \langle H^\text{free} \rangle_{\rho^\text{in}}^2 \big{)} $ with $H^\text{free}=J^x,J^y$, 
  we obtain $ I_\text{QF} \left[  \rho^\text{out}_{j,m,\vartheta}   \right]  =
 \frac{N(N+2)}{2} -m^2,$ where $\rho^\text{out}_{j,m,\vartheta}=U^\text{free}_\vartheta|j,m \rangle(U^\text{free}_\vartheta|j,m \rangle)^\dagger$ \cite{pezze2018quantum}.
\begin{figure*}[!t] 
\centering
\includegraphics[width=0.9\textwidth]{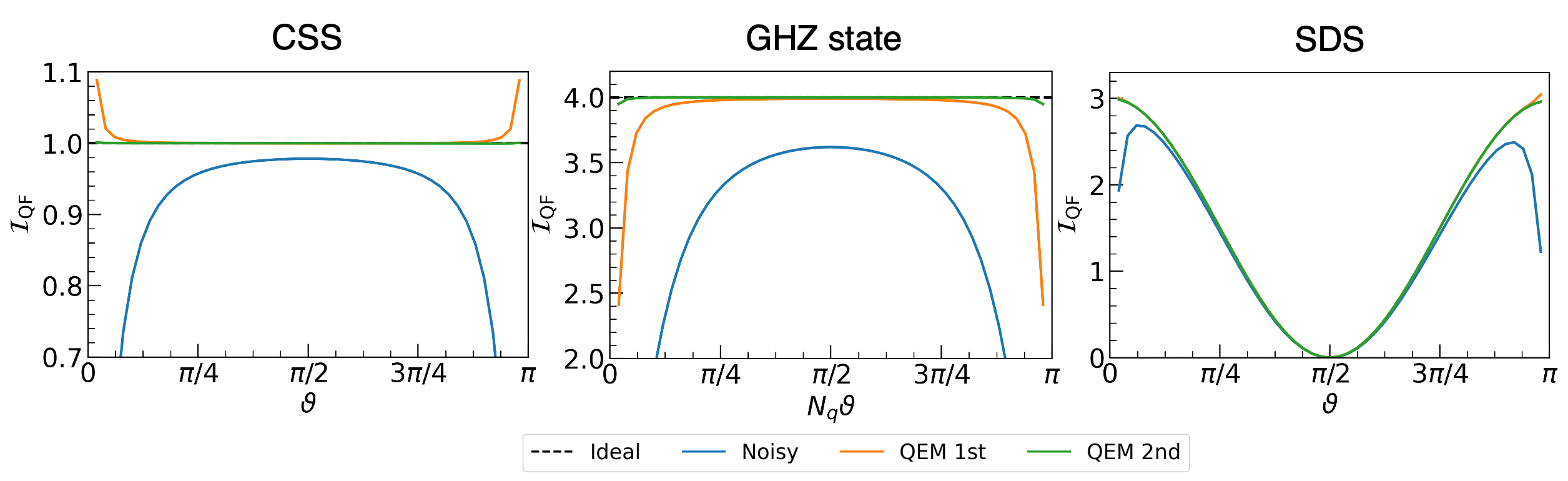}
\caption{Simulation results of magnetic field  quantum sensing for (a) CSS, (b) GHZ state, and (c) SDS with $N_q=4.$ 
For all these figures, the vertical axes describe CFI while for (a) and (c) the horizontal axes describe $\vartheta$ and for (b) the horizontal axis describes $N_q\vartheta$ such that $T^\text{free}=1$. The black dashed lines, the blue curves, the  orange curves, and the green curves represent the ideal CFI (QFI), the noisy CFI, the first-order CFI (QFI), and   the second-order CFI (QFI),  respectively. We choose the quantum noise channel to be the MPD.     
}  
\label{resultsOmegaMPD} 
\end{figure*} 

Let us discuss our simulation results in Fig. \ref{resultsSDS}.
 We take the same numerical settings in terms of $\tau^\text{PD},\tau^\text{AD},$ and $N_q$ with the ones used in the simulations for the CSS and the GHZ states while taking $\vartheta=\frac{\pi}{100}$.
The plots for the results obtained under the same numerical settings with those for the CSS as well as the GHZ states are labeled by the same alphabets from (a) to (r). 
In these simulations the Toffoli gates in Fig. \ref{SDScircuit} 
are treated not as a composite of single- and two-qubit gates but as a single gate operation in order to save the computational times. 
We see that for the MPD and the MAD although the trace distances of the quantum-error mitigated states and the ideal states are smaller than those of the noisy states and the ideal states the quantum-error-mitigated CFI  show radically the different behaviors from those of the ideal CFI except for $N_q=2$ and 
$\tau^\text{PD}=10^{-4}$ and $\tau^\text{AD}=10^{-4}$. 
For instance, as we see in Figs. \ref{resultsSDS}(a) and (c) the second-order-QEM CFI
$ \mathcal{I}_{\text{CF}} \left[  \rho^\text{out}_{\text{QEM}_{\text{2nd}},\vartheta}; P_{|x\rangle, |x\rangle \in \{ |0\rangle, |1\rangle  \}^{\otimes N}}   \right]$ exhibit increasing behaviors with respect to $\tau^\text{PD}$ and $N_q$, respectively, while the first-order-QEM CFI $ \mathcal{I}_{\text{CF}} \left[  \rho^\text{out}_{\text{QEM}_{\text{1st}},\vartheta}; P_{|x\rangle, |x\rangle \in \{ |0\rangle, |1\rangle  \}^{\otimes N}}   \right]$ in Figs. \ref{resultsSDS}(a) and (c) and the second-order-QEM CFI $ \mathcal{I}_{\text{CF}} \left[  \rho^\text{out}_{\text{QEM}_{\text{2nd}},\vartheta}; P_{|x\rangle, |x\rangle \in \{ |0\rangle, |1\rangle  \}^{\otimes N}}   \right]$ in Fig. \ref{resultsSDS}(i) (the result for $\tau^\text{AD}=10^{-2}$) show the sign changing from positive to negative. Moreover, in Fig. \ref{resultsSDS}(i) we observe that $ \mathcal{I}_{\text{CF}} \left[  \rho^\text{out}_{\text{QEM}_{\text{1st}},\vartheta}; P_{|x\rangle, |x\rangle \in \{ |0\rangle, |1\rangle  \}^{\otimes N}} \right]$
vanishes for every $N_q$ while  we see the vanishing of $ \mathcal{I}_{\text{CF}} \left[  \rho^\text{out}_{\text{QEM}_{\text{2nd}},\vartheta}; P_{|x\rangle, |x\rangle \in \{ |0\rangle, |1\rangle  \}^{\otimes N}} \right]$ 
at $N_q=6$. 
In the case of the NMPD the noise effects are small enough that our protocol is effective as presented in in Fig. \ref{resultsSDS} (n) and (p), i.e., the ratios $\text{RT}_\text{QEM,QFI}$ exceed 14. 
The analysis of understanding such singular characteristics of the quantum-error-mitigated CFI for the MPD and MAD
are complicated and we explain the details of them in Appendix \ref{Dickeappendix}.  
We consider that such radical differences between the quantum-error mitigated CFI and the ideal CFI  occur since the depth of the quantum circuit of $U^\text{in}$ is too deep for large $N_q$ ($d^\text{in}=14$ and $34$ for $N_q=4$ and $6$, respectively) and the associated quantum noise effects due to large $\tau^\text{PD}$ and $\tau^\text{AD}$ are too big. 
Consequently, our QEM protocol fails except for $N_q=2$ and $N_q\geq4$ with $\tau^\text{PD}=10^{-4}$ and $\tau^\text{AD}=10^{-4}$,
and in order to leverage our protocol for the cases of large $\tau^\text{PD}$ and $\tau^\text{AD}$ with big $N_q$  much higher-order perturbative calculations are necessary.   

\section{Discussions}\label{discussions}
In this section, we discuss the following three themes of our QEM protocol,  $\vartheta$ dependence and advantages, comparison with other methods, and the efficacy of our protocol under NISQ-device parameters.
\subsection{$\vartheta$ Dependence and Advantages}\label{thetadependence}
The simulation results presented in Sec. \ref{nss} are the results obtained by the simulations with the values of  $\vartheta$ being fixed: 
$\vartheta=\frac{\pi}{2},\frac{\pi}{2N_q},$ and $\frac{\pi}{100}$ for the CSS, the GHZ states, and the SDS, respectively. 
Here we perform simulations by varying $\vartheta = \omega T^\text{free}$ such that  the value of $T^\text{free}$ is fixed to be one while  $\omega$ is varied as  
$\omega=\frac{\pi}{50}\times i_\omega$ and $\omega=\frac{\pi}{50 N_q}\times i_\omega$ for the CSS and the SDS, and the GHZ state, respectively while taking $N_q=4,\mu=1, \gamma_\text{PD}=10^{-2}$ ($\tau^\text{PD}=10^{-3}$), and examine the  $\vartheta$ dependence of the CFI and the QFI, in particular the quantum-error-mitigated QFI.
Recall that $\omega = \gamma_\text{g}B$ and the physical operators $O$  are taken to be $J^x, X^{\otimes N_q},$ and $(J^z)^2$ for the CSS, the GHZ state, and the SDS, respectively.
Namely, such simulations are the simulations of CFI and QFI to examine how the magnetic field sensitivity changes with respect to the magnitude of the applied magnetic field $B.$ For all the simulations quantum noise is chosen to be MPD.
First, let us analyze from the case of quantum metrology with an initial state taken to be the CSS. The simulation result is presented in Fig. \ref{resultsOmegaMPD}(a).
The ideal CFI or QFI $ \mathcal{I}_{\text{QF}} \left[ \rho^{\text{out,CSS}}_{\text{ideal},\vartheta} \right]$ is always constant which is equal to one.
In contrast, the noisy CFI  $\mathcal{ I}_{\text{CF}} \left[\rho^{\text{out,CSS}}_{\text{MPD},\vartheta}; P_{|x\rangle, |x\rangle \in \{ |+\rangle, |-\rangle  \}} \right] $ shows the oscillatory behavior (damping oscillation) which originates from the trigonometric functions $\cos\vartheta$ and $\sin\vartheta$ (see Eq. \eqref{CFIMPDCSS}) and takes maximum at $\vartheta=\frac{\pi}{2}$.
The quantum-error-mitigated CFI or QFI $\mathcal{ I}_{\text{QF}} \left[\rho^{\text{out,CSS}}_{\text{QEM}_{\text{1st}},\vartheta} \right]$ and  
$\mathcal{ I}_{\text{QF}} \left[\rho^{\text{out,CSS}}_{\text{QEM}_{\text{2nd}},\vartheta} \right]$
show the similar behaviors with the behavior of  $ \mathcal{I}_{\text{QF}} \left[ \rho^{\text{out,CSS}}_{\text{ideal},\vartheta} \right]$, and the behavior of $\mathcal{ I}_{\text{QF}} \left[\rho^{\text{out,CSS}}_{\text{QEM}_{\text{2nd}},\vartheta} \right]$ is closer to
that of  $ \mathcal{I}_{\text{QF}} \left[ \rho^{\text{out,CSS}}_{\text{ideal},\vartheta} \right]$ compared to the behavior of $\mathcal{ I}_{\text{QF}} \left[\rho^{\text{out,CSS}}_{\text{QEM}_{\text{1st}},\vartheta} \right]$ as expected.
Like the ideal case, the quantum-error-mitigated QFI $\mathcal{ I}_{\text{QF}} \left[\rho^{\text{out,CSS}}_{\text{QEM}_{\text{1st/2nd}},\vartheta} \right]$ are flat in almost all the values of $\vartheta$: 
$\mathcal{ I}_{\text{QF}} \left[\rho^{\text{out,CSS}}_{\text{QEM}_{\text{1st}},\vartheta} \right]$ is approximately flat in the range $\left(\frac{\pi}{8}, \frac{7\pi}{8}\right)$. This  is the indication such that the quantum-error-mitigated states are closer to the ideal state compared to the noisy state. Thus, our QEM protocol  works well.   
Next, let us analyze the simulation results of quantum metrology with an initial state taken to be the GHZ state which are shown in Fig. \ref{resultsOmegaMPD}(b). 
Like the case of the CSS, $ \mathcal{I}_{\text{QF}} \left[ \rho^{\text{out,GHZ}_{N_q}}_{\text{ideal},\vartheta} \right]$ is always constant and is equal to $N_q$ (=4).
On the other hand, $\mathcal{ I}_{\text{CF}} \left[\rho^{\text{out,GHZ}_{N_q}}_{\text{MPD},\vartheta}; P_{|x\rangle, |x\rangle \in \{ |+\rangle, |-\rangle  \}^{\otimes N_q}} \right] $ shows the rapid oscillation originating from 
the trigonometric functions $\cos(4 \vartheta)$ and $\sin(4\vartheta)$ (see Eq. \eqref{CFIMPDGHZ}) and takes maximum at $\vartheta=\frac{\pi}{2\cdot4}$. Such a shortening of the period given by the factor $N_q=4$ is due to the quantum entanglement which is the attribution of the GHZ state. The quantum-error-mitigated QFI $\mathcal{ I}_{\text{QF}} \left[\rho^{\text{out,GHZ}_{N_q}}_{\text{QEM}_{\text{1st/2nd}},\vartheta} \right]$ also show the oscillatory behaviors since the quantum-error-mitigated states are the quantities obtained from the noisy state $\rho^{\text{out,GHZ}_{N_q}}_{\text{MPD},\vartheta}$. The periods are, however, much longer than the period of $\mathcal{ I}_{\text{QF}} \left[\rho^{\text{out,GHZ}_{N_q}}_{\text{MPD},\vartheta} \right] $ and can considered to be flat in almost entire area of $\vartheta$, $\vartheta \in \left(\frac{\pi}{8N_q}, \frac{7\pi}{8N_q} \right)$, and their values are approximately equal to four. Thus, such a result implies that our QEM protocol also works for the case of the GHZ state with the various values of $\vartheta$. 
The deviation of $\mathcal{ I}_{\text{QF}} \left[\rho^{\text{out,GHZ}}_{\text{QEM}_{\text{1st}},\vartheta} \right]$ from  $\mathcal{ I}_{\text{QF}} \left[\rho^{\text{out,GHZ}}_{\text{ideal},\vartheta} \right]$ 
  in the ranges $\left[0,\frac{\pi}{8N_q}\right]$ and $\left[\frac{7\pi}{8N_q}, \frac{\pi}{N_q}\right]$ can be understood as follows: the analysis for the case of the CSS can be done by taking $N_q\to1$.  
We set $t^\text{PD}(N_q) = 2\big{(} N^\text{GHZ}_\text{PD}(N) +N_\text{free}N \big{)}\tau^\text{PD}$ and expand $\mathcal{ I}_{\text{QF}} \left[\rho^{\text{out,GHZ}}_{\text{noisye},\vartheta} \right]$ given in Eq.  \eqref{CFIMPDGHZ}
and we obtain $\mathcal{ I}_{\text{QF}} \left[\rho^{\text{out,GHZ}}_{\text{noisy},\vartheta} \right]= N_q +c^\text{MPD,GHZ}_1t^\text{PD}(N_q)+\frac{1}{2}c^\text{MPD,GHZ}_2\big{(}t^\text{PD}(N_q)\big{)}^2 +\mathcal{O}\left(t^\text{PD}(N_q) \right)^3.$ 
The coefficients $c^\text{MPD,GHZ}_1,c^\text{MPD,GHZ}_2$ are $c^\text{MPD,GHZ}_1 \propto \sin^{-2}N_q\vartheta, c^\text{MPD,GHZ}_2  \propto \sin^{-4}N_q\vartheta$, and thus 
$\mathcal{ I}_{\text{QF}} \left[\rho^{\text{out,GHZ}}_{\text{QEM}_{\text{1st}},\vartheta} \right]$ largely deviates from  $\mathcal{ I}_{\text{QF}} \left[\rho^{\text{out,GHZ}}_{\text{ideal},\vartheta} \right]$ as $\vartheta$
approaches zero. 
As a result, our QEM protocol is effective for the various values of $\vartheta$ and the quantum-error-mitigated QFI take the values which are approximately equal to the ideal QFI for these two cases. 
Finally, we discuss the results of the $\vartheta$ dependence for the SDS. In this case, the quantum-error-mitigated CFI exhibit approximately the same behavior with the ideal CFI in the entire region and thus our QEM protocol also works for the SDS with any value of $\vartheta$.    

Consequently, our protocol is effective for all the initial states with the various values of $\vartheta$. 
It is able to broaden the range of $\vartheta$ where our protocol is valid provided that much higher-order perturbative calculations are done or when $\tau^\text{PD}$ gets smaller. 
Let us comment two advantages of our protocol according to the result in Fig. \ref{resultsOmegaMPD}.
First, in the cases of noisy quantum metrology the analysis of
 CFI, QFI, or the optimization of sensitivities are much more cumbersome than ideal cases \cite{toth2014quantum,pezze2018quantum,degen2017quantum,huelga1997improvement,escher2011general,matsuzaki2011magnetic,demkowicz2012elusive,chin2012quantum,chaves2013noisy,kolodynski2013efficient,demkowicz2014using,alipour2014quantum,alipour2014quantum,ozaydin2014phase,jeske2014quantum,macieszczak2015zeno,brask2015improved,smirne2016ultimate,sekatski2017quantum,demkowicz2017adaptive,hou2017quantum,matsuzaki2018quantum,koczor2020variational,he2021quantum,long2022entanglement}.  
As we can see from Figs. \ref{resultsCSS}, \ref{resultsGHZ}, \ref{resultsSDS}, and \ref{resultsOmegaMPD}, the quantum-error-mitigated states are sufficiently close to the ideal states and the optimal value of $\vartheta$ for the quantum-error-mitigated CFI is (almost) equal to that for the ideal CFI.  
This implies that the optimization of quantum-error-mitigated CFI can be done with the same computational cost of the optimization of ideal CFI.
Second, the quantum-error-mitigated CFI for all the three states are equal to the ideal QFI at  small $\vartheta$. Such a result  
is a great benefit for  weak magnetic field sensing.
We expect that by using our QEM protocol we can perform the  quantum  sensing whose sensitivity is given approximately by the ideal QFI.
We insist that these two advantages become powerful tools in quantum metrology with initial states taken to be quantum entangled states. 
This is because owing to our QEM protocol we can conduct quantum metrology with high sensitivities assisted by the quantum entanglement which are described (approximately) by ideal QFI, which is $\mathcal{O}(N_q^2)$, 
even under the presence of quantum noise.  
Before ending, we note that the similar simulations can be done for noisy quantum metrology of $\omega$, and the CFI as well as the QFI of $\omega$ can be easily calculated by those of $\vartheta$ with using the property of a variance $\text{Var}(aX)=a^2\text{Var}(X)$ ($a$ is a real constant and $X$ is a random variable)
or the error-propagation formula \eqref{errorpropagation} with using $\frac{d}{d\vartheta}=\frac{1}{T^\text{free}}\frac{d}{d\omega}$. 
Since we use the same quantum-error-mitigated states which are used for the noisy quantum metrology of $\vartheta$,
we expect from  Figs.  \ref{resultsCSS}, \ref{resultsGHZ}, \ref{resultsSDS}, and \ref{resultsOmegaMPD} that our QEM protocol also works well for the noisy quantum metrology of $\omega$. 
\subsection{Comparison with Other Methods} 
QEM is one of the important issues in the research and development of quantum computing and has been investigated  intensively \cite{QCchemistryRMP2020,hybridQCalgorithmJPSJ2021,EMPRL2017,EMNature2019,EMPRX2017,EMPRX2018,EMarxiv2018,PhysRevA.98.062339,song2019quantum,zhang2020error,mcardle2019error,jattana2020general,xiong2020sampling,zlokapa2020deep,EMPRA2021,CandSQEMPRAp2021,OttenGrayQEM1,OttenGrayQEM2,QSEQEM, CliffordQEM,LearningBasedQEM,VirtualDistillationQEM,koczor2021exponential,PRXQuantum.2.010316,piveteau2021error,lostaglio2021error,suzuki2022quantum,piveteau2022quasiprobability, pascuzzi2022computationally, takagi2021optimal, larose2022mitiq, koczor2021dominant,hama2022quantum,cai2022quantum}.
Up to now various kinds of schemes have been established  including zero-noise extrapolation (ZNE) \cite{QCchemistryRMP2020,hybridQCalgorithmJPSJ2021,EMPRL2017,EMNature2019,EMPRX2017,EMPRX2018,pascuzzi2022computationally,larose2022mitiq,cai2022quantum,zhao2021error},  a purification based method called exponential suppression by derangements or virtual distillation (ESD/VD)   \cite{koczor2021exponential,koczor2021dominant,VirtualDistillationQEM,cai2022quantum,yamamoto2022error}, and probabilistic error cancellation (PEC)  \cite{QCchemistryRMP2020,hybridQCalgorithmJPSJ2021,EMPRL2017,EMPRX2018,song2019quantum,zhang2020error,xiong2020sampling,CandSQEMPRAp2021,LearningBasedQEM,piveteau2021error,piveteau2022quasiprobability,takagi2021optimal,larose2022mitiq,cai2022quantum}.
 ZNE is one of the powerful method for the reduction of error in noisy quantum computing and also has been applied to that in noisy quantum metrology \cite{zhao2021error}.  
At present, however,  it only works for time-independent quantum noise channels \cite{EMPRL2017,EMNature2019,zhao2021error} whereas our protocol can be applied to time-dependent quantum noise (non-Markovian quantum noise).
  ESD/VD has been recently applied to noisy quantum metrology for the reduction of systematic errors which describe the noise fluctuation with respect to the experimental runs of quantum metrology (e.g., a $T_1$ time  is different  for every experimental run) but not to the reduction of decoherence effects \cite{yamamoto2022error}.
In this work, we make a comparison with  PEC. 
The reason we do so is because the PEC is conceptually similar to our protocol in the sense that  PEC is conducted by additional quantum circuits which are constructed by insertions of additional quantum computational operations (gate operations and resets) into original quantum circuits (quantum circuits describing the quantum algorithms to be ran).   

First, let us briefly explain the formalism of PEC.   In this method, error mitigation is performed by constructing the inverse operations of noisy gate operations which consist erroneous original circuits called recovery operations. The recovery operations act on single-qubit states for every layer of  quantum algorithms (let us assume that every qubit is influenced by the same quantum noise effect independently) and are constructed by sixteen basis operations $\mathcal{B}_{a}$ (superoperators acting on density matrices) with $a=1,\ldots,16$ and quasiprobabilities $\eta_b$ $(b=1,\ldots,N_\text{quasp})$,  where $N_\text{quasp}$ is the number of finite quasiprobabilities and they can be either positive or negative values. 
By denoting the recovery operation acting on the $j$-th qubit in the $k$-th layer of the quantum algorithm ($i=1,\ldots,d^\text{tot}$) by 
$ \mathcal{E}^{-1}_{j;k} = \sum_{a_{j;k}=1}^{N_\text{quasp}}  \eta_{a_{j;k}} \mathcal{B}_{a_{j;k}}$, the total recovery operation to be performed is described as
$ \mathcal{E}^{-1}_\text{tot} =  \prod_{k=1}^{d^\text{tot}} \bigotimes_{j=1}^{N_q} \left[\sum_{a_{j;k}=1}^{N_\text{quasp}}  \eta_{a_{j;k}} \mathcal{B}_{a_{j;k}} \right]=
\prod_{k=1}^{d^{\text{tot}}}  \mathcal{E}^{-1}_{k}$.
Thus, the total recovery operation $ \mathcal{E}^{-1}_\text{tot} $ is composed of $N_\text{quasp}^{N_qd^\text{tot}}$ types of quantum circuits and the associated $N_\text{quasp}^{N_qd^{\text{tot}}}$ quasiprobabilities. 
Note that  the $N_\text{quasp}^{N_qd^\text{tot}}$ quantum circuits include the original circuit.
 To write these formulas compactly hereafter we rewrite $ \mathcal{E}^{-1}_\text{tot} $ as 
 $ \mathcal{E}^{-1}_\text{tot} = \sum_{\alpha=1}^{N_\text{quasp}^{N_qd^\text{tot}}}  \eta_{\alpha} \mathcal{B}_{\alpha} =  
 \prod_{k=1}^{d^\text{tot}} \sum_{\alpha_k=1}^{N_\text{quasp}^{N_q}}  \eta_{\alpha_k} \mathcal{B}_{\alpha_k} $. 
 Namely, PEC is performed by $N_\text{quasp}^{N_qd^\text{tot}}$ types of  quantum circuits which are constructed by the basis operations $ \mathcal{B}_{\alpha}$ and the associated quasiprobabilities  $\eta_{\alpha}$.
 Let us denote an expectation value of a physical operator $O$ with respect to a quantum state $\rho$ by $\langle O \rangle_\rho(=\text{Tr}(O\rho))$. 
 By writing the ideal quantum state generated by the unitary operation $\prod_{k=1}^{d^\text{tot}} U_k$ and the one created by  the recovery operation $ \mathcal{E}^{-1}_\text{tot} $
 by $\rho_{d^\text{tot}\cdots1}$ and $\rho^\text{PEC}_{d^\text{tot}\cdots1}$,  respectively,
  the two expectation values $\langle O \rangle_{\rho_{d^\text{tot}\cdots1}}$ and $\langle O \rangle_{\rho^\text{PEC}_{d^\text{tot}\cdots1}}$ are  equivalent and it is described such that 
\begin{align}
\rho_{d^\text{tot}\cdots1} = \rho^\text{PEC}_{d^\text{tot}\cdots1}= \prod_{k=1}^{d^\text{tot}}  \sum_{\alpha_k=1}^{N_\text{quasp}^{N_q}}   \eta_{\alpha_k} \mathcal{B}_{\alpha_k} \left[  \mathcal{E}_{k} \left[\rho_{k\cdots1}\right] \right].
\label{PECformula1}
\end{align}
\begin{figure*}[!t] 
\centering
\includegraphics[width=0.8\textwidth]{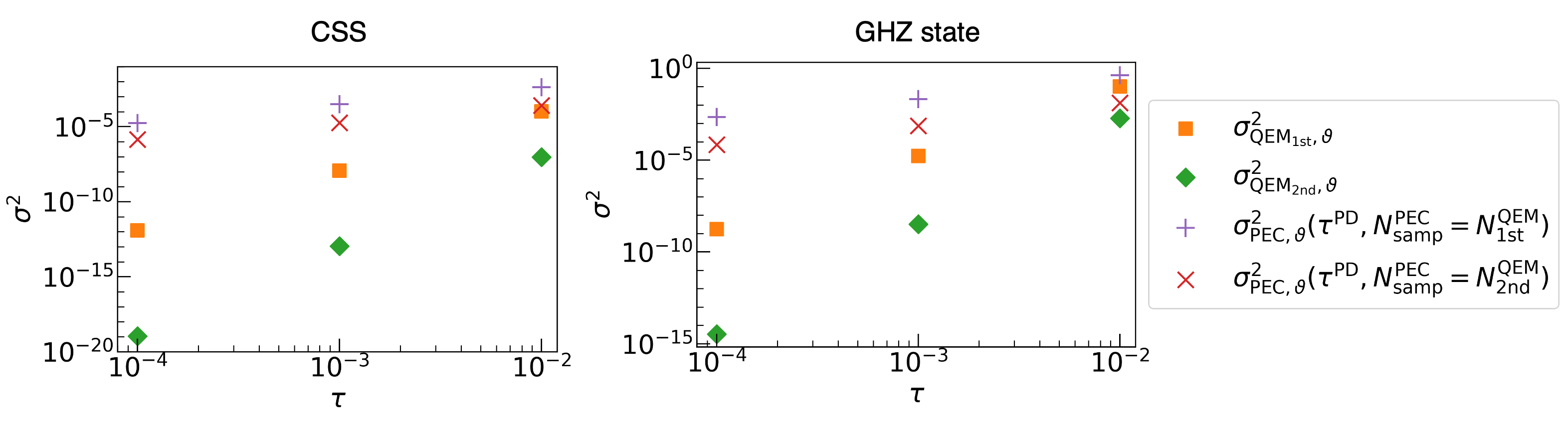}
\caption{Numerical comparisons between our protocol and PEC for (a) the CSS, (b) the GHZ state with $N_q=3$ and  $N_\text{rept}=10^2$.
The variances $ \sigma^2_{\text{QEM}_\text{1st},\vartheta}$ and $ \sigma^2_{\text{QEM}_\text{2nd},\vartheta}$ are plotted by orange squares and green diamonds, respectively, 
while $\sigma^2_{\text{PEC},\vartheta}(\tau^\text{PD},N^\text{PEC}_\text{samp}=N^\text{QEM}_\text{1st})$ and $\sigma^2_{\text{PEC},\vartheta}(\tau^\text{PD},N^\text{PEC}_\text{samp}=N^\text{QEM}_\text{2nd})$ 
are plotted by purple crosses and red saltires, respectively.    }  
\label{QEMandPEC} 
\end{figure*} 
On the other hand, in order to numerically  perform  PEC we need to introduce a sampling number associated with the quasiprobability and we denote it by $N^\text{PEC}_\text{samp}:$
we call it PEC sampling number.
PEC is numerically performed by generating $N^\text{PEC}_\text{samp}$ types of quantum circuits such that each quantum circuit is generated with the probability 
$\frac{1}{N^\text{PEC}_\text{samp}}$ and let us denote the original circuit, an $i$th generated quantum circuit, and the quasiprobability associated with the appearance of the $i$th generated quantum circuit by $C_\text{org}$, $C^\text{PEC}_i$, and $\eta_{C^\text{PEC}_i},$ respectively. By writing the expectation value of $O$ obtained by the quantum circuit $C^\text{PEC}_i$ by $\langle O \rangle_{C^\text{PEC}_i}$,  a numerical expression of the PEC-performed expectation value  $\langle O \rangle_{ \rho^\text{PEC}_{d\cdots1} }$, which we denote by $\langle O \rangle_{[N^\text{PEC}_\text{samp};\rho_{d\cdots1}]}$, is given by \cite{EMPRL2017}
\begin{align}
\langle O \rangle_{[N^\text{PEC}_\text{samp};\rho_{d\cdots1}]} = \sum_{i=1}^{N^\text{PEC}_\text{samp}} \frac{\text{sgn}(\eta_{C^\text{PEC}_i})}{N^\text{PEC}_\text{samp}}\langle O \rangle_{C^\text{PEC}_i}.
\label{PECformula2}
\end{align}
In the limit $N^\text{PEC}_\text{samp} \to \infty$, we obtain $\langle O \rangle_{[N^\text{PEC}_\text{samp};\rho_{d\cdots1}]} \to \langle O \rangle_{ \rho^\text{PEC}_{d\cdots1} }.$
Let us introduce accuracy $\epsilon$ defined by $\epsilon=\Big{|}   \langle O \rangle_{\rho_{d\cdots1}}-\langle O \rangle_{[N^\text{PEC}_\text{samp};\rho_{d\cdots1}]}     \Big{|}$. 
The order of $N^\text{PEC}_\text{samp}$ in $\epsilon$ is  $\mathcal{O}\left(\epsilon^{-2}\right)$ \cite{EMPRL2017,EMPRX2018,CandSQEMPRAp2021,takagi2021optimal,larose2022mitiq,cai2022quantum}, i.e., in the limit $N^\text{PEC}_\text{samp}\to\infty$ we obtain $\epsilon\to 0$, which describes the exact error mitigation. 

Let us explain the numerical setting for the comparison between our protocol and PEC. 
We do this for quantum metrology under the MPD where initial states are taken to be the CSS and the GHZ  state for $N_q=3$ by using the error-propagation formula \eqref{errorpropagation}.
For the parameters, we take $\Delta t=0.1$, $\tau^\text{PD}=10^{-n^\text{PD}}$ ($n^\text{PD}=2,3,4$) and $d^\text{free}=10$.
To make a fair comparison, we perform the simulations of these two methods such that each simulation is executed under the same amount of quantum computational resource.
In this case, we consider it to be the number of quantum circuits for performing error mitigation since both methods are conducted by ensembles of quantum circuits. 
In these simulations we take $N^\text{PEC}_\text{samp}=N_qd^\text{tot}+1$ and 
$N_qd^\text{tot}\left(1+\frac{N_q(d^\text{tot}+1)}{2} \right)+1$, which are the numbers of quantum circuits to perform the first-order and the second-order QEM based on our protocol, respectively. 
Such a comparison can be described as the numerical comparison under the same amount of quantum computational resource (the number of ensembles of quantum circuits) 
to examine whether error-mitigated sensitivities $\left[\text{Var}[\vartheta]\right]^{\text{QEM}_\text{1st/2nd}}$ (quantum-error-mitigated sensitivity in the first-order/second-order perturbation) are smaller than $\left[\text{Var}[\vartheta]\right]^\text{PEC}$ (quantum-error-mitigated sensitivity obtained by PEC) or not. 
In other words, we verify whether the quantum-error-mitigated CFI 
$\mathcal{ I}_{\text{CF}} \left[\rho^{\text{out,CSS}}_{\text{QEM}_{\text{1st/2nd}},\vartheta}; P_{|x\rangle, |x\rangle \in \{ |+\rangle, |-\rangle  \}} \right] = 
\left(\left[\text{Var}[\vartheta]\right]^{\text{QEM}_\text{1st/2nd}}_\text{CSS}\right)^{-1}$
and $\mathcal{ I}_{\text{CF}} \left[\rho^{\text{out,GHZ}_{N_q}}_{\text{QEM}_{\text{1st/2nd}},\vartheta}; P_{|x\rangle, |x\rangle \in \{ |+\rangle, |-\rangle  \}^{\otimes N_q}} \right]= 
\left(\left[\text{Var}[\vartheta]\right]^{\text{QEM}_\text{1st/2nd}}_{\text{GHZ}_{N_q}}\right)^{-1} $
are smaller than the PEC-performed CFI which are obtained by Eq. \eqref{PECformula2} and the error-propagation formula \eqref{errorpropagation}. 
We write the PEC-performed CFI for the CSS and the GHZ state by $\mathcal{ I}^\text{PEC,CSS}_{\text{CF},\vartheta} \left[ P_{|x\rangle, |x\rangle \in \{ |+\rangle, |-\rangle  \}} \right] =  
\left(\left[\text{Var}[\vartheta]\right]^{\text{PEC}}_\text{CSS}\right)^{-1}$
and $\mathcal{ I}^{\text{PEC,GHZ}_{N_q}}_{\text{CF},\vartheta} \left[ P_{|x\rangle, |x\rangle \in \{ |+\rangle, |-\rangle  \}^{\otimes N_q}} \right] = 
\left(\left[\text{Var}[\vartheta]\right]^{\text{PEC}}_{\text{GHZ}_{N_q}}\right)^{-1}$, respectively.
In this comparison, we need to take into account that the sensitivities $\left[\text{Var}[\vartheta]\right]^{\text{QEM}_\text{1st/2nd}}$ are deterministic whereas 
$\left[\text{Var}[\vartheta]\right]^\text{PEC}_{\text{CSS}/\text{GHZ}_{N_q}}$ are probabilistic. Let us say that we perform $N_\text{rept}$ times the computation of PEC with the sampling number $N^\text{PEC}_\text{samp}$
and write the variance of $\vartheta$ obtained in the $l$th round ($l=1,\ldots,N_\text{rept}$) by  
$\left[\text{Var}[\vartheta]\right]^\text{PEC}_{\text{CSS}/\text{GHZ}_{N_q}}(\tau^\text{PD},N^\text{PEC}_\text{samp},l)$.
 In general $\left[\text{Var}[\vartheta]\right]^\text{PEC}_{\text{CSS}/\text{GHZ}_{N_q}}(\tau^\text{PD},N^\text{PEC}_\text{samp},l_1) \neq \left[\text{Var}[\vartheta]\right]^\text{PEC}_{\text{CSS}/\text{GHZ}_{N_q}}(\tau^\text{PD},N^\text{PEC}_\text{samp},l_2)$ for $l_1\neq l_2.$
To quantify how close the variances $\left[\text{Var}[\vartheta]\right]^\text{PEC}_{\text{CSS}/\text{GHZ}_{N_q}}(\tau^\text{PD},N^\text{PEC}_\text{samp},l)$ 
are to the ideal variances $\left[\text{Var}[\vartheta]\right]^\text{ideal}_{\text{CSS}/\text{GHZ}_{N_q}}$,
we introduce a variance defined by 
 \begin{widetext} \begin{align}
\sigma^2_{\text{PEC},\vartheta}(\tau^\text{PD},N^\text{PEC}_\text{samp}) = \sum_{l=1}^{N_\text{rept}} \frac{1}{N_\text{rept}} \Big{(} 
\left[\left[\text{Var}[\vartheta]\right]^\text{PEC}(\tau^\text{PD},N^\text{PEC}_\text{samp},l)\right]^{-1}  -  
\left[\left[\text{Var}[\vartheta]\right]^\text{ideal}\right]^{-1} \Big{)}^2.
\label{PECformula3}
\end{align}  \end{widetext}
In the above equation, we have not written the subscript $\text{CSS}/\text{GHZ}_{N_q}$ and  used $\left[\text{Var}[\vartheta]\right]^{\text{QEM}_\text{1st/2nd}}$  and $\left[\text{Var}[\vartheta]\right]^\text{PEC}$
instead of CFI so as to write it shortly. 
On the other hand, we rewrite the variance $\text{Var}\left[\vartheta]\right]^{\text{QEM}_\text{1st/2nd}}$ as $\text{Var}\left[\vartheta]\right]^{\text{QEM}_\text{1st/2nd}}(\tau^\text{PD})$ to emphasize its $\tau^\text{PD}$ dependency and introduce a variance defined by $ \sigma^2_{ \text{QEM}_\text{1st/2nd},\vartheta}(\tau^\text{PD}) = 
\left(\left[\left[\text{Var}[\vartheta]\right]^{\text{QEM}_\text{1st/2nd}}(\tau^\text{PD})\right]^{-1}  
-  \left[\left[\text{Var}[\vartheta]\right]^\text{ideal}\right]^{-1} \right)^2 $, where we have omitted the the subscript $\text{CSS}/\text{GHZ}_{N_q}$.
The variance $ \sigma^2_{ \text{QEM}_\text{1st/2nd},\vartheta}(\tau^\text{PD})$ describes  how close the variances $\left[\text{Var}[\vartheta]\right]^{\text{QEM}_\text{1st/2nd}}$ are  to $\left[\text{Var}[\vartheta]\right]^\text{ideal}$:
In contrast to PEC, $\text{Var}\left[\vartheta]\right]^{\text{QEM}_\text{1st/2nd}}(\tau^\text{PD})$ is deterministic and the variance  $ \sigma^2_{\text{QEM}_\text{1st/2nd},\vartheta}$ is equivalent to the square of the error (accuracy) of our protocol. We compare the qualities of these two QEM methods by examining numerically  which variance is smaller, i.e., the method exhibiting the smaller variance has the better quality. In Fig. \ref{QEMandPEC}, we present the numerical results of  $ \sigma^2_{\text{QEM}_\text{1st/2nd},\vartheta}(\tau^\text{PD})$  and $ \sigma^2_{\text{PEC},\vartheta}(\tau^\text{PD},N^\text{PEC}_\text{samp})$: Figs. \ref{QEMandPEC} (a) and (b) are the results for the CSS and the GHZ state, respectively, and we take $N_\text{rept}=10^2$.
Here we plot  $ \sigma^2_{\text{QEM}_\text{1st},\vartheta}$ and $ \sigma^2_{\text{QEM}_\text{2nd},\vartheta}$ by orange squares and green diamonds, respectively, 
while we plot $\sigma^2_{\text{PEC},\vartheta}(\tau^\text{PD},N^\text{PEC}_\text{samp}=N^\text{QEM}_\text{1st})$ 
and $\sigma^2_{\text{PEC},\vartheta}(\tau^\text{PD},N^\text{PEC}_\text{samp}=N^\text{QEM}_\text{2nd})$ by purple crosses and red saltires, respectively. 
The two numbers $N^\text{QEM}_\text{1st}$ and $N^\text{QEM}_\text{2nd}$ are the number of quantum circuits which we need for performing the first- and second-order QEM, respectively, ($N^\text{QEM}_\text{1st}=N_qd^\text{tot}+1,N^\text{QEM}_\text{2nd}=1+N_qd^\text{tot}\left(1+\frac{N_q(d^\text{tot}+1)}{2} \right)$),
and for the CSS $N^\text{QEM}_\text{1st}=37, N^\text{QEM}_\text{2nd}= 739$ whereas for the GHZ states $N^\text{QEM}_\text{1st}=43, N^\text{QEM}_\text{2nd}= 988$.
From Fig. \ref{QEMandPEC}, we see that the relation $ \sigma^2_{\text{QEM}_\text{2nd},\vartheta}(\tau^\text{PD}) <   \sigma^2_{\text{QEM}_\text{1st},\vartheta}(\tau^\text{PD}) < \sigma^2_{\text{PEC},\vartheta}(\tau^\text{PD},N^\text{PEC}_\text{samp}=N^\text{QEM}_\text{2nd})<\sigma^2_{\text{PEC},\vartheta}(\tau^\text{PD},N^\text{PEC}_\text{samp}=N^\text{QEM}_\text{1st})$ is satisfied except for the GHZ state with $\tau^\text{PD} =10^{-2}$, and in this case $\sigma^2_{\text{QEM}_\text{2nd},\vartheta}=1.9\times 10^{-3},\sigma^2_{\text{PEC},\vartheta}(\tau^\text{PD},N^\text{PEC}_\text{samp}=N^\text{QEM}_\text{2nd})= 1.3 \times 10^{-2} $. 
In addition to the above simulations, we have also performed the simulations for $N^\text{PEC}_\text{rept}=10^3$ and we have numerically verified that the same relation of the four variances is satisfied.
As a result, our QEM protocol outperforms PEC.

At the end, let us make some comments on advantages of our protocol compared to PEC.
Firstly, PEC is a probabilistic theory and every time we perform it the quantum-error-mitigated expectation values and sensitivities we obtain are different in general while in our protocol we always obtain the same ones. This means that  PEC requires additional computational cost given by the repetition number $N_\text{rept}$ to check its accuracy as we have demonstrated above whereas it is not needed for performing our protocol. 
Secondly, for both methods the operations which become erroneous in real devices are not only the unitary operations $ U_k$ ($k=1,\ldots,d^\text{tot}$) but also the additional operations which constitute the ensembles of quantum circuits for error mitigation. In PEC the errors associated with these additional operations are not mitigated 
and the error-cancelled quantum states and the associated expectation values described by Eq. \eqref{PECformula1} cannot be obtained in realistic circumstances.
On the other hand, in our protocol such additional errors are mitigated: see the discussion above Eq. \eqref{appendixQEMformula} in Appendix \ref{appendix1}. 
Such a self-consistent treatment of error mitigation is one of the advantage of our protocol compared to PEC and we consider that this is going to be crucial for quantum metrology as well as quantum computing for large $N_q$ and $d^\text{tot}$.

\subsection{NISQ-Device Platforms for Our QEM Protocol}
At the end, let us discuss the effectiveness of our QEM protocol under NISQ-device parameters. 
Let us consider three types of quantum hardware,  trapped-ion qubit systems \cite{linke2017experimental,trappedionNISQ2019,leibfried2004toward,degen2017quantum,dorner2012quantum} and superconducting circuits (superconducting qubit systems)\cite{linke2017experimental,SCQRPP2017,SCQNISQ20191,SCQARCMP2020,ZhugroupSQC2020,TsaigroupSCCQC2021,degen2017quantum}, 
which are mainly used nowadays as hardware of quantum computers, 
and nitrogen-vacancy (NV) centers in diamond \cite{rondin2014magnetometry,degen2017quantum}.
First, we discuss from the case of  trapped-ion qubit systems. 
The trapped-ion qubit systems are one of the promising candidates for harnessing quantum technologies including quantum computing and quantum metrology, and by using them
the physical quantities such as the relative phase  \cite{leibfried2004toward}, a frequency \cite{huelga1997improvement,dorner2012quantum}, magnetic field \cite{kotler2011single,baumgart2016ultrasensitive}, and 
electric field \cite{maiwald2009stylus,gilmore2021quantum} have been measured. The major source of decoherence is considered to be PD \cite{trappedionNISQ2019,degen2017quantum,huelga1997improvement,dorner2012quantum}.
The noise strength of PD is characterized by a $T_2$ time which is the inverse of the decay rate $\gamma_\text{PD}.$
For instance, $T_2$ times of hyperfine-state ion trap qubits are around $10^4$ (msec) \cite{trappedionNISQ2019} and $\gamma_\text{PD}= 10^{-4}$ (msec)$^{-1}$.
On the other hand, two-qubit gate times are about  0.1 (msec)  \cite{trappedionNISQ2019}  and by taking $\Delta t = 0.1$ (msec) the dimensionless time $\tau_\text{PD}$ is estimated to be $10^{-5}$. From Figs. \ref{resultsCSS}(b), \ref{resultsGHZ}(b), and \ref{resultsSDS}(b) we see that the ratios $\text{RT}_\text{QEM$_\text{1st}$,QFI}$ exceed $2.8 \times 10^{2}$ for  $\tau_\text{PD} = 10^{-4}$ and therefore we expect that our QEM protocol works well for noisy quantum metrology conducted by NISQ trapped-ion qubit systems. 

Next, let us examine the case of superconducting qubit systems. 
As similar to the trapped-ion qubits, the superconducting qubits are one of the promising solid-state platforms for the implementation of quantum computing as well as quantum metrology, and by using them the quantum sensing, for instance,
the magnetic field sensing \cite{bal2012ultrasensitive} has been conducted.
In the superconducting qubit systems AD emerges as decoherence and is characterized by a $T_1$ time and is described in terms of the two quantities,
a Bose-Einstein distribution function $\bar{n}(\omega_\text{SCQ}) = \frac{1}{e^{\beta \hbar \omega_\text{SCQ}}}$ and the decay rate $\gamma_\text{AD}$: $T_1=\frac{1}{2(2\bar{n}(\omega_\text{SCQ}) +1)\gamma_\text{AD}}$. Here $\beta$ is the inverse temperature and $ \omega_\text{SCQ}$ is the frequency separation between $|0\rangle$ and $|1\rangle$ and is about 5 (GHz) \cite{SCQNISQ20191,TsaigroupSCCQC2021}.
The temperature of superconducting qubit is about $10$ mK \cite{SCQNISQ20191,TsaigroupSCCQC2021} and the Bose-Einstein distribution function for this case is estimated to be $\bar{n} \sim 10^{-11}$, and thus $\gamma_\text{AD} \simeq (2T_1)^{-1}.$ Then,  $T_1$ times are estimated to be 100 ($\mu$sec) \cite{SCQARCMP2020,ZhugroupSQC2020} and $\gamma_\text{AD}= 0.5\times10^{-2}$ ($\mu$sec)$^{-1}$. 
On the other hand, two-qubit gate times are around 0.1 ($\mu$sec) \cite{SCQARCMP2020,ZhugroupSQC2020} and by setting $\Delta t = 0.1$ ($\mu$sec)  
$\tau_\text{AD}$ is estimated to be $0.5\times 10^{-3}$. 
 From Figs. \ref{resultsCSS}(h), \ref{resultsGHZ}(h), and \ref{resultsSDS}(h) we see that the ratios $\text{RT}_\text{QEM$_\text{1st}$,QFI}$ is larger than fifty for  $\tau_\text{AD} = 0.5\times10^{-3}$ except for the SDS and thus our QEM protocol is considered to be valid also for NISQ superconducting qubit systems.  

Finally, let us discuss the case of noisy quantum metrology using NV centers in diamond. In NV centers in diamond there exist many elements which can be used for quantum technology applications (e.g., quantum information processing tasks, quantum memory, and quantum sensing) such as ${}^{13}$C nuclear spins, ${}^{15}$N nuclear spins, and electron spin states of negatively charged vacancy denoted by NV$^-$. In particular, the advantages of utilizing the electron spin states of NV$^-$ are (i) the controlling of electron spin states including initialization and readout can be done at room (ambient) temperature and (ii) there are many types of means to control the electron spin states of NV$^-$ including electromagnetically and optically ones (one of the representative scheme is called optically-detected  magnetic resonance (ODMR)) \cite{rondin2014magnetometry}. Owing to these benefits the electron spin systems of NV$^-$ are considered to be promising candidates to perform quantum metrology and up to now various types of physical quantities have been measured by using single-NV$^-$ systems or  ensemble-NV$^-$ systems, for instance, magnetic field \cite{rondin2014magnetometry,balasubramanian2008nanoscale,maze2008nanoscale,taylor2008high,fang2013high,grinolds2013nanoscale,masuyama2018extending}, electric field \cite{dolde2011electric,michl2019robust,barson2021nanoscale,bian2021nanoscale,qiu2022nanoscale}, and temperature \cite{kucsko2013nanometre,neumann2013high,toyli2013fluorescence,wang2015high}. 
Although the electron spin states of NV$^-$ possess such great potentials for applications to various quantum technologies, NV$^-$ is surrounded by many other systems including  ${}^{13}$C and ${}^{12}$C nuclear spins, ${}^{15}$N nuclear spins, and spin impurities of ${}^{14}$N so-called P1 centers and exhibits a complex coupling structure and an associated decoherence  \cite{dobrovitski2008decoherence,wang2013spin,scharfenberger2014coherent,chou2015optimal,ajisaka2016decoherence,burkard2017designing,bauch2020decoherence,park2022decoherence}. 
The study of decoherence of  NV$^-$ is called a central spin problem and has been intensively investigated up to now \cite{dobrovitski2008decoherence,wang2013spin,scharfenberger2014coherent,chou2015optimal,ajisaka2016decoherence,burkard2017designing,bauch2020decoherence,park2022decoherence}.
 The NV$^-$ systems show both PD and AD which are characterized by $T_2$ ($T^\ast_2$) and $T_1$ times, respectively. 
 Here let us focus on the PD caused by the electron spins of P1 centers and examine the validity of our QEM protocol. 
 Experimentally, the PD of NV$^-$ systems is studied by two types of measurement schemes, free induction decay (FID) or Ramsey interference and spin (Hahn)-echo (SE). In the former scheme a $T^\ast_2$ time is measured while in the latter one a $T_2$ time is. 
 It has been analyzed that the magnitudes of both $T^\ast_2$ and $T_2$ times are proportional to the inverse of the P1 center concentration described by the unit of ppm:
 for instance, a $T^\ast_2$ ($T_2$) time for 1ppm and 10 ppm are measured or theoretically estimated to be 10 (100) $\mu$sec and 1 (10) $\mu$sec, respectively \cite{bauch2020decoherence,park2022decoherence}.
 Furthermore, the decaying behaviors of a single NV$^-$ and an ensemble of NV$^-$ are different.   
 The decay behavior of a single NV$^-$ (NV$^-$ ensemble) induced by PD is observed to be the functional form, $\exp\left( -\left(\frac{t}{T_2} \right)^2 \right)$ ($\exp\left( -\left(\frac{t}{T_2} \right) \right)$), while in SE a single NV$^-$ (NV$^-$ ensemble) exhibits the decay behavior expressed by  $\exp\left( -\frac{t}{T_2} \right)^3$ ($\exp\left( -\left(\frac{t}{T_2} \right)^{\frac{3}{2}} \right)$) \cite{dobrovitski2008decoherence,bauch2020decoherence}. 
 Since we are interested in the quantum metrology with initial states taken to be quantum entangled states let us focus on NV$^-$ ensemble systems. The PD in FID is expressed as Markovian PD while that in SE can be effectively represented as NMPD with using the function $F_\text{NMPD}(t)$ in Eq. \eqref{NMPDQME2} by taking $F_\text{NMPD}(t)=(\gamma_\text{PD}t^3)^{\frac{1}{2}}$. On the other hand, two-qubit gate operations such as controlled-rotational gate, CPHASE, and CNOT, can be implemented, for instance, by using the NV$^-$ electron spin state and the nuclear spin states of ${}^{13}$C or ${}^{15}$N  \cite{scharfenberger2014coherent,chou2015optimal,rong2015experimental} as two qubits or by using two NV$^-$ electron spins as two qubits which are exploited in a cavity QED system \cite{burkard2017designing}. For both setups the two-qubit gate times are estimated or experimentally measured to be around 0.1 $\mu$sec.
By taking $\Delta t$ to be 0.1 $\mu$sec and $\gamma_\text{PD}=10^{-2},10^{-1},$ and $1$ MHz, the dimensionless times $\tau_\text{PD}$ become $10^{-3},10^{-2},$ and $10^{-1},$ respectively. 
First, let us consider the cases of the CSS and the GHZ states
with $N_q=5$ and the quantum noise is the MPD since it is more harmful than NMPD. When the initial state is taken to be the CSS (GHZ),  for $10^{-3}$ and $10^{-2}$ the ratios  $ \text{RT}_\text{QEM$_\text{1st}$,QFI}$ 
are about $3.0 \times 10^{3} ~(2.0 \times 10^{2})$ and $9.2 \times 10 ~(6.3)$, respectively and the first-order QEM works for both cases. On the other hand, for $\tau_\text{PD}=0.1$ the ratios  $ \text{RT}_\text{QEM$_\text{1st}$,QFI}$
are around one for both cases and  the second-order QEM  is necessary to be performed. 
Finally, let us analyze the case of the SDS with $N_q=6$ and the quantum noise is the MPD by examining the numerical values of $ \text{RT}_\text{QEM$_\text{2nd}$,QFI}$. 
For $\tau_\text{PD}=10^{-3},10^{-2},$ and $10^{-1},$ the values of the ratio $ \text{RT}_\text{QEM$_\text{2nd}$,QFI}$ are $1.9\times 10^{5}$, $1.3\times 10$, and 1.0, respectively. Our protocol works by performing higher-order QEM or when $\tau_\text{PD}$ satisfies $\tau_\text{PD} \geq 10^{-2}$.

\section{ Conclusions}\label{conclusions} 
In this paper, we have established our QEM protocol for noisy quantum metrology by considering the various kinds of quantum noise channels as well as the initial states.
Here we have focused on the mitigation of decoherence which occurs during the initialization and  the free evolution. 
These two procedures are mathematically described as the unitary operations and the product of them can be regarded as an unitary operation describing a quantum algorithm of quantum computing.
By taking account of this,  we have developed our QEM protocol composed by gate operations and quantum measurements on ancilla bits which is described as the ensembles of the quantum circuits called quantum-error-mitigation circuit groups.
The quantum-error-mitigation circuit groups enable us to theoretically evaluate the decoherence effects and we have done this by constructing the perturbation theories where the quantum noise strengths are treated as the perturbative parameters.
After the establishment of our QEM protocol we have performed the numerical simulations by considering three types of initial states and quantum noise channels in order to demonstrate and examine its efficacy and we have done them up to the second-order perturbation regimes. Consequently, we have numerically verified that our QEM is effective for every simulation of noisy quantum metrology except for the SDS when $N_q\geq4$ with $\tau^\text{AD} \geq10^{-2}$: In order to improve our results for the SDS, we need to conduct higher-order QEM or $\tau^\text{AD} \leq10^{-3}$. 
As similar to our QEM scheme for noisy quantum computing \cite{hama2022quantum}, 
the benefits of our QEM protocol are (i) it can be performed with any type of quantum device, (ii) it can be applied to noisy quantum metrology for any type of initial state as well as any type of Hamiltonian including a single parameter or multiparameters to be sensed, and (iii)  it can be applied to any type of decoherence with an arbitrary strength provided that a higher-order perturbation calculation is done.  Note that in the context of quantum computing the benefit (ii) corresponds to the applicability of our QEM protocol to any type of quantum algorithm.   
All these benefits originate from the way our protocol is constructed, i,e.,  it is composed of quantum gates and quantum measurements on ancilla bits. 

Besides the above benefits the big and intrinsic advantage of applying our QEM protocol to noisy quantum metrology is its effectiveness for various values of $\vartheta$ as discussed in Sec. \ref{discussions}. As studied previously \cite{degen2017quantum,huelga1997improvement},  the optimization of the sensitivity in noisy quantum metrology  is very complicated.
 In contrast, as we have seen in Fig. \ref{resultsOmegaMPD} owing to our QEM scheme the quantum-error-mitigated CFI exhibit approximately the same behaviors with those of the ideal CFI for almost every $\vartheta$.
 This means that the computational cost of the optimization of the quantum-error-mitigated CFI is equivalent to that of the ideal CFI, and we can realize the quantum sensing with high sensitivities given by ideal QFI which are attributed to quantum entanglement even under the influence of quantum noise. Such an characteristic is going to be a powerful tool for the application to weak magnetic field sensing which is an important issue in material science, biology, and medical science \cite{rondin2014magnetometry,taylor2008high}.

In this work, we have described the dynamical processes of the noisy qubits induced by the non-Markovian quantum noise by Eq. \eqref{noisyoutput}. 
As our future work, we would like to do an elaborate analysis on noisy quantum dynamics induced by the interaction between qubits and an environment and gate operations with taking into account the associated non-Markovianity  \cite{white2020demonstration,white2022non,white2022filtering}. In addition, we develop the improved QEM methods.

We expect that our QEM protocol given by the quantum-noise-effect circuit groups is applicable not only to quantum sensing and quantum computing \cite{hama2022quantum} but also to many other types of quantum technologies such as quantum communication and quantum network \cite{acin2018quantum,gisin2007quantum,chen2021review,wei2022towards}. 
As mentioned above, our protocol are described by quantum gates and measurements and can be conducted as a software manipulation. 
We expect that our QEM protocol paves the way for realizing programmable and high quality quantum technologies which are going to be a basic building block for industrial development and solving societal problems.  
\acknowledgements
 We thank all the other members of Quemix Inc. for giving us the fruitful comments and reading this manuscript carefully.  
 Y.H. thanks Gregory Anthony Liam White for fruitful comments on non-Markovianity in noisy quantum computing.  
 This work was supported by MEXT as ”Program for Promoting Researches on the Supercomputer Fugaku” (JP-MXP1020200205) and JSPS KAKENHI as ”Grant-in- Aid for Scientific Research(A)” Grant Number 21H04553. This study was carried out using the TSUBAME3.0 supercomputer at Tokyo Institute of Technology.
 \begin{widetext}
 \appendix 
\section{Second-order QEM }\label{appendix1}   
In this section, we present the formulas for the second-order QEM: see also the appendix in \cite{hama2022quantum}. 
To perform the second-order QEM, we need to calculate two quantities,  $\Delta^\text{QN}_2 (\rho_{d\cdots1})$ and  $\Delta^\text{QN}_1 \big{(} \Delta^\text{QN}_1(\rho_{d\cdots1}) \big{)}$, 
where the superscript ``QN" stands for quantum noise such as the MPD and the MAD. 
Recall that the density matrix $\rho_{d\cdots1}$ is given by $\rho_{d\cdots1} = \left(\prod_{l=1}^d U_l \right) \cdot |0\rangle^{\otimes N_q}\langle0|\cdot \left(\prod_{l=1}^d U_l \right)^\dagger$.
In the case of quantum metrology $d=d_\text{tot}$. The quantity $\Delta^\text{QN}_2 (\rho_{d\cdots1})$ is the intrinsic second-order quantum noise effect on $\rho_{d\cdots1}$ 
whereas $\Delta^\text{QN}_1 \big{(} \Delta^\text{QN}_1(\rho_{d\cdots1}) \big{)}$ is the first-order quantum noise effect on $\Delta^\text{QN}_1(\rho_{d\cdots1})$.
As we can understand from the term $ - \tau_\text{AD}\cdot \Delta^{\text{AD}}_1 \big{(} \rho^\text{out}_{\vartheta,\text{ideal}} \big{)} $ in Eq. \eqref{QEMDM},
we can neglect the error $\delta^\text{QN}_1 \big{(} \Delta^\text{QN}_1(\rho_{d\cdots1}) \big{)}$ for performing QEM in the first-order regime but in the second-order regime we have to take into account it.
A basic idea of our second-order QEM is given by
\begin{footnotesize}
\begin{align}
  \rho^\text{out}_{\vartheta,\text{QEM}}  &=\rho^\text{out,real}_{\vartheta,\text{noisy}} 
  - \tau_\text{QN}\cdot \Delta^{\text{QN}}_1 \big{(} \rho^\text{out}_{\vartheta,\text{ideal}} \big{)}   - \frac{(\tau_\text{QN})^2}{2}\cdot \Delta^{\text{QN}}_2 \big{(} \rho^\text{out}_{\vartheta,\text{ideal}} \big{)} +
(\tau_\text{QN})^2 \cdot \Delta^\text{QN}_1 \big{(} \Delta^\text{QN}_1\rho_{d\cdots1} \big{)}   \notag\\
 &=   \rho^\text{out}_{\vartheta,\text{ideal}}
  + \tau_\text{QN} \left( \delta^{\text{QN}}_{1} \big{(}   \rho^\text{out}_{\vartheta,\text{ideal}}   \big{)} 
 -\Delta^{\text{QN}}_{1} \big{(}   \rho^\text{out}_{\vartheta,\text{ideal}}   \big{)}   \right) \notag\\
 &+ \frac{(\tau_\text{QN})^2}{2}  \left( \delta^{\text{QN}}_{2} \big{(}   \rho^\text{out}_{\vartheta,\text{ideal}}   \big{)} 
 -\Delta^{\text{QN}}_{2} \big{(}   \rho^\text{out}_{\vartheta,\text{ideal}}   \big{)}   \right)
  + (\tau_\text{QN})^2  \left(\delta^\text{QN}_1 \big{(} \Delta^\text{QN}_1\rho_{d\cdots1} \big{)}-\Delta^\text{QN}_1 \big{(} \Delta^\text{QN}_1\rho_{d\cdots1} \big{)}  \right)
 + \mathcal{O}\big{(}(\tau_\text{QN})^3\big{)},
\label{appendixQEMformula}
\end{align}\end{footnotesize}
where $\tau_\text{QN}$ is the dimensionless perturbative parameter, e.g.,  $\tau_\text{QN}= \tau_\text{PD}(=\frac{\gamma_\text{PD}\Delta t}{2})$ and $\tau_\text{QN}= \tau_\text{AD}(=\gamma_\text{AD}\Delta t)$ for the MPD and the MAD, respectively. 
The terms in the second line of the right-hand side describes the first-order QEM (see also Eq. \eqref{QEMDM}) whereas the ones in the third line represent the second-order QEM.
In the following, we discuss separately the formulas of  $\Delta^\text{QN}_2 \rho_{d\cdots1}$ and  $\Delta^\text{QN}_1 \big{(} \Delta^\text{QN}_1\rho_{d\cdots1} \big{)}$ for the MPD, the MAD, and the NMPD.
\subsection{Second-order QEM  for MPD}\label{appendixMPD}
The quantum master equation describing the MPD process is given by Eq.  \eqref{MPDQME1} or 
 \begin{align}
\frac { \partial  \rho_\text{MPD} (t)}{\partial t }  & =    \frac{\gamma _\text{PD}}{2} \mathcal{L} ^{ \text{PD}} [\rho_\text{MPD} (t)]  =   \frac{\gamma _\text{PD}}{2} \sum_{j=0}^{N_q-1} \left[  Z_j     \rho_\text{MPD} (t)  Z_j -  \rho_\text{MPD} (t)   \right], 
 \label{appendixMPDQME}
\end{align}
and we obtain $\rho_\text{MPD} (t+\Delta t) = \rho_\text{MPD} (t) +  \tau_\text{PD}\mathcal{L} _{ \text{PD}} [\rho_\text{MPD} (t)] + \frac{(\tau_\text{PD})^2}{2} \mathcal{L} ^2_{ \text{PD}} [\rho_\text{MPD} (t)]+\mathcal{O}\big{(}(\tau_\text{PD})^3\big{)}$.
Up to $\mathcal{O}\big{(}(\tau_\text{PD})^2\big{)}$, we have $\rho_\text{MPD} (d\Delta t) = \rho_{d\cdots1}+ \tau_\text{PD} \Delta^{\text{MPD}}_{1} (\rho_{d\cdots1})  + \frac{(\tau_\text{PD})^2}{2}  \Delta^{\text{MPD}}_{2} (\rho_{d\cdots1})+\mathcal{O}\big{(}(\tau_\text{PD})^3\big{)}$, 
where
\begin{footnotesize} \begin{align}
\Delta^{\text{MPD}}_{1} (\rho_{d\cdots1}) &= -dN_q \rho_{d\cdots1}  + \sum_{j=0}^{N_q-1}  \sum_{k=1}^{d}   \left(\prod_{l=k+1}^d U_l \right)Z_j\rho_{k\cdots1}Z_j \left(\prod_{l=k+1}^d U_l \right)^\dagger, \notag\\
\Delta^{\text{MPD}}_{2} (\rho_{d\cdots1}) &= (dN_q)^2 \rho_{d\cdots1}  -2dN_q \sum_{j=0}^{N_q-1}  \sum_{k=1}^{d}   \left(\prod_{l=k+1}^d U_l \right) Z_j\rho_{k\cdots1}Z_j \left(\prod_{l=k+1}^d U_l \right)^\dagger \notag\\
&+\sum_{j_1,j_2=0}^{N_q-1} \sum_{k=1}^d   \left( \prod_{l=k+1}^d U_l  \right)Z_{j_1 }Z_{j_2 }  \rho_{k\ldots1} Z_{j_2 }  Z_{j_1 } \left( \prod_{l=k+1}^d U_l  \right)^\dagger  \notag\\
&+  2\sum_{k_1=2}^{d-1} \sum_{k_2=1}^{k_1-1}  \sum_{j_1,j_2=0}^{N_q-1}    \left( \prod_{l_1=k_1+1}^{d} U_{l_1} \right) Z_{j_1 }  \left( \prod_{l_2=k_2+1}^{k_1} U_{l_2} \right) Z_{j_2} 
\rho_{k_2\cdots1}      Z_{j_2 }   \left( \prod_{l_2=k_2+1}^{k_1} U_{l_2} \right)^\dagger Z_{j_1 }  \left( \prod_{l_1=k_1+1}^{d} U_{l_1} \right)^\dagger,
\label{appendixPDeffect1}
\end{align}\end{footnotesize}
where $\rho_{k\cdots1} = \left(\prod_{l=1}^k U_l \right) \cdot |0\rangle^{\otimes N_q}\langle0|\cdot \left(\prod_{l=1}^k U_l \right)^\dagger$ and  
$ \prod_{l=d+1}^d U_l = \boldsymbol{1}.$
On the other hand, $\Delta^{\text{MPD}}_1 \left( \Delta^{\text{MPD}}_1(\rho_{d\ldots1})    \right)$ is given by
\begin{align}
\Delta^{\text{MPD}}_1 \left( \Delta^{\text{MPD}}_1 (\rho_{d\ldots1})    \right) &= (dN_q)^2\rho_{d\ldots1} - (2d+1)N_q  \sum_{j_1 =0}^{N_q-1} \sum_{k=1}^d   \left( \prod_{l=k+1}^d U_l  \right)Z_{j_1 } \rho_{d\ldots1} Z_{j_1 } \left( \prod_{l=k+1}^d U_l  \right)^\dagger \notag\\
 &+2\sum_{j_1,j_2}^{N_q-1} \sum_{k=1}^d   \left( \prod_{l=k+1}^d U_l  \right)Z_{j_1 }Z_{j_2 }  \rho_{d\ldots1} Z_{j_2 }  Z_{j_1 } \left( \prod_{l=k+1}^d U_l  \right)^\dagger 
 \notag\\
&+  2\sum_{k_1=2}^{d} \sum_{k_2=1}^{k_1-1}  \sum_{j_1,j_2}^{N_q-1}    \left( \prod_{l_1=k_1+1}^{d} U_{l_1} \right) Z_{j_1 }  \left( \prod_{l_2=k_2+1}^{k_1} U_{l_2} \right) Z_{j_2} 
\rho_{k_2\cdots1}      Z_{j_2 } \notag\\
&\times \left( \prod_{l_2=k_2+1}^{k_1} U_{l_2} \right)^\dagger Z_{j_1 }  \left( \prod_{l_1=k_1+1}^{d} U_{l_1} \right)^\dagger.
\label{appendixPDeffect2}
\end{align}
\subsection{Second-order QEM  for MAD}\label{appendixMAD}
The quantum master equation for the MAD process is given by  \eqref{MADQME1} or
\begin{align}
\frac { \partial \rho_\text{MAD} (t)}{\partial t } & =  \gamma_\text{AD}  \mathcal{L} ^\text{AD}  [\rho_\text{MAD}(t)] 
=  \gamma_\text{AD}  \sum_{j_1 =0}^{N_q-1}\mathcal{L} ^\text{AD}_j [\rho_\text{MAD}(t)]  =  \gamma_\text{AD}  \left[  \tilde{\sigma}^-_j    \rho_\text{MAD} (t)   \tilde{\sigma}^+_j 
 - \frac{1}{2} \big{\{}   \tilde{\sigma}^+_j  \tilde{\sigma}^-_j ,  \rho_\text{MAD} (t)   \big{\}} \right] .
 \label{appendixMADQME}
\end{align}
As similar to the analysis done in the case of the MPD, we solve  $\rho_\text{MAD} (t+\Delta t) $ as $\rho_\text{MAD} (t+\Delta t) = \rho_\text{MAD} (t) +  
\tau_\text{AD}\mathcal{L} _{ \text{AD}} [\rho_\text{MAD} (t)] + \frac{(\tau_\text{AD})^2}{2} \mathcal{L} ^2_{ \text{AD}} [\rho_\text{MAD} (t)]+\mathcal{O}\big{(}(\tau_\text{AD})^3\big{)}$ and 
we obtain $\rho_\text{MAD} (d\Delta t) = \rho_{d\cdots1}+ \tau_\text{AD} \Delta^{\text{MAD}}_{1} (\rho_{d\cdots1})  + \frac{(\tau_\text{AD})_2}{2}  \Delta^{\text{MAD}}_{2} (\rho_{d\cdots1})+\mathcal{O}\big{(}(\tau_\text{AD})^3\big{)}$, where
 \begin{align} 
 \Delta^\text{MAD}_1 (\rho  _{d\cdots1}) & = - \frac{dN_q}{4}\rho_{d\ldots1} + \sum_{j_1 =0}^{N_q-1} \sum_{k=1}^d   \left( \prod_{l=k+1}^d U_l  \right)
 \left(\frac{Z_{j_1 } \rho_{d\ldots1} Z_{j_1 } }{4}+ \tilde{\sigma}^-_{j_1 } \rho_{d\ldots1} \tilde{\sigma}^+_{j_1 } - P^1_{j_1 } \rho_{d\ldots1} P^1_{j_1 }
 \right)
 \left( \prod_{l=k+1}^d U_l  \right)^\dagger, \notag\\
 \Delta^\text{MAD}_2 (\rho  _{d\cdots1}) & =   \sum_{k=1}^{d} \sum_{j_1,j_2=0}^{N_q-1}  \sum_{p_1, p_2} c_{p_1} c_{p_2}  \left( \prod_{l=k+1}^d U_l  \right) s_{j_1,p_1} s_{j_2,p_2}  \rho_{k\cdots1} s_{j_2,p_2}^\dagger   s_{j_1,p_1} ^\dagger
  \left( \prod_{l=k+1}^d U_l  \right)^\dagger  \notag\\
&+ 
 2\sum_{k_1=2}^{d}  \sum_{k_2=1}^{k_1-1}
 \sum_{j_1,j_2=0}^{N_q-1}  \sum_{p_1, p_2} c_{p_1} c_{p_2} 
  \left( \prod_{l_1=k_1+1}^{d} U_{l_1} \right) s_{j_1,p_1}  \left( \prod_{l_2=k_2+1}^{k_1} U_{l_2} \right) s_{j_2,p_2} 
\rho_{k_2\cdots1} s_{j_2,p_2} ^\dagger  \notag\\
  & \times      \left( \prod_{l_2=k_2+1}^{k_1} U_{l_2} \right)^\dagger s_{j_1,p_1} ^\dagger  \left( \prod_{l_1=k_1+1}^{d} U_{l_1} \right)^\dagger.
   \label{appendixsecondorderADeffect2}
  \end{align} 
Let us explain the notations used for describing $ \Delta^\text{MAD}_2 (\rho  _{d\cdots1})$ in the above equation. 
The subscripts $p_1,p_2$ appearing in the coefficients  $c_{p_1},c_{p_2}$ and the operators $s_{j_1,p_1}, s_{j_2,p_2}$
are the ones for labeling the four operators $\big{\{} \boldsymbol{1}, Z, \tilde{\sigma}^-, P^1 \big{\}}$: the operator $s_{j_1(2),p_1(2)}$ stands for the operator $p_{1(2)}$ acting on the qubit $Q_{j_{1(2)}}$.
Writing these quantities explicitly,  $ \big{\{} c_{\boldsymbol{1}}, c_{Z}, c_{\tilde{\sigma}^-}, c_{P^1} \big{\}}=\big{\{} -\frac{1}{4}, \frac{1}{4}, 1,-1 \big{\}}$
and $ \big{\{} s_{j_{1(2)},\boldsymbol{1}}, s_{j_{1(2)},Z}, s_{j_{1(2)},\tilde{\sigma}^-}, s_{j_{1(2)},P^1} \big{\}}  
=\big{\{} \boldsymbol{1}_{j_{1(2)}}, Z_{j_{1(2)}},  \tilde{\sigma}^-_{j_{1(2)}},      P^1_{j_{1(2)}}       \big{\}}$.
For $\Delta^\text{MAD}_1 (\rho  _{d\cdots1}) $ see also Eq. \eqref{evalMADeffect}.
In contrast to the case of the MPD, the MAD effects $ \Delta^\text{MAD}_1 (\rho  _{d\cdots1}) $ and $ \Delta^\text{MAD}_2 (\rho  _{d\cdots1}) $ include the non-unitary operators $\tilde{\sigma}^-, P^1$ 
and to calculate these quantities we need  ancilla bits.
We introduce a single ancilla bit $Q_{\text{a1}}$  to calculate $ \Delta^\text{MAD}_1 (\rho  _{d\cdots1})$ while for calculating $ \Delta^\text{MAD}_2 (\rho  _{d\cdots1})$ we use two ancilla bits $Q_{\text{a1}},Q_{\text{a2}}$. 
Next, we reformulate the unitary operations $U_{l}$ as $U_{l}\otimes \boldsymbol{1}_{\text{a1}}$ for calculating $ \Delta^\text{MAD}_1 (\rho  _{d\cdots1})$ 
whereas for calculating $ \Delta^\text{MAD}_2 (\rho  _{d\cdots1})$ we reformulate $U_{l}$ as $U_{l}\otimes \boldsymbol{1}_{\text{a1a2}}$.
Then, we substitute   $\tilde{\sigma}^-_{j_{1(2)}}$ with $\text{C}R_y(\pi)[Q_{\text{r}_{j_{1(2)}}};Q_{\text{a1}}] \cdot \text{C}X[Q_\text{a1(2)};Q_{\text{r}_{j_{1(2)}}}]$ while we substitute 
$P^1_{Q_{j_{1(2)}}}$ with $\text{C}R_y(\pi)[Q_{\text{r}_{j_{1(2)}}};Q_{\text{a1(2)}}] \cdot X_{Q_{\text{a1(2)}}}    \cdot \text{C}X[Q_\text{a1(2)};Q_{\text{r}_{j_{1(2)}}}]$.
Next, let us explain how to calculate  $\Delta^\text{MAD}_1 \left( \Delta^\text{MAD}_1 (\rho_{d\ldots1})    \right).$ 
Like the calculation of  $\Delta^\text{MAD}_2 (\rho  _{d\cdots1}) $, we have to express  $\tilde{\sigma}^-_{j_{1(2)}}$ and $P^1_{j_{1(2)}}$ in terms of the control unitary operations
$\text{C}R_y(\pi)[Q_{\text{r}_{j_{1(2)}}};Q_{\text{a}_{1(2)}}]$  and $\text{C}X[Q_\text{a};Q_{\text{r}_{j_{1(2)}}}]$ and the $X$-gate operations on the ancilla bits. 
To accomplish this, we rewrite $ \Delta^\text{MAD}_1 (\rho  _{d\cdots1})$ as 
 \begin{align} 
 \Delta^\text{MAD}_1 (\rho  _{d\cdots1})  & = - \frac{dN_q}{4}\rho_{d\ldots1} + \frac{1}{4}\sum_{j_1 =0}^{N_q-1} \sum_{k_1=1}^{d}   \left( \prod_{l=k_1+1}^{d+1} U^{Z_{j_1},k_1}_l  \right) \rho_{k_1\ldots1}
 \left( \prod_{l=k_1+1}^{d+1} U^{Z_{j_1},k_1}_l   \right)^\dagger \notag\\
& +\text{Tr}_{Q_\text{a1}}\left[P^1_{Q_\text{a1}} \left( \prod_{l=k_1+1}^{d+2} U^{\tilde{\sigma}^-_{j_1},k_1}_l  \right) \hat{\rho}_{k_1\ldots1}  \left( \prod_{l=k_1+1}^{d+2} U^{\tilde{\sigma}^-_{j_1},k_1}_l   \right)^\dagger \right] \notag\\
& -\text{Tr}_{Q_\text{a1}}\left[P^0_{Q_\text{a1}}\left( \prod_{l=k_1+1}^{d+3} U^{P^1_{j_1},k_1}_l  \right) \hat{\rho}_{k_1\ldots1}  \left( \prod_{l=k_1+1}^{d+3} U^{P^1_{j_1},k_1}_l   \right)^\dagger\right],  \notag\\
 & =  \sum_{j_1 =0}^{N_q-1} \sum_{k_1=1}^{d}  
 \left(  - \frac{\rho_{d\ldots1}}{4} + \frac{\rho_{d+1\ldots1}^{Z_{j_1},k_1} }{4} +  \rho_{d+2\ldots1}^{\tilde{\sigma}^-_{j_1},k_1} - \rho_{d+3\ldots1}^{P^1_{j_1},k_1}     \right),
   \label{appendixsecondorderMADeffect3}
  \end{align} 
 where
\begin{footnotesize}  \begin{align} 
\rho_{d+1\ldots1}^{Z_{j_1},k}  &= \left( \prod_{l=1}^{d+1} U_l^{Z_{j_1},k}  \right) |0\rangle^{\otimes N_q}\langle0|  \left( \prod_{l=1}^{d+1} U_l^{Z_{j_1},k}  \right)^\dagger, \notag\\
\Big{\{}  U_l^{Z_{j_1},k_1}   \Big{\}}_{l=1}^{d+1} & = \left[ U_1,U_2,\ldots, U_{k_1}, Z_{j_1}, U_{k_1+1},\ldots, U_d     \right] := \left[ U_1^{Z_{j_1},k_1}, \ldots, U_{d+1}^{Z_{j_1},k_1} \right], \notag\\
\hat{\rho}_{k_1\ldots1} &= \left(\prod_{l=1}^k U_l\otimes \boldsymbol{1}_\text{a1} \right) \cdot |0\rangle^{\otimes N_q+1}\langle0|\cdot \left(\prod_{l=1}^k U_l\otimes \boldsymbol{1}_\text{a1} \right)^\dagger , \notag\\
 \rho_{d+2\ldots1}^{\tilde{\sigma}^-_{j_1},k_1}  &= 
 \text{Tr}_{Q_{\text{a1}}}\left[ P^1_{Q_{\text{a1}}}\cdot  \left( \prod_{l=1}^{d+2} U_l^{\tilde{\sigma}^-_{j_1},k_1}  \right) |0\rangle^{\otimes N_q+1}\langle0|  \left( \prod_{l=1}^{d+2} U_l^{\tilde{\sigma}^-_{j_1},k_1}  \right)^\dagger   \right], \notag\\
 \Big{\{}  U_l^{\tilde{\sigma}^-_{j_1},k_1}   \Big{\}}_{l=1}^{d+2} & = \left[ U_1\otimes\boldsymbol{1}_{Q_{\text{a1}}},\ldots, U_{k_1}\otimes\boldsymbol{1}_{Q_{\text{a1}}}, U_{\text{C}R_y(\pi)}[Q_{j_1};Q_{\text{a}2}],
U_{\text{C}X}[Q_{\text{a}2};Q_{j_1}], U_{k_1+1}\otimes\boldsymbol{1}_{Q_{\text{a1}}}, \ldots,  U_{d}\otimes\boldsymbol{1}_{Q_{\text{a1}}}      \right] \notag\\ & := \left[ U_1^{\tilde{\sigma}^-_{j_1},k_1}, \ldots, U_{d+2}^{\tilde{\sigma}^-_{j_1},k_1} \right], \notag\\
\rho_{d+3\ldots1}^{P^1_{j_1},k_1}  &= 
 \text{Tr}_{Q_{\text{a1}}}\left[ P^0_{Q_{\text{a1}}}\cdot \left( \prod_{l=1}^{d+3} U_l^{P^1_{j_1},k_1}  \right) |0\rangle^{\otimes N_q+1}\langle0|  \left( \prod_{l=1}^{d+3} U_l^{P^1_{j_1},k_1}  \right)^\dagger
 \right], \notag\\
 \Big{\{}  U_l^{P^1_{j_1},k_1}   \Big{\}}_{l=1}^{d+3} & = \left[ U_1\otimes\boldsymbol{1}_{Q_{\text{a1}}},\ldots, U_{k_1}\otimes\boldsymbol{1}_{Q_{\text{a1}}}, U_{\text{C}R_y(\pi)}[Q_{j_1};Q_{\text{a1}}], X_{Q_{\text{a1}}},
U_{\text{C}X}[Q_{\text{a1}};Q_{j}], U_{k_1+1}\otimes\boldsymbol{1}_{Q_{\text{a1}}}, \ldots,  U_{d}\otimes\boldsymbol{1}_{Q_{\text{a1}}}     \right]   \notag\\&:= \left[ U_1^{\tilde{\sigma}^-_{j_1},k_1}, \ldots, U_{d+3}^{\tilde{\sigma}^-_{j_1},k_1} \right],
 \label{additionalunitaries1}
  \end{align}\end{footnotesize}
  where $P^{(0)1}_{Q_{\text{a1}}}$ is the projection measurement operator of $|0(1)\rangle_{Q_{\text{a1}}}.$ 
 By using Eqs. \eqref{appendixsecondorderMADeffect3} and \eqref{additionalunitaries1} we obtain 
\begin{footnotesize} \begin{align} 
 \Delta^\text{MAD}_1 \left( \Delta^\text{MAD}_1 (\rho_{d\ldots1})    \right) &=  - \frac{dN_q}{4} \Delta^\text{MAD}_1 (\rho_{d\ldots1}) 
 + \sum_{j_1 =0}^{N_q-1} \sum_{k_1=1}^d \frac{1}{4}\Delta^\text{MAD}_1\left(   \rho_{d+1\ldots1}^{Z_{j_1},k_1} \right) 
  +\Delta^\text{MAD}_1 \left( \rho_{d+2\ldots1}^{ \tilde{\sigma}^-_{j_1},k_1} \right)  -\Delta^\text{MAD}_1 \left( \rho_{d+3\ldots1}^{P^1_{j_1},k_1}   \right)  , \label{delta1delta1AD1}
  \end{align}\end{footnotesize}
 where
 \footnotesize  \begin{align}  
\Delta^\text{MAD}_1\left(   \rho_{d+1\ldots1}^{Z_{j_1},k_1} \right)  &=  \sum_{j_2=0}^{N_q-1} \sum_{k_2=1}^{d+1}\sum_p  c_p 
   \left( \prod_{l=k_2+1}^{d+1} U_l^{Z_{j_1},k_1}  \right)s_{j_2,p } \rho_{k_2\ldots1}^{Z_{j_1},k_1}  s_{j_2,p }^\dagger  \left( \prod_{l=k_2+1}^{d+1} U_l^{Z_{j_1},k_1}  \right)^\dagger,  \notag\\
      \rho_{k_2\ldots1}^{Z_{j_1},k_1}  &= \left( \prod_{l=1}^{k_2} U_l^{Z_{j_1},k_1}  \right) |0\rangle^{\otimes N_q}\langle0|  \left( \prod_{l=1}^{k_2} U_l^{Z_{j_1},k_1}  \right)^\dagger, \notag\\
  \Delta^\text{MAD}_1\left(   \rho_{d+2\ldots1}^{\tilde{\sigma}^-_{j_1},k_1} \right)  &=  \sum_{j_2=0}^{N_q} \sum_{k_2=1}^{d+2} \sum_p c_p 
  \text{Tr}_{Q_{\text{a1}}}\left[ P^1_{Q_{\text{a1}}}\cdot
   \left( \prod_{l=k_2+1}^{d+2} U_l^{\tilde{\sigma}^-_{j_1},k_1}  \right)s_{j_2,p } \rho_{k_2\ldots1}^{\tilde{\sigma}^-_{j_1},k_1}  s_{j_2,p }^\dagger  \left( \prod_{l=k_2+1}^{d+2} U_l^{\tilde{\sigma}^-_{j_1},k_1}  \right)^\dagger\right],  \notag\\
      \rho_{k_2\ldots1}^{\tilde{\sigma}^-_{j_1},k_1}  &= \left( \prod_{l=1}^{k_2} U_l^{\tilde{\sigma}^-_{j_1},k_1}  \right) |0\rangle^{\otimes N_q+1}\langle0|  \left( \prod_{l=1}^{k_2} U_l^{\tilde{\sigma}^-_{j_1},k_1}  \right)^\dagger, \notag\\
\Delta^\text{MAD}_1\left(   \rho_{d+3\ldots1}^{P^1_{j_1},k_1} \right)  &=  \sum_{j_2=0}^{N_q} \sum_{k_2=1}^{d+3} \sum_p c_p 
 \text{Tr}_{Q_{\text{a1}}}\left[ P^0_{Q_{\text{a1}}}\cdot  \left( \prod_{l=k_2+1}^{d+3} U_l^{P^1_{j_1},k_1}  \right)s_{j_2,p } \rho_{k_2\ldots1}^{P^1_{j_1},k_1}  s_{j_2,p }^\dagger  \left( \prod_{l=k_2+1}^{d+3} U_l^{P^1_{j_1},k_1}  \right)^\dagger\right],  \notag\\
      \rho_{k_2\ldots1}^{P^1_{j_1},k_1}  &= \left( \prod_{l=1}^{k_2} U_l^{P^1_{j_1},k_1}  \right) |0\rangle^{\otimes N_q+1}\langle0|  \left( \prod_{l=1}^{k_2} U_l^{P^1_{j_1},k_1}  \right)^\dagger.
  \label{delta1delta1MAD2}
  \end{align}  \normalsize 
 For creating the operations $\tilde{\sigma}^-_{j_2 } $ and $P^1_{j_2 }$ in the above equation we use the other ancilla bit $Q_{\text{a2}}$ and they are given by 
 $ \text{Tr}_{Q_{\text{a2}}}\left[P^1_{Q_{\text{a2}}} \cdot \text{C}R_y(\pi)[Q_{\text{r}_{j_2}};Q_{\text{a2}}]     \cdot \text{C}X[Q_\text{a2};Q_{\text{r}_{j_2}}] \right] = \tilde{\sigma}^-_{j_2 } $ and
 $  \text{Tr}_{Q_{\text{a2}}}\left[P^0_{Q_{\text{a2}}} \cdot \text{C}R_y(\pi)[Q_{\text{r}_{j_2}};Q_{\text{a2}}] \cdot X_{Q_{\text{a2}}}    \cdot \text{C}X[Q_\text{a2};Q_{\text{r}_{j_2}}]\right]=P^1_{j_2 }$.
 Note that both the measurements on the ancilla bits $Q_{\text{a1}}$ and $Q_{\text{a2}}$ and those on the $N_q$ system qubits are performed at the same time and are done after the execution of all the unitary operations is completed.   
\subsection{Second-order QEM  for NMPD}\label{appendixNMPD}
The quantum master equation describing the NMPD is given by Eq.  \eqref{NMPDQME1} or  
  \begin{align}
 \frac { \partial \rho_\text{NMPD} (t)}{\partial t }  =  \frac{\gamma_\text{PD}}{2} f_\text{NMPD}(t) \mathcal{L} _{\text{PD}} [\rho (t)] 
  =   \frac{\gamma_\text{PD}}{2} f_\text{NMPD}(t) \sum_{j=0}^{N_q-1}  \Big{[} Z^-_j    \rho_\text{NMPD}(t)   Z^-_j  -  \rho_\text{NMPD} (t)   \big{\}} \Big{]},
 \label{appendixNMPDQME}
\end{align} where $f_\text{NMPD}(t) = \frac{1 - e^{\gamma_\text{C}t} }{2}.  $
From Eq. \eqref{NMPDQS1}, the time evolution of the quantum state from $\rho_\text{NMPD}(t_0)$ to $\rho_\text{NMPD}(t_d)$ is described in terms of the map
 $ \mathcal{E}^{\text{NMPD}}_{t_{l+1},t_{l}}: \rho_\text{NMPD} (t_l) \to \rho_\text{NMPD} (t_{l+1})$ as
 \begin{align}
 \rho_\text{NMPD}(t_d) = \prod_{l=1}^{d}
 \exp\left(\frac{\gamma_\text{PD} }{2} \left(F_\text{NMPD}(t_l)-F_\text{NMPD}(t_{l-1})\right)  \mathcal{L} _\text{PD}  \right)  \rho_\text{NMPD} (t_0).
 \label{TENMPD1appendix}
  \end{align}
By taking account of the structure of the noisy time evolution represented by Eq. \eqref{TENMPD1appendix}, we construct our QEM protocol for the NMPD by using a dimensionless time defined by
 \begin{align}
 \tau_\text{NMPD}(t_{l}) = \frac{\gamma_\text{PD}}{2} \left( F_\text{NMPD}(t_{l+1})-F_\text{NMPD}(t_{l}) \right),
 \label{TENMPD2appendix}
  \end{align}
and use it as the perturbative parameter. By calculating $\rho_\text{NMPD}(t_{l+1}) = \left[\boldsymbol{1}+ \tau_\text{NMPD}(t_l)\mathcal{L} _\text{PD}+\frac{\tau^2_\text{NMPD}(t_l)}{2!}\mathcal{L} ^2_\text{PD} \right]\rho_\text{NMPD}(t_{l}) + \mathcal{O}\big{(}(\tau_\text{NMPD}(t_l))^3\big{)},$ we obtain $\rho_\text{NMPD} (d\Delta t) = \rho_{d\cdots1}+  \Delta^{\text{NMPD}}_{1} (\rho_{d\cdots1})  + \frac{1}{2}  \Delta^{\text{NMPD}}_{2} (\rho_{d\cdots1})+\mathcal{O}\big{(}(\tau^\text{NMPD}(t_k))^3\big{)}$, where
 \begin{footnotesize}\begin{align}
\Delta^{\text{NMPD}}_{1} (\rho_{d\cdots1}) &=   \sum_{k=1}^{d}\sum_{j=0}^{N_q-1} \tau^\text{NMPD}(t_{k-1})  \left(\prod_{l=k+1}^d U_l \right)
  Z_j\rho_{k\cdots1}Z_j \left(\prod_{l=k+1}^d U_l \right)^\dagger-N_q\sum_{k=1}^{d} \tau^\text{NMPD}(t_{k-1}) \rho_{d\cdots1}, \notag\\
\Delta^{\text{NMPD}}_{2} (\rho_{d\cdots1})  &=\left[\sum_{k=1}^{d}   \left(N_q \tau^\text{NMPD}(t_{k-1})\right)^2  +2N_q^2\sum_{k_1=2}^{d} \sum_{k_2=1}^{k_1-1} \tau^\text{NMPD}(t_{k_1-1}) \tau^\text{NMPD}(t_{k_2-1})\right]\rho_{d\cdots1} \notag\\
&- 2N_q \sum_{k=1}^{d} \sum_{j=0}^{N_q-1}  \big{(} \tau^\text{NMPD}(t_{k-1})  \big{)}^2  \left(\prod_{l=k+1}^d U_l \right)  Z_j\rho_{k\cdots1}Z_j \left(\prod_{l=k+1}^d U_l \right)^\dagger\notag\\
&+  \sum_{k=1}^{d} \sum_{j_1,j_2=0}^{N_q-1}  \left( \tau^\text{NMPD}(t_{k-1}) \right)^2   \left(\prod_{l=k+1}^d U_l \right)  Z_{j_1}Z_{j_2}\rho_{k\cdots1}Z_{j_2}Z_{j_1} \left(\prod_{l=k+1}^d U_l \right)^\dagger\notag\\
&+2\sum_{k_1=2}^{d} \sum_{k_2=1}^{k_1-1} \sum_{j_1,j_2=0}^{N_q-1} \tau^\text{NMPD}(t_{k_1-1}) \tau^\text{NMPD}(t_{k_2-1}) \left( \prod_{l_1=k_1+1}^{d} U_{l_1} \right) Z_{j_1} \left( \prod_{l_2=k_2+1}^{k_1} U_{l_2} \right) Z_{j_2}\rho_{k_2\cdots1}Z_{j_2}
\left( \prod_{l_2=k_2+1}^{k_1} U_{l_2} \right)^\dagger \notag\\ &\times Z_{j_1} \left(\prod_{l_1=k_1+1}^d U_{l_1} \right)^\dagger
-2N_q\sum_{k_1=2}^{d} \sum_{k_2=1}^{k_1-1}\sum_{j=0}^{N_q-1} \tau^\text{NMPD}(t_{k_1-1}) \tau^\text{NMPD}(t_{k_2-1}) \notag\\
&\times \left[ 
 \left(\prod_{l=k_1+1}^d U_l \right) Z_j\rho_{k_1\cdots1}Z_j \left(\prod_{l=k_1+1}^d U_l \right)^\dagger +\left(\prod_{l=k_2+1}^d U_l \right) Z_j\rho_{k_2\cdots1}Z_j \left(\prod_{l=k_2+1}^d U_l \right)^\dagger
\right],
\label{appendixNMPDeffect1}
\end{align}\end{footnotesize}
Next,  $\Delta^\text{NMPD}_1 \left( \Delta^\text{NMPD}_1 (\rho_{d\ldots1})    \right)$ is given by
\begin{footnotesize}\begin{align}
\Delta^\text{NMPD}_1 \left( \Delta^\text{NMPD}_1 \rho_{d\ldots1}    \right) &= N_q^2\sum_{k_1=1}^{d} \sum_{k_2=1}^{d} \tau^\text{NMPD}(t_{k_1-1}) \tau^\text{NMPD}(t_{k_2-1}) \rho_{d\ldots1}\notag\\
& -N_q \sum_{k_1=1}^{d} \sum_{k_2=1}^{d} \sum_{j=0}^{N_q-1} \tau^\text{NMPD}(t_{k_1-1}) \tau^\text{NMPD}(t_{k_2-1})
 \left(\prod_{l=k_2+1}^{d} U_l \right)Z_{j}\rho_{k_2\ldots1} Z_{j}\left(\prod_{l=k_2+1}^{d} U_l \right)^\dagger\notag\\
& -N_q \sum_{k_1=1}^{d} \sum_{k_2=1}^{d+1}\sum_{j=0}^{N_q-1} \tau^\text{NMPD}(t_{k_1-1})   \tau^\text{NMPD}(t_{k_2-1}) \rho_{d+1\ldots1}^{Z_{j_1},k_1} \notag\\
&+\sum_{k_1=1}^{d} \sum_{k_2=1}^{d+1} \sum_{j_1,j_2=0}^{N_q-1} \tau^\text{NMPD}(t_{k_1-1}) \tau^\text{NMPD}(t_{k_2-1})
 \left(\prod_{l=k_2+1}^{d+1} U_l^{Z_{j_1},k_1}\right)Z_{j_2}\rho_{k_2\ldots1}^{Z_{j_1},k_1} Z_{j_2}\left(\prod_{l=k_2+1}^{d+1} U_l^{Z_{j_1},k_1}\right)^\dagger,
\label{appendixNMPDeffect2}
\end{align}\end{footnotesize}
where we have used Eq. \eqref{additionalunitaries1} for describing the third and fourth lines in the above equation.  
We represent a $p$th-order NMPD effect  in $\tau^\text{NMPD}(t_{l})$  in a real device by a symbol $\delta^\text{NMPD}_p $ and use it to express a real quantum state $\rho_{d\cdots1}^\text{real}$ as
$\rho_{d\cdots1}^\text{real} =\sum_{p=0}^\infty  \frac{1}{p!}\delta^\text{NMPD}_p(\rho_{d\cdots1}). $ 
From Eqs. \eqref{appendixQEMformula}, \eqref{appendixNMPDeffect1}, and \eqref{appendixNMPDeffect2} the formula of QEM for the NMPD effect  is given by \begin{footnotesize}
\begin{align}
  \rho^\text{out}_{\vartheta,\text{QEM}}  &=\rho^\text{out,real}_{\vartheta,\text{NMPD}} 
  -\Delta^{\text{NMPD}}_1 \big{(} \rho^\text{out}_{\vartheta,\text{ideal}} \big{)}   - \frac{1}{2} \Delta^{\text{NMPD}}_2 \big{(} \rho^\text{out}_{\vartheta,\text{ideal}} \big{)} +
 \Delta^\text{NMPD}_1 \big{(} \Delta^\text{NMPD}_1\rho_{d\cdots1} \big{)}   \notag\\
 &=   \rho^\text{out}_{\vartheta,\text{ideal}}
  +  \left( \delta^{\text{NMPD}}_{1} \big{(}   \rho^\text{out}_{\vartheta,\text{ideal}}   \big{)} 
 -\Delta^{\text{NMPD}}_{1} \big{(}   \rho^\text{out}_{\vartheta,\text{ideal}}   \big{)}   \right) \notag\\
 &+ \frac{1}{2}  \left( \delta^{\text{NMPD}}_{2} \big{(}   \rho^\text{out}_{\vartheta,\text{ideal}}   \big{)} 
 -\Delta^{\text{NMPD}}_{2} \big{(}   \rho^\text{out}_{\vartheta,\text{ideal}}   \big{)}   \right)
  +   \left(\delta^\text{NMPD}_1 \big{(} \Delta^\text{NMPD}_1\rho_{d\cdots1} \big{)}-\Delta^\text{NMPD}_1 \big{(} \Delta^\text{NMPD}_1\rho_{d\cdots1} \big{)}  \right)
+ \mathcal{O}\big{(}( \tau^\text{NMPD}(t_{k}) )^3\big{)}.
\label{appendixQEMformulaforNMPD}
\end{align} \end{footnotesize}
\section{Derivations of Eqs. \eqref{CSSQFIMPD} and \eqref{GHZQFI2}}\label{appendix2}   
In this section we demonstrate the derivation of Eq.  \eqref{CSSQFIMPD} and \eqref{GHZQFI2}.
To do this we show that the output density matrix of the GHZ state under the MPD effect is given by
\begin{footnotesize}\begin{align} 
	\rho^{\text{out,GHZ}_{N_q}}_{\text{MPD},\vartheta}  = 
	\frac{1}{2}\left (
		\begin{array}{ccccc} 
		 1 &0 & \cdots & 0 & \exp\left( -N_qi \vartheta - \big{(} N^\text{GHZ}_\text{PD}(N_q)+N^\text{free}N_q \big{)}\tau^\text{PD} \right)  \\ 
		 0 & \cdots & \cdots & \cdots & 0 \\ 
		 \vdots & \vdots & \vdots & \vdots & \vdots  \\ 
		 0 & \cdots &\cdots  &\cdots &0 \\ 
		 \exp\left( N_qi \vartheta - \big{(} N^\text{GHZ}_\text{PD}(N_q)+N^\text{free}N_q \big{)}\tau^\text{PD} \right) & 0& \cdots & 0 & 1 
		\end{array}
	\right ).  \label{outputGHZMPD1}
\end{align} \end{footnotesize}
Once we have the above density-matrix formula we can straightforwardly derive Eq. \eqref{GHZQFI2} by using Eqs. \eqref{QFIexpression1} or \eqref{QFIexpression2}.
In the following we derive Eq. \eqref{outputGHZMPD1}  based on induction with using Eq. \eqref{QFIexpression1} since using Eq. \eqref{QFIexpression1} is simpler to derive Eq. \eqref{outputGHZMPD1} .
First for $N_q=1$, which corresponds to the quantum metrology with an initial state taken to be the CSS  under the MPD effect, we can verify that the output density matrix is
\begin{align} 
	{\rho}^\text{out,CSS}_{\text{MPD},\vartheta}  = 
	\frac{1}{2}\left (
		\begin{array}{ccc} 
		 1 & \exp\left( -i \vartheta - \big{(}\tau^\text{PD}+N_\text{free}\tau^\text{PD} \big{)} \right)  \\ 
		 \exp\left( i \vartheta - \big{(}\tau^\text{PD} +N_\text{free}\tau^\text{PD} \big{)} \right) &  1 
		\end{array}
	\right ), \label{outputCSSMPD}
\end{align} 
where the superscript of the above density matrix is written by not ``GHZ" but by ``CSS".
The two eigenvalues of the density matrix ${\rho}^\text{out,CSS}_{\text{MPD},\vartheta} $ in Eq.  \eqref{outputCSSMPD} are
$\lambda_{1,1} = \frac{1+e^{-\tau^\text{PD}}}{2}, \lambda_{1,2} = \frac{1-e^{-\tau^\text{PD}}}{2}$ and the two eigenvectors are
$| \lambda_{1,1} \rangle = \left( \frac{e^{-i\vartheta}}{\sqrt{2}}, \frac{1}{\sqrt{2}}      \right)^\text{T}, | \lambda_{1,2} \rangle = \left( \frac{-e^{-i\vartheta}}{\sqrt{2}}, \frac{1}{\sqrt{2}}      \right)^\text{T}$
with the superscript ``T" denoting the transpose. Note that the eigenvector $| \lambda_{1,a} \rangle $ ($a=1,2$) is the association of the eigenvalue $ \lambda_{1,a}$.
By using these eigenvalues and eigenvectors with Eq. \eqref{QFIexpression1}  we obtain $I_{\text{QF}} \left[ \rho^{\text{out,CSS}}_{\vartheta, \text{MPD}} \right] $ in Eq.  \eqref{CSSQFIMPD} or 
\begin{align} 
I_{\text{QF}} \left[ \rho^{\text{out,CSS}}_{\vartheta, \text{PD}} \right] = e^{-(1+N_\text{free}) \tau^\text{PD} } .
\label{appCSSQFIPD}
\end{align} 
Next, let us analyze for $N_q=2$ with setting the grouping number $\mu=1$.
In this case, the output density matrix becomes
\begin{align} 
	\rho^{\text{out,GHZ}_N}_{\text{MPD},\vartheta}  = 
	\frac{1}{2}\left (
		\begin{array}{cccc} 
		 1 &0 &0 & \exp\left( -2i \vartheta - \big{(} 3\tau^\text{PD} +N_\text{free}\tau^\text{PD} \big{)} \right)  \\ 
		 0 &0 &0 &0  \\ 
		 0 &0 &0 &0  \\ 
		 \exp\left( 2i \vartheta - \big{(} 3\tau^\text{PD} +N_\text{free}\tau^\text{PD} \big{)} \right) &0 &0 & 1 
		\end{array}
	\right ), \label{outputGHZMPD2}
\end{align} 
 and  we have the four eigenvalues $\lambda_{2,1} =   \frac{1+e^{-(3+N_\text{free})\tau_\text{PD} }}{2},  \lambda_{2,2} =   \frac{1-e^{-(3+N_\text{free})\tau_\text{PD}}}{2},  \lambda_{2,3} =  \lambda_{2,4} = 0$ and the eigenvectors
$| \lambda_{2,1} \rangle = \left( \frac{e^{-2i\vartheta}}{\sqrt{2}}, 0,0,\frac{1}{\sqrt{2}}   \right)^\text{T}, | \lambda_{2,2} \rangle = \left( \frac{-e^{-2i\vartheta}}{\sqrt{2}}, 0, 0, \frac{1}{\sqrt{2}}      \right)^\text{T},
| \lambda_{2,3} \rangle = \left(0,1,0,0\right)^\text{T}, | \lambda_{2,4} \rangle =  \left(0,0,1,0\right)^\text{T}.$  By using them with Eq. \eqref{QFIexpression1} we obtain \eqref{GHZQFI2} for $N_q=2:$  
 \begin{align}
 \mathcal{I}_{\text{QF}} \left[  \rho^{\text{out,GHZ}_{N_q}}_{\text{MPD},\vartheta}  \right] =  2 e^{- (3+2N_\text{free}) \tau_\text{PD} } .   \label{appGHZQFIPD1} 
\end{align} 
Let us now suppose that the output density matrix is described by Eq. \eqref{outputGHZMPD1} for $N_q=k$ and show that the output density matrix for $N_q=k+1$ is also described by Eq. \eqref{outputGHZMPD1}. 
To do this, first we analyze the formula of the density matrix which describes the quantum state after the erroneous initialization operation which creates the noisy GHZ state.  
By using Eq. \eqref{outputGHZMPD1} with setting $\vartheta=N^\text{free}=0$, let us rewrite the tensor-product state of the noisy GHZ state of the $k$ qubits and  $|0\rangle$ of the $k+1$ qubit as
 \begin{align}  
	\rho^{\text{in,GHZ}_{k}}_{\text{MPD}}\otimes | 0\rangle_{ k} \langle 0|  = \left[\frac{1}{2}( | 0\rangle^{\otimes k} \langle 0| +  | 1\rangle^{\otimes k} \langle 1|) 
	+  \frac{e^{ - \frac{k(k+1)}{2} \tau^\text{PD} } }{2} \left(  | 0\rangle^{\otimes k} \langle 1| +  | 1\rangle^{\otimes k} \langle 0|   \right)\right]\otimes | 0\rangle_{ k+1} \langle 0|.  \label{inputGHZMPD1}
\end{align} 
To create the GHZ state of the $k+1$ qubits we operate the CNOT $\text{C}X[Q_{k-1};Q_{k}]$ after the operation of $\text{C}X[Q_{k-2};Q_{k-1}]$.
When we operate $\text{C}X[Q_{k-1};Q_{k}]$ on $ \rho^{\text{in,GHZ}_{k}}_{\text{MPD}}\otimes | 0\rangle_{ k+1} \langle 0| $ we ideally obtain
\begin{align} 
\check{\rho}^{\text{in,GHZ}_{2l+1}}_{\vartheta,\text{MPD}}  &\equiv 	 
 \text{C}X[Q_{k-1};Q_{k}] \cdot  \rho^{\text{in,GHZ}_{k}}_{\text{MPD}}\otimes | 0\rangle_{ k} \langle 0|    \cdot \left(      \text{C}X[Q_{k-1};Q_{k}]   \right)^\dagger \notag\\
 & = \left[ \frac{1}{2}( | 0\rangle^{\otimes k+1} \langle 0| +  | 1\rangle^{\otimes k+1} \langle 1|) 
	+   \frac{e^{ - \frac{k(k+1)}{2} \tau^\text{PD} } }{2}  \left( | 0\rangle^{\otimes k+1} \langle 1| +  | 1\rangle^{\otimes k+1} \langle 0|   \right) \right] .
	 \label{inputGHZPD2}
\end{align} 
Next let us analyze the noise effect which the $k+1$ qubits experience after the operation of $\text{C}X[Q_{k-1};Q_{k}]$.
Such a noise effect is represented by the Kraus operators $ \mathcal{M}^\text{PD}_{\vec{\alpha}} = \mathcal{M}^\text{PD}_{\alpha_0} \otimes \cdots \otimes \mathcal{M}^\text{PD}_{\alpha_{k}}$, 
where $\alpha_j=0,1$ with $j=0,1,\ldots,k$: the Kraus operators of the MPD acting independently on single-qubit states are given by $ \mathcal{M}^\text{PD}_{0} = \sqrt{\frac{1+e^{-\tau^\text{PD}}}{2}} \boldsymbol{1}_{2\times2} 
= \sqrt{\lambda_{1,1}}\boldsymbol{1}_{2\times2}$ and $\mathcal{M}^\text{PD}_{1} = \sqrt{\frac{1-e^{-\tau^\text{PD}}}{2}}Z=\sqrt{\lambda_{1,2}}Z. $
We can easily verify that the diagonal components  $| 0\rangle^{\otimes k+1} \langle 0|$ and  $| 1\rangle^{\otimes k+1} \langle 1|$ are invariant under the action of $ \mathcal{M}^\text{PD}_{\vec{\alpha}} $  and we only need to consider the 
off-diagonal terms $| 0\rangle^{\otimes k+1} \langle 1|$ and  $| 1\rangle^{\otimes k+1} \langle 0|$. The action of the Kraus operators $ \mathcal{M}^\text{PD}_{\vec{\alpha}}$ on  $| 0\rangle^{\otimes k+1} \langle 0|$  is expressed as
\begin{align}
\sum_{\vec{\alpha}}   \mathcal{M}^\text{PD}_{\vec{\alpha}}  | 0\rangle^{\otimes k+1} \langle 1|      \left(  \mathcal{M}^\text{PD}_{\vec{\alpha}}  \right)^\dagger =
\bigotimes_{j=0}^k   \sum_{\alpha_j=0}^1    \mathcal{M}^\text{PD}_{\alpha_j}  | 0\rangle_j \langle 1|      \left(  \mathcal{M}^\text{PD}_{\alpha_j}  \right)^\dagger = e^{ - (k+1)\tau^\text{PD} } | 1\rangle^{\otimes k+1} \langle 0|.
\label{inputGHZPD3}
\end{align} 
The similar analysis can be done for  $| 1\rangle^{\otimes k+1} \langle 0|$, and as a result we obtain
\begin{align}
\rho^{\text{in,GHZ}_{k+1}}_{\text{MPD},\vartheta} &\equiv 
\sum_{\vec{\alpha}}   \mathcal{M}^\text{MPD}_{\vec{\alpha}} \cdot \check{\rho}^{\text{in,GHZ}_{2l+1}}_{\vartheta,\text{MPD}} \cdot \left( \mathcal{M}^\text{MPD}_{\vec{\alpha}} \right)^\dagger \notag\\
& =  \frac{1}{2}( | 0\rangle^{\otimes k+1} \langle 0| +  | 1\rangle^{\otimes k+1} \langle 1|) 
	+  \frac{e^{ -\frac{(k+1)(k+2)}{2}} \tau^\text{PD} }{2} \left( | 0\rangle^{\otimes k+1} \langle 1| +  | 1\rangle^{\otimes k+1} \langle 0|   \right) 	 .
\label{inputGHZPD4}
\end{align} 
Finally, we analyze the noisy quantum state generated by the erroneous unitary operation $U^\text{free}_\vartheta= \prod_{l=1}^{N^\text{free}} \bigotimes_{j=0}^{k+1} R_z(-\omega \Delta t)  $.
Since the diagonal components are invariant under this erroneous operation we just have to analyze its effect on the off-diagonal components. 
By using Eqs. \eqref{inputGHZPD3}  and \eqref{inputGHZPD4}  the noisy GHZ state of the $k+1$ qubits generated by the erroneous unitary operations $U^\text{in,GHZ}$ and $U^\text{free}_\vartheta$ is expressed as
\begin{footnotesize} \begin{align}
\rho^{\text{out,GHZ}_{k+1}}_{\text{MPD},\vartheta} 
& = \frac{1}{2}( | 0\rangle^{\otimes k+1} \langle 0| +  | 1\rangle^{\otimes k+1} \langle 1|) 
+  \frac{\exp\left((k+1)i\vartheta -\left(N^\text{GHZ}_\text{PD}(k+1)+(k+1)N^\text{free}
\right) \tau^\text{PD}\right) }{2}  | 0\rangle^{\otimes k+1} \langle 1|
\notag\\
&	+  \frac{\exp\left(-(k+1)i\vartheta -\left(N^\text{GHZ}_\text{PD}(k+1)+(k+1)N^\text{free}
\right) \tau^\text{PD}\right) }{2}    | 1\rangle^{\otimes k+1} \langle 0|   	 .
\label{outGHZPD1}
\end{align} \end{footnotesize}
Consequently, we have proved Eq. \eqref{outputGHZMPD1} owing to the induction. Finally, let us derive Eq. \eqref{GHZQFI2}.
The two finite eigenvalues (all the rest are zero) of $\rho^{\text{out,GHZ}_{N_q}}_{\vartheta,\text{MPD}} $ in Eq. \eqref{outputGHZMPD1} are 
$\lambda_{N_q,1} = \frac{1+e^{-\left(N^\text{GHZ}_\text{PD}(N_q)+N^\text{free}N_q \right)\tau_{\text{PD}}} }{2}, \lambda_{N_q,2} = \frac{1+e^{-\left(N^\text{GHZ}_\text{PD}(N_q)-N^\text{free}N_q \right)\tau_{\text{PD}}} }{2}$ and the associated eigenvectors are 
$| \lambda_{N_q,1} \rangle = \left( \frac{e^{-N_qi\vartheta}}{\sqrt{2}}, 0,\ldots,0,\frac{1}{\sqrt{2}}   \right)^\text{T}, | \lambda_{N_q,2} \rangle = \left( \frac{-e^{-N_qi\vartheta} }{\sqrt{2}}, 0, \ldots,0, \frac{1}{\sqrt{2}}      \right)^\text{T}$.
By using them with Eq. \eqref{QFIexpression1} we obtain Eq. \eqref{GHZQFI2}.
\section{Derivation of Eq. \eqref{CSSQFIMAD} }\label{appendix3}   
In this section, we show the derivation of Eq. \eqref{CSSQFIMAD}. 
First, the output density matrix $\rho^{\text{out,CSS}}_{\vartheta,\text{MAD}} $ is given by 
\begin{align} 
	{\rho}^\text{out,CSS}_{\vartheta,\text{MAD}}  = 
	\left (
		\begin{array}{cc} 
		 1 - \frac{\exp\big{(}-  d_\text{tot} \tau_\text{AD} \big{)} }{2} &\frac{ 1 }{2}\exp\left( -i \vartheta - \frac{d_\text{tot}\tau_\text{AD} }{2} \right)  \\ 
	\frac{ 1 }{2}\exp\left( i \vartheta  - \frac{d_\text{tot}\tau_\text{AD} }{2} \right)	 & \frac{\exp\big{(} -d_\text{tot}\tau_\text{AD}\big{)} }{2}
		\end{array}
	\right ). \label{outputCSSAD}
\end{align} 
The eigenvalues $\Lambda_a$ and the associated eigenvectors $| \Lambda_a \rangle$ $(a=1,2)$ are 
\begin{align} 
& \Lambda_1 = \frac{1+\sqrt{e^{-2d_\text{tot}\tau_\text{AD}}-e^{-d_\text{tot}\tau_\text{AD}} +1 }}{2}, \quad \Lambda_2 = \frac{1-\sqrt{e^{-2d_\text{tot}\tau_\text{AD}}-e^{-d_\text{tot}\tau_\text{AD}} +1 }}{2}, \notag\\
& |\Lambda_1  \rangle =L^{-1}_{\Lambda_1} \left ( \begin{array}{c} 
 e^{-i\vartheta}\left[  2\sinh\left( \frac{d_\text{tot}\tau_\text{AD}}{2}\right) +\sqrt{2\cosh \left(d_\text{tot}\tau_\text{AD}\right)-1}   \right] \\
 1
\end{array}
	\right ), \notag\\
&	 |\Lambda_2  \rangle =L^{-1}_{\Lambda_2} \left ( \begin{array}{c} 
 e^{-i\vartheta}\left[  2\sinh\left( \frac{d_\text{tot}\tau_\text{AD}}{2}\right) - \sqrt{2\cosh \left(d_\text{tot}\tau_\text{AD}\right) -1}   \right] \\
 1
\end{array}
	\right ), \notag\\
& L_{\Lambda_1} = \left[   2\left(2\cosh \left(d_\text{tot}\tau_\text{AD}\right)-1\right)^{\frac{1}{2}} \left[
\left(2\cosh \left(d_\text{tot}\tau_\text{AD}\right)-1\right)^{\frac{1}{2}}+2\sinh \left( \frac{d_\text{tot}\tau_\text{AD}}{2} \right)
\right]  \right]^{-\frac{1}{2}},  \notag\\
& L_{\Lambda_2} = \left[   2\left(2\cosh \left(d_\text{tot}\tau_\text{AD}\right)-1\right)^{\frac{1}{2}} \left[
\left(2\cosh \left(d_\text{tot}\tau_\text{AD}\right)-1\right)^{\frac{1}{2}} - 2\sinh \left( \frac{d_\text{tot}\tau_\text{AD}}{2} \right)
\right]  \right]^{-\frac{1}{2}}.
 \label{CSSADeigsystems}
\end{align} 
From Eqs. \eqref{QFIexpression1} and \eqref{CSSADeigsystems} we obtain Eq. \eqref{CSSQFIMAD}.

\section{Analysis of Dicke-State Simulations }\label{Dickeappendix}  
In this section, we analyze in detail the singular behaviors of the quantum-error-mitigated CFI in Fig. \ref{resultsSDS} by using the plots in Fig. \eqref{figDickeappendix}, which are the results for $N_q=4$ and the quantum noise channel chosen to be the MPD. We investigate such behaviors by separating the numerator and the denominator in the right-hand side of the error-propagation formula \cite{toth2014quantum} which are $\left(\partial_\vartheta \langle (J^z)^2 \rangle_{\rho^\text{out}_{\text{QEM}_{\text{1st/2nd}},\vartheta}}\right)^2,$ and $\left[\Delta^2(J^z)^2\right]_{\rho^\text{out}_{\text{QEM}_{\text{1st/2nd}},\vartheta}} = \langle (J^z)^4 \rangle_{\rho^\text{out}_{\text{QEM}_{\text{1st/2nd}},\vartheta}}-\left( \langle (J^z)^2 \rangle_{\rho^\text{out}_{\text{QEM}_{\text{1st/2nd}},\vartheta}}\right)^2, $ respectively, and study each characteristic. Let us explain from the plots in Figs. \ref{figDickeappendix} (a), (b), and (c) which are the results of the $\tau^\text{PD}$-dependencies of  $ \mathcal{I}_{\text{CF}} \left[  \rho^\text{out}_{\text{QEM}_{\text{1st/2nd}},\vartheta}   \right]$,
$\left(\partial_\vartheta \langle (J^z)^2 \rangle_{\rho^\text{out}_{\text{QEM}_{\text{1st/2nd}},\vartheta}}\right)^2,$
and $\left[\Delta^2(J^z)^2\right]_{\rho^\text{out}_{\text{QEM}_{\text{1st/2nd}},\vartheta}} $, respectively. We have set $\vartheta=\frac{\pi}{100}$. 
In Fig. \ref{figDickeappendix} (a) the singular characteristic, the sign changing from positive to negative, are observed for both $ \mathcal{I}_{\text{CF}} \left[  \rho^\text{out}_{\text{QEM}_{\text{1st}},\vartheta}   \right]$ and $ \mathcal{I}_{\text{CF}} \left[  \rho^\text{out}_{\text{QEM}_{\text{2nd}},\vartheta}   \right]$.
In Fig. \ref{figDickeappendix} (b) we see that $\left(\partial_\vartheta \langle (J^z)^2 \rangle_{\rho^\text{out}_{\text{QEM}_{\text{1st/2nd}},\vartheta}}\right)^2$ decay as $\tau^\text{PD}$ gets increased, and in Fig. \ref{figDickeappendix} (c) we see the sign changing of $\left[\Delta^2(J^z)^2\right]_{\rho^\text{out}_{\text{QEM}_{\text{1st/2nd}},\vartheta}}$ as similar to $ \mathcal{I}_{\text{CF}} \left[  \rho^\text{out}_{\text{QEM}_{\text{1st/2nd}},\vartheta}   \right]$. This sign changing can be understood as follows.
The quantum-error-mitigated density matrices $\rho^\text{out}_{\text{QEM}_{\text{1st/2nd}},\vartheta}$ are Hermitian operators and can be expressed in spectral decomposition forms as $\rho^\text{out}_{\text{QEM}_{\text{1st/2nd}},\vartheta}=\sum_{\alpha=0}^{2^{N_q}-1}\lambda^\alpha_{\text{QEM}_{\text{1st/2nd}}} |\lambda^\alpha_{\text{QEM}_{\text{1st/2nd}}}\rangle \langle \lambda^\alpha_{\text{QEM}_{\text{1st/2nd}}}|,$ where the eigenvalues $\lambda^\alpha_{\text{QEM}_{\text{1st/2nd}}}$ can be either positive or negative,
and the variance $\left[\Delta^2(J^z)^2\right]_{\rho^\text{out}_{\text{QEM}_{\text{1st/2nd}},\vartheta}}$ is expressed in terms of them.
At sufficiently small $\tau^\text{PD}$ the contribution from the positive eigenvalues to $\left[\Delta^2(J^z)^2\right]_{\rho^\text{out}_{\text{QEM}_{\text{1st/2nd}},\vartheta}}$ dominates since $\rho^\text{out}_{\text{QEM}_{\text{1st/2nd}},\vartheta}$ are close to the ideal quantum states
and they take  positive values. At some point of $\tau^\text{PD}$  the contribution from the negative eigenvalues starts to become larger than that from the positive eigenvalues, and as a result $\left[\Delta^2(J^z)^2\right]_{\rho^\text{out}_{\text{QEM}_{\text{1st/2nd}},\vartheta}}$ become zero and further get negative. For sufficiently large $\tau^\text{PD}$,
we expect that the behaviors of the density matrices $\rho^\text{out}_{\text{QEM}_{\text{1st/2nd}},\vartheta}$
get close to those of (no QEM) noisy density matrices  and the contribution from the positive eigenvalues becomes dominant and $\left[\Delta^2(J^z)^2\right]_{\rho^\text{out}_{\text{QEM}_{\text{1st/2nd}},\vartheta}}$ get positive. Meanwhile, the derivative terms $\left(\partial_\vartheta \langle (J^z)^2 \rangle_{\rho^\text{out}_{\text{QEM}_{\text{1st/2nd}},\vartheta}}\right)^2$ approaches zero.
This is because the PD effects are very strong for $\tau^\text{PD} \gg1$
such that the off-diagonal elements of the quantum-error-mitigated density matrices which are described by the phase factors of $\vartheta$ get sufficiently small and their derivative terms approach zero.
As a result, $ \mathcal{I}_{\text{CF}} \left[  \rho^\text{out}_{\text{QEM}_{\text{1st/2nd}},\vartheta}   \right]$ exhibit two characteristics, the sign changing from positive to negative  and the vanishment at sufficiently large $\tau^\text{PD}$.    
Next, let us explain the results in Figs. \ref{figDickeappendix} (d), (e), and (f). 
Here we have examined the behaviors of $ \mathcal{I}_{\text{CF}} \left[  \rho^\text{out}_{\text{QEM}_{\text{1st/2nd}},\vartheta}   \right]$ in $\vartheta$ for both the ranges $\vartheta < \frac{\pi}{100}$ and $\vartheta > \frac{\pi}{100}$ ($\vartheta = \frac{\pi}{1600} \times i_{\vartheta}$ with $i_{\vartheta}= 1, 2, \ldots, 100$)
by taking  $\tau^\text{PD}=10^{-2}$. 
We see in Fig. \ref{figDickeappendix} (d) that  the quantum-error-mitigated CFI $  \mathcal{I}_{\text{CF}} \left[  \rho^\text{out}_{\text{QEM}_{\text{1st/2nd}},\vartheta}   \right]$  are negative for small-$\vartheta$ region ($\vartheta \leq \frac{\pi}{1600} \times 57$ for the first-order QEM while $\vartheta \leq \frac{\pi}{1600} \times 43$ for the second-order QEM) whereas for the larger region of $\vartheta$ ($\vartheta \geq \frac{\pi}{1600} \times 58$ and $\vartheta \geq \frac{\pi}{1600} \times 44$ for the first-order and the second-order QEM, respectively)
they become positive. Similarly, this sign changing is observed for the variances 
$\left[\Delta^2(J^z)^2\right]_{\rho^\text{out}_{\text{QEM}_{\text{1st/2nd}},\vartheta}}$ and we show them in Fig. \ref{figDickeappendix} (f).
In addition, as presented in Fig. \ref{figDickeappendix} (e) the derivative terms $\left(\partial_\vartheta \langle (J^z)^2 \rangle_{\rho^\text{out}_{\text{QEM}_{\text{1st/2nd}},\vartheta}}\right)^2$ behave as linear functions in $\vartheta$. 
As a result, our QEM protocol fails for the SDS with large $N_q$ and $\tau^\text{PD}$ and $\vartheta=\frac{\pi}{100}$ owing to the above things. The same singular characteristics (the sign changing and the vanishing at $\tau^\text{PD}\gg1$) emerge in quantum metrology for $N_q=6$ under the MPD effect and that influenced by the MAD as shown in  Fig. \ref{resultsSDS}(a), (c), (g), and (i). The origins of them can be understood by using the same argument as the one given above.   

\begin{figure*}[!t] 
\centering
\includegraphics[width=0.8\textwidth]{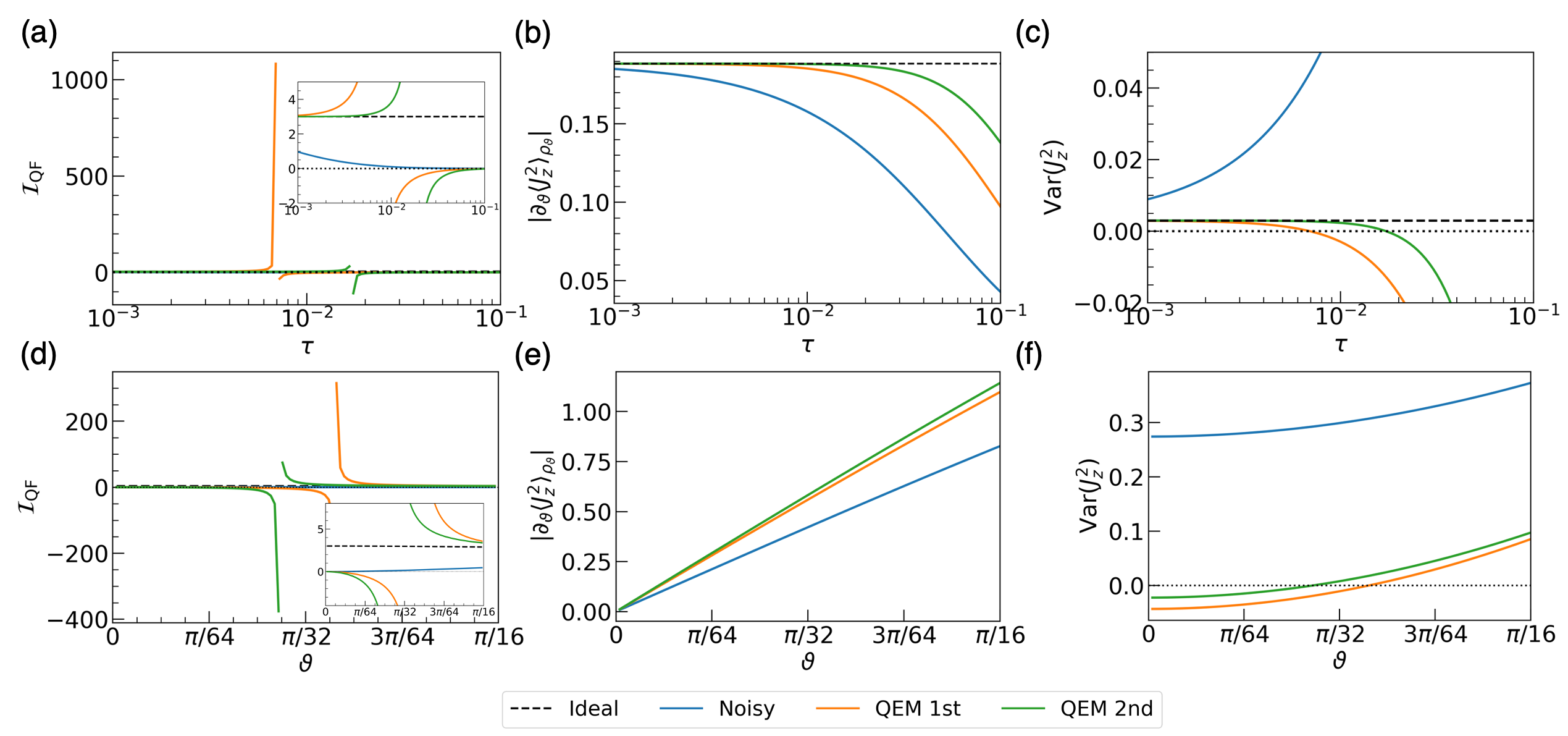}
\caption{Numerical results for $N_q=4$ with the quantum noise channel taken to be the MPD. 
(a) and (d), (b) and (e), and (c) and (f) are the plots of $ \mathcal{I}_{\text{CF}} \left[  \rho^\text{out}_{\text{QEM}_{\text{1st/2nd}},\vartheta}   \right]$,
$\left(\partial_\vartheta \langle (J^z)^2 \rangle_{\rho^\text{out}_{\text{QEM}_{\text{1st/2nd}},\vartheta}}\right)^2,$
and $\left[\Delta^2(J^z)^2\right]_{\rho^\text{out}_{\text{QEM}_{\text{1st/2nd}},\vartheta}} $, respectively. The horizontal axes in (a), (b), and (c) are taken to be $\tau^\text{PD}$ whereas those in (d), (e), and (f)  to be $\vartheta$. The plots in (a), (b), and (c) are for $\vartheta=\frac{\pi}{100}$ while those in (d), (e), and (f)  are for $\tau^\text{PD}=10^{-2}$.  }  
\label{figDickeappendix} 
\end{figure*}

\end{widetext}
\bibliographystyle{apsrev4-1} 
\bibliography{QEM_Qmetrologyref.bib}
\end{document}